\documentclass[reqno,11pt]{amsart}
\usepackage{fullpage}
\usepackage{amsfonts}
\usepackage{amssymb}
\usepackage{graphicx}
\usepackage{arydshln}
\usepackage{latexsym}
\usepackage[mathscr]{eucal}
\usepackage{cite}
\usepackage{color}
\usepackage{bm}
\usepackage{mathrsfs}
\usepackage{amsmath,amsthm,accents}
\numberwithin{equation}{section}
\DeclareMathOperator{\sech}{sech}
\DeclareMathOperator{\csch}{csch}
\usepackage{subfigure}
\numberwithin{equation}{section}

\newtheorem{Thm}{Theorem}
\def \wh#1{\widehat{#1}}
\def \wt#1{\widetilde{#1}}
\def \wb#1{\overline{#1}}

\def \dwh{\underaccent{{\cc@style\widehat{\mskip10mu}}}}
\def \dt#1{\underaccent{\tilde}{#1}}

\def \cdd#1{\accentset{\bullet}{#1}}

\newcommand{\nn}{\nonumber}

\newcommand{\Ga}{\Gamma}

\newcommand{\al}{\alpha}
\newcommand{\be}{\beta}
\newcommand{\st}{\hbox{\tiny\it{T}}}

\begin{document}
	
\title{Solutions and continuum limits to nonlocal discrete sine-Gordon equations: bilinearization reduction method}	
\author{Xiao-bo Xiang$^{1}$,~~Song-lin Zhao$^{1*}$,~~Ying Shi$^{2}$\\
\\\lowercase{\scshape{
$^{1}$ Department of Applied Mathematics, Zhejiang University of Technology,
Hangzhou 310023, P.R. China}}
\\\lowercase{\scshape{
$^{2}$ Department of Mathematics, Zhejiang University of Science and Technology,
Hangzhou 310023, P.R. China}}}
\email{Corresponding Author: songlinzhao@zjut.edu.cn}

\begin{abstract}

In this paper, we investigate local and nonlocal reductions of a discrete negative order Ablowitz-Kaup-Newell-Segur equation.
By the bilinearization reduction method, we construct exact solutions in double Casoratian form to the reduced nonlocal discrete sine-Gordon equations.
Then, nonlocal semi-discrete sine-Gordon equations and their solutions are obtained through the continuum limits.
The dynamics of soliton solutions are analyzed and illustrated by asymptotic analysis.
The research ideas and methods in this paper can be generalized to other nonlocal discrete integrable systems.

\end{abstract}


\maketitle

\section{Introduction}
\label{sec-1}

The nonlocal integrable systems, dating back to the appearance of
nonlocal $\mathcal {PT}$-symmetric nonlinear Schr\"{o}dinger (NLS) equation \cite{AbMu-2013},
exert important roles in mathematics and physics.
In mathematics, they usually possess many interesting properties, such as multi-soliton solutions,
Lax integrability and an infinite number of conservation laws \cite{Fokas}.
As for physics, the nonlocal integrable systems have possible applications in multi-place events \cite{Lou-JMP, Lou-CTP}
and other areas of physics, such as quantum chromodynamics \cite{MPW}, electric circuits \cite{LSEK}, optics \cite{RMGCSK, MMGC},
Bose-Einstein condensates \cite{DGPS}, and so forth.
Following the establishment of nonlocal $\mathcal {PT}$-symmetric NLS equation, many other new nonlocal integrable equations have been proposed.
Examples include the nonlocal Korteweg-de Vries equation \cite{Lou-SR}, the reverse space-time modified Korteweg-de Vries
(mKdV) equation \cite{JZ-2017}, the reverse space-time sine-Gordon (sG) equation \cite{AFLM-SAPM},
the reverse space-time NLS equation \cite{AFLM-2018}, the nonlocal derivative NLS equation \cite{SSZ-2019},
the fully $\mathcal {PT}$-symmetric and partially $\mathcal {PT}$-symmetric Davey-Stewartson equations \cite{Zhou-SAPM, RZFH}, and many others.
In addition, many researchers have successfully extended certain known effective approaches
to solve the nonlocal integrable systems,
such as the inverse scattering transform \cite{AFLM-2018, AM-Nonl-2016}, Darboux transformations \cite{LiX-PRE-2015, SXZ},
the bilinear method \cite{XU-AML-2016, CDLZ, GP-CNSNS} and the Cauchy matrix approach \cite{FZ-ROMP}.

To date, many interesting results on nonlocal integrable systems have been obtained.
Inspired by Ablowitz and Musslimani's work, for example, a new unified two-parameter wave model,
connecting integrable local and nonlocal vector NLS equations,
was investigated. This model is in possession of a Lax pair and an infinite number of conservation laws and is $\mathcal {PT}$-symmetric \cite{Yan}.
In light of the relations between nonlocal and local integrable equations, not only the integrability of nonlocal
equations was established immediately but also their Lax pairs and analytical solutions were constructed from those of the local equations \cite{YY-SAPM}.
With the assumption of stationary solution, general Jacobi elliptic-function and hyperbolic-function
solutions were obtained for a nonlocal NLS equation \cite{XCLM}, in which the bounded cases obey either the $\mathcal {PT}$- or
anti-$\mathcal {PT}$-symmetric relation.
It turns out that the focusing nonlocal NLS equation has four types of bounded Jacobi elliptic-function solutions,
as well as the bright- and dark-soliton solutions.
Besides the nonlocal continuous integrable systems, there is also some progress in the nonlocal semi-discrete integrable systems.
For instance, by discretizing the transverse coordinate $x$ into the discrete lattice sites, a semi-discrete version of the
$\mathcal {PT}$-symmetric NLS equation was introduced in \cite{SMMC}, where its integrability and $\mathcal {PT}$-symmetry breaking conditions were identified.
By the Ablowitz-Ladik scattering problem, Ablowitz and Musslimani proposed
an integrable nonlocal semi-discrete NLS equation \cite{AM-2014}, whose
linear Lax pair formulation, Hamiltonian properties and action-angle variables were investigated in \cite{Ger}.
Furthermore, various aspects of nonlocal semi-discrete integrable models on exact solutions and gauge equivalence
have been performed \cite{DLZ-AMC, MZ-JMP, MSZ-AML}.

Since the information of the whole hierarchy of integrable partial differential equations can be encoded into an integrable partial
difference equation in an implicit way,
the integrable partial difference equations (namely discrete integrable systems) have much richer structures (see \cite{N-2004-math}).
Most of the mathematical methods in the theory of integrable systems are invented based on the differential
operators, so usually they are no longer directly effective for the studies of discrete equations (relating to difference operators).
Therefore, to deeply understand the intrinsic structure of nonlocal discrete systems, it is essential to develop classic methods.
For the first paper on nonlocal discrete integrable systems, one can refer to \cite{ZKZ}, where two types of nonlocal discrete
integrable equations were investigated as reductions of two-component Adler-Bobenko-Suris systems (cf. \cite{Bri}).
In particular, a reverse-$(n,m)$ nonlocal H1 equation and a reverse-$n$ nonlocal H1 equation were shown explicitly.
Quite recently, nonlocal complex reduction of a discrete negative order Ablowitz-Kaup-Newell-Segur (AKNS) equation was studied in \cite{XFZ-TMPH},
where a nonlocal complex discrete sG equation, as well as its exact solutions in Cauchy matrix type, was constructed.
We refer the reader to the recent survey \cite{XZ-TMPH} and
the references therein for more details on the Cauchy matrix solutions of nonlocal complex integrable equations.

The AKNS system is usually indispensable to the studies of nonlocal integrable systems.
Among all discrete versions of the AKNS system, one is the following discrete (non-potential) negative order AKNS (dnAKNS) system
\begin{subequations}
\label{dnAKNS}
\begin{align}
& 4\big(u+\wh{\wt{u}}-(\wt{u}+\wh{u})w\big)
=\delta\epsilon\big(u+\wh{\wt{u}}+(\wt{u}+\wh{u})w\big),\\
& 4\big(v+\wh{\wt{v}}-(\wt{v}+\wh{v})w\big)
=\delta\epsilon\big(v+\wh{\wt{v}}+(\wt{v}+\wh{v})w\big),\\
\label{dnAKNS-c}
& (1+\epsilon^{2}\wh{u}\wh{v})w=(1+\epsilon^{2}uv)\dt{w}.
\end{align}
\end{subequations}
The notations adopted in \eqref{dnAKNS} are as follows: all dependent variables $u,~v$ and $w$ are functions of discrete
coordinates $(n,m)\in \mathbb{Z}^2$, e.g., $u=u_{n, m}$; the difference operations $u\mapsto \wt{u}$ and $u\mapsto \wh{u}$
denote elementary shifts in the two directions of the lattice, i.e., $\wt{u}=u_{n+1,m}, \wh{u}=u_{n,m+1}$, while for the down
and combined shifts we have $\dt{u}=u_{n-1,m}$ and $\wh{\wt{u}}=u_{n+1,m+1}$.
Aside from that, $\epsilon$ and $\delta$ are continuous lattice parameters associated with the grid size in the directions
of the lattice given by the independent variables $n$ and $m$, respectively.
System \eqref{dnAKNS} \cite{Yu-SAPM} was established by discretizing the bilinear operators of the usual negative order AKNS equation \cite{ZJZ-PD,KK},
which can be transformed back to the negative order AKNS equation under appropriate continuum limits.

Motivated by the understanding of relations between discrete and continuous integrable systems, we shall explore nonlocal
reductions of the dnAKNS equation \eqref{dnAKNS} through the so-called bilinearization reduction method,
which was originally proposed in \cite{DLZ-AMC, CDLZ} to solve the nonlocal integrable models reduced from the AKNS
hierarchy \cite{AKNS-1973, AKNS-1974}.
This method involves, first taking appropriate reductions to get the nonlocal
integrable systems, and second solving the matrix equation algebraically to derive the exact solutions.
By employing the bilinearization reduction method, the present paper will be devoted to deriving some nonlocal
discrete sG equations, as well as their exact solutions in double Casoratian form, from the dnAKNS equation \eqref{dnAKNS}.

The paper is organized as follows.
Section 2 presents double Caosratian solutions to the dnAKNS equation \eqref{dnAKNS}.
In Section 3, we firstly consider local and nonlocal reduction of the dnAKNS equation \eqref{dnAKNS} in the real case.
Next, multi-soliton solutions and Jordan-block solutions for the resulting real nonlocal discrete sG equation
along with their basic analytical features are exhibited.
And then, continuum limits, including semi-continuous limits and continuous limits, are discussed emphatically.
In Section 4, complex nonlocal discrete sG equation reduced from the dnAKNS equation \eqref{dnAKNS} is investigated, as well as the
corresponding one-soliton solution, dynamics and continuum limits.
Section 5 is for conclusions and some remarks.

\section{Double Casoratian solutions to the dnAKNS equation \eqref{dnAKNS}}

In this section, we show how to construct double Casoratian solutions for the dnAKNS equation \eqref{dnAKNS}.
To begin we introduce some notations on the double Wronskian/Casoratian, which will be
used in the rest part of the present paper.

For the basic column vectors $\phi=(\phi_1, \phi_2,\cdots,\phi_{N+M+2})^{\st}$ and $\psi=(\psi_1, \psi_2,\cdots,\psi_{N+M+2})^{\st}$
with continuous spatial variable $x$, the $(N+M+2)\times (N+M+2)$ double Wronskian is defined as
\begin{align}
\label{Wro-def}
W^{(N,M)}(\phi,\psi)& = |\phi,\phi^{(1)},\cdots,\phi^{(N)};\psi,\psi^{(1)},\cdots,\psi^{(M)}| \nn \\
& =|0,1,\cdots,N;0,1,\cdots,M|=|\wh{N};\wh{M}|,
\end{align}
where $\phi^{(l)}=\frac{\partial^l \phi}{\partial x^l}$, $\psi^{(l)}=\frac{\partial^l \psi}{\partial x^l}$, and we use the compact
form as given in \cite{FN-1983}.
Here $\wh{N}$ indicates the set of consecutive columns $0,1,\cdots,N$.
The Casoratian is a discrete version of the Wronskian.
Let $E$ be a shift operator defined as
\begin{align}
\label{E-shift}
E^jf_{n,m}=f_{n+j,m}, \quad (j \in \mathbb{Z}).
\end{align}
For the basic column vectors $\Phi=(\Phi_1, \Phi_2,\cdots,\Phi_{N+M+2})^{\st}$ and
$\Psi=(\Psi_1, \Psi_2,\cdots,\Psi_{N+M+2})^{\st}$ with the discrete independent variable $n$, the double Casoratian is defined as
\begin{align}
\mbox{C}^{(N,M)}(\Phi, \Psi)& =|\Phi, E^2\Phi, \ldots, E^{2N}\Phi; \Psi, E^2\Psi, \ldots, E^{2M}\Psi| \nn \\
& =|0, 1,\ldots, N;0,1,\ldots, M|=|\wh{N};\wh{M}|.
\end{align}

\subsection{Bilinearization and double Casoratian solutions}

To proceed, we first recall bilinearization of the system \eqref{dnAKNS} as shown in \cite{Yu-SAPM}.
Through the transformation of dependent variables
\begin{align}
\label{dnAKNS-uvw-tran}
u=g/f,\quad v=h/f,\quad
w=\wt{f}\wh{f}/(f\wh{\wt{f}}),
\end{align}
the system \eqref{dnAKNS} can be transformed into
\begin{subequations}
\label{dnAKNS-bili}
\begin{align}
& 4(\wh{\wt{g}}f-\wt{g}\wh{f}-\wh{g}\wt{f}+g \wh{\wt{f}})=\delta\epsilon
  (\wh{\wt{g}}f+\wt{g}\wh{f}+\wh{g}\wt{f}+g\wh{\wt{f}}), \\
& 4(\wh{\wt{h}}f-\wt{h}\wh{f}-\wh{h}\wt{f}+h\wh{\wt{f}})=\delta\epsilon
(\wh{\wt{h}}f+\wt{h}\wh{f}+\wh{h}\wt{f}+h\wh{\wt{f}}), \\
& \wt{f}\dt{f}-f^2=\epsilon^{2}gh.
\end{align}
\end{subequations}
And then, it can be directly checked that the bilinear equations \eqref{dnAKNS-bili} are invariant under the gauge transformation
\begin{align}
f\rightarrow f~\mbox{exp}(\alpha_0 n+\beta_0 m), \quad
g\rightarrow g~\mbox{exp}(\alpha_0 n+\beta_0 m), \quad
h\rightarrow h~\mbox{exp}(\alpha_0 n+\beta_0 m),
\end{align}
where $\alpha_0$ and $\beta_0$ are two constants.

The crucial point of applying the Casoratian technique is that the condition equations hold for each
Casoratian entry. Thus let us focus on the following condition equation set (CES)
\begin{subequations}
\label{dnAKNS-CES}
\begin{align}
& \wt{\Phi}=A\Phi,\quad \wh{\Phi}=[((4+\delta\epsilon)E^{2}-(4-\delta\epsilon))
((4-\delta\epsilon)E^{2}-(4+\delta\epsilon))^{-1}]^{\frac{1}{2}}\Phi,\\
& \wt{\Psi}=A^{-1}\Psi,\quad
\wh{\Psi}=[((4+\delta\epsilon)E^{2}-(4-\delta\epsilon))
((4-\delta\epsilon)E^{2}-(4+\delta\epsilon))^{-1}]^{\frac{1}{2}}\Psi,
\end{align}
\end{subequations}
where $A$ is an arbitrary invertible complex constant matrix of order $N+M+2$ and $E$ is the shift operator defined by \eqref{E-shift}.
Then double Casoratian solutions of the bilinear equations \eqref{dnAKNS-bili} can be described in the following theorem.
\begin{Thm}
The double Casorati determinants
\begin{align}
\label{dnAKNS-dCs}
f=|\wh{N};\wh{M}|,\quad g=(1/\epsilon)|\wh{N+1};\wh{M-1}|, \quad h=(1/\epsilon) |\wh{N-1};\wh{M+1}|,
\end{align}	
solve the bilinear system \eqref{dnAKNS-bili}, provided that the basic column vectors $\Phi$ and $\Psi$ are given by the CES \eqref{dnAKNS-CES}.
\end{Thm}
By solving the CES \eqref{dnAKNS-CES}, we know that the basic column vectors $\Phi$ and $\Psi$ are expressed as
\begin{subequations}
\label{dnKANS-Phsi-A}
\begin{align}
& \Phi=A^n[((4+\delta\epsilon)A^{2}-(4-\delta\epsilon)I)
((4-\delta\epsilon)A^{2}-(4+\delta\epsilon)I)^{-1}]^{\frac{m}{2}}C^{+},\\
& \Psi=A^{-n}[((4+\delta\epsilon)A^{2}-(4-\delta\epsilon)I)
((4-\delta\epsilon)A^{2}-(4+\delta\epsilon)I)^{-1}]^{-\frac{m}{2}}C^{-},
\end{align}
\end{subequations}
where $C^{\pm}=(c_{1}^{\pm},c_{2}^{\pm},\dots c_{N+1}^{\pm};d_{1}^{\pm},d_{2}^{\pm},\dots d_{M+1}^{\pm})^{\st}$ are constant column vectors.
Here and hereafter $I$ is the unit matrix whose index indicating the size is omitted.
We denote \eqref{dnAKNS-dCs} by $f(\Phi,\Psi)$, $g(\Phi,\Psi)$ and $h(\Phi,\Psi)$ when their entries are taken as \eqref{dnKANS-Phsi-A}.
For the convenience of calculation, \eqref{dnKANS-Phsi-A} is written by replacing $A$ by $e^{K}$ as
\footnote{
The exponential of a matrix is always an invertible matrix.
The inverse matrix of $e^X$ is given by $e^{-X}$. The matrix exponential then gives a map
\[\displaystyle \exp \colon M_{s}(\mathbb {C})\to \mathrm {GL} (s,\mathbb {C})\]
from the space of all $s\times s$ matrices to the general linear group of degree $s$, i.e. the group of all $s\times s$
invertible matrices. In fact, this map is surjective which means that every invertible complex matrix can be written as the exponential
of some other matrix (cf. \cite{Hall}).}
\begin{subequations}
\label{dnKANS-Phsi-K}
\begin{align}
& \Phi=e^{Kn}[(4+\delta\epsilon\coth K)(4-\delta \epsilon \coth K)^{-1}]^{\frac{m}{2}}C^{+}, \\
& \Psi=e^{-Kn}[(4+\delta\epsilon\coth K)(4-\delta \epsilon \coth K)^{-1}]^{-\frac{m}{2}}C^{-}.
\end{align}
\end{subequations}

\subsection{Similarity invariance of exact solutions}

Note that in the basic column vectors \eqref{dnKANS-Phsi-K}, $K$ is an arbitrary constant matrix.
To show the similarity invariance of exact solutions, we suppose $\wb{K}$ is
any matrix that is similar to $K$, i.e.,
\begin{align}
\label{Ga-K}
\wb{K}=TKT^{-1},
\end{align}
where $T$ is the transform matrix. By taking \eqref{Ga-K} into \eqref{dnKANS-Phsi-K},
and denoting $\wb{C}^{\pm}=TC^{\pm}$, we naturally get the following two new basic column vectors
\begin{subequations}
\label{dnKANS-Phsi-wbK}
\begin{align}
& \wb{\Phi}=T\Phi=e^{\wb{K} n}[(4+\delta\epsilon\coth \wb{K})(4-\delta \epsilon \coth \wb{K})^{-1}]^{\frac{m}{2}}\wb{C}^{+}, \\
& \wb{\Psi}=T\Psi=e^{-\wb{K} n}[(4+\delta\epsilon\coth \wb{K})(4-\delta \epsilon \coth \wb{K})^{-1}]^{-\frac{m}{2}}\wb{C}^{-}.
\end{align}
\end{subequations}
Since for the arbitrariness of matrix $A$, it is apparent that $\wb{\Phi}$ and $\wb{\Psi}$ still satisfy the CES \eqref{dnAKNS-CES}
with $A=e^{\wb{K}}$.
Noticing the relations $f(\wb{\Phi},\wb{\Psi})=|T|f(\Phi,\Psi)$, $g(\wb{\Phi},\wb{\Psi})=|T|g(\Phi,\Psi)$
and $h(\wb{\Phi},\wb{\Psi})=|T|h(\Phi,\Psi)$, as well as the transformation \eqref{dnAKNS-uvw-tran},
one can easily find that $(f(\wb{\Phi},\wb{\Psi}),g(\wb{\Phi},\wb{\Psi}),h(\wb{\Phi},\wb{\Psi}))$
and $(f(\Phi,\Psi),g(\Phi,\Psi),h(\Phi,\Psi))$ lead to same solutions for the dnAKNS equation \eqref{dnAKNS}.
Now let us start from the following $\Phi$ and $\Psi$:
\begin{subequations}
\label{dnKANS-Phsi-Ga}
\begin{align}
& \Phi=e^{\Ga n}[(4+\delta\epsilon\coth \Ga)(4-\delta \epsilon \coth \Ga)^{-1}]^{\frac{m}{2}}C^{+}, \\
& \Psi=e^{-\Ga n}[(4+\delta\epsilon\coth \Ga)(4-\delta \epsilon \coth \Ga)^{-1}]^{-\frac{m}{2}}C^{-},
\end{align}
\end{subequations}
where $K$ in \eqref{dnKANS-Phsi-K} is replaced by its Jordan canonical form $\Ga$.
Then various exact solutions, including soliton solutions, Jordan-block solutions, rational solutions and mixed solutions,
for the dnAKNS system \eqref{dnAKNS} can be derived in terms of different eigenvalue structures of matrix $\Ga$.

\section{Real reduction: solutions and continuum limits}

In this section, we adopt the bilinearization reduction method to reduce the dnAKNS equation \eqref{dnAKNS}
in the real sense. By taking appropriate reduction, the real local and nonlocal discrete sG equation along with
some exact solutions is derived.
Moreover, the continuum limits are investigated to construct two real
local and nonlocal semi-discrete sG equations as well as a
real local and nonlocal continuous sG equation.
In what follows, for the function $f:=f(x_1,x_2)$, notation $f_\sigma$ means $f_\sigma=f(\sigma x_1, \sigma x_2)$,
where $\sigma=\pm 1$.
For $\sigma=1$, the notation $f_{\sigma}=f$ should not be confused with the component $f_1$ in matrix.
When both the independent variables $x_1$ and $x_2$ are discrete, referred to as
the discrete case, it is necessary to figure out
$\wt{f}_{\sigma}=f(\sigma x_1+\sigma,\sigma x_2)$, $\wh{f}_{\sigma}=f(\sigma x_1,\sigma x_2+\sigma)$
and $\wh{\wt{f}}_{\sigma}=f(\sigma x_1+\sigma,\sigma x_2+\sigma)$, respectively.

\subsection{Real reduction}

The system \eqref{dnAKNS} admits real reduction
\begin{align}
\label{real-Re}
v=\eta u_\sigma, \quad \eta, \sigma=\pm 1.
\end{align}	
It is worth noting that equation \eqref{dnAKNS-c} implies
\begin{align}
\label{dnAKNS-c-rw}
w(n,m)=\prod_{j=n_0}^{n-1}\frac{1+\epsilon^{2}u(j,m)v(j,m)}{1+\epsilon^{2}\wh{u}(j,m)\wh{v}(j,m)},
\quad n_0\in \mathbb{Z},
\end{align}
which possesses the nonautonomous structure (see \cite{SRH, GR}).
Imposing \eqref{real-Re} into \eqref{dnAKNS-c-rw} gives rise to $w=w_{\sigma}$.
Thus the real local and nonlocal discrete sG (rndsG) equation reads
\begin{subequations}
\label{rndsG}
\begin{align}
& 4\big(u+\wh{\wt{u}}-(\wt{u}+\wh{u})w\big)
=\delta\epsilon\big(u+\wh{\wt{u}}+(\wt{u}+\wh{u})w \big), \\
\label{rndsG-rb}
& (1+\eta\epsilon^{2}\wh{u}\wh{u}_{\sigma})w=(1+\eta\epsilon^{2}uu_{\sigma})\dt{w}.
\end{align}
\end{subequations}
When $\sigma=1$, equation \eqref{rndsG} is the real local discrete sG equation,
while when $\sigma=-1$, equation \eqref{rndsG} is the real nonlocal discrete sG equation.
It is obvious that the equation \eqref{rndsG} is preserved under transformation $u\rightarrow -u$.
Besides, equation \eqref{rndsG} with $(\sigma,\eta)=(\pm 1,1)$ and with $(\sigma,\eta)=(\pm 1,-1)$
can be transformed into each other by taking $u\rightarrow iu$.

For the sake of constructing exact solutions to equation \eqref{rndsG}, it is necessary to
impose suitable constraint on the pair $(\Phi,\Psi)$ in double Casoratian \eqref{dnAKNS-dCs} so that the
transformation \eqref{dnAKNS-uvw-tran} coincides with the reduction \eqref{real-Re}.
To this end, we take $M=N$. Due to arbitrariness of vectors $C^{\pm}$,
we replace $C^{\pm}$ by $e^{\mp N\Ga}C^{\pm}$ in \eqref{dnKANS-Phsi-Ga} and pay attention to the following double Casorati determinants:
\begin{align*}
f=|e^{-N\Ga}\widehat{\Phi^{(N)}};e^{N\Ga}\widehat{\Psi^{(N)}}|,\;	
g=\frac{1}{\epsilon} |e^{-N\Ga}\widehat{\Phi^{(N+1)}};e^{N\Ga}\widehat{\Psi^{(N-1)}}|,\;
h=\frac{1}{\epsilon}
|e^{-N\Ga}\widehat{\Phi^{(N-1)}};e^{N\Ga}\widehat{\Psi^{(N+1)}}|.
\end{align*}

The following theorem shows double Casoratian solutions to the rndsG equation \eqref{rndsG}.
\begin{Thm}
\label{Thm-rndsG-solu}
The functions $u=g/f$ and $w=\wt{f}\wh{f}/(f\wh{\wt{f}})$ with
\begin{align}
\label{rndsG-solu}
\quad f=|e^{-N\Ga}\widehat{\Phi^{(N)}};e^{N\Ga}\widehat{\Psi^{(N)}}|,\quad
g=(1/\epsilon)|e^{-N\Ga}\widehat{\Phi^{(N+1)}};e^{N\Ga}\widehat{\Psi^{(N-1)}}|,
\end{align}
solve the rndsG equation \eqref{rndsG}, if the $(2N+2)$-th order column vectors
$\Phi$ and $\Psi$ defined by \eqref{dnKANS-Phsi-Ga} satisfy the following relation
\begin{align}
\label{rndsG-Phsi-T}
\Psi=T\Phi_\sigma,
\end{align}
where $T\in\mathbb{C}^{(2N+2)\times(2N+2)}$ is a constant matrix satisfying
\begin{align}
\label{rndsG-AT}
\Ga T+\sigma T\Ga=\bm 0,\quad T^{2}=\left\{
\begin{array}{l}
-\eta I, \quad \mbox{with} \quad \sigma=1,\\
\eta|e^{\Ga}|^{2}I, \quad \mbox{with} \quad \sigma=-1,
\end{array}\right.
\end{align}
and $C^{-}=TC^{+}$.
\end{Thm}

The first equation in \eqref{rndsG-AT} is nothing but the famous Sylvester equation (cf. \cite{Syl,BR}).
The idea of the verification for Theorem \ref{Thm-rndsG-solu} is using the constraints \eqref{rndsG-Phsi-T} and
\eqref{rndsG-AT} to determine the relationship between $(f,h)$ and $(f_{\sigma},g_{\sigma})$, leading to
\eqref{real-Re}. Here we just present the verification for the case of $\sigma=-1$.

\begin{proof}
To begin, we substitute \eqref{rndsG-Phsi-T} into $f$, yielding
\begin{align}
f=|e^{-N\Ga}\Phi,e^{(-N+2)\Ga}\Phi,\dots,e^{N\Ga}\Phi;e^{N\Ga}T\Phi_{-1},
e^{(N-2)\Ga}T\Phi_{-1},\dots,e^{-N\Ga}T\Phi_{-1}|.	
\end{align}
In terms of \eqref{rndsG-AT}, one can rearrange $f$ as
\begin{align*}
f &=|e^{-N\Ga}\Phi,e^{(-N+2)\Ga}\Phi,\dots,e^{N\Ga}\Phi;Te^{N\Ga}\Phi_{-1},
Te^{(N-2)\Ga}\Phi_{-1},\dots,Te^{-N\Ga}\Phi_{-1}| \nn\\
& =(\eta|e^{\Ga}|^{-2})^{N+1}|T||Te^{-N\Ga}\Phi,Te^{(-N+2)\Ga}\Phi,\ldots,Te^{N\Ga}\Phi;
e^{N\Ga}\Phi_{-1},e^{(N-2)\Ga}\Phi_{-1},\ldots,e^{-N\Ga}\Phi_{-1}|\nn\\
& =(-\eta|e^{\Ga}|^{-2})^{N+1}|T||e^{N\Ga}\Phi_{-1},e^{(N-2)\Ga}\Phi_{-1},\ldots,e^{-N\Ga}\Phi_{-1};
Te^{-N\Ga}\Phi,Te^{(-N+2)\Ga}\Phi,\ldots,Te^{N\Ga}\Phi|\nn\\
& =(-\eta|e^{\Ga}|^{-2})^{N+1}|T||e^{N\Ga}\Phi_{-1},e^{(N-2)\Ga}\Phi_{-1},\ldots,e^{-N\Ga}\Phi_{-1};
e^{-N\Ga}T\Phi,e^{(-N+2)\Ga}T\Phi,\ldots,e^{N\Ga}T\Phi|\nn\\
& =(-\eta|e^{\Ga}|^{-2})^{N+1}|T||e^{-N\Ga}\Phi_{-1},e^{(-N+2)\Ga}\Phi_{-1},\ldots,e^{N\Ga}\Phi_{-1};
e^{N\Ga}T\Phi,e^{(N-2)\Ga}T\Phi,\ldots,e^{-N\Ga}T\Phi|\nn\\
& =(-\eta|e^{\Ga}|^{-2})^{N+1}|T|f_{-1}.
\end{align*}	
A similar calculation leads to $h=\eta^{-1}(-\eta|e^{\Ga}|^{-2})^{N+1}|T|g_{-1}$.
With the help of transformation \eqref{dnAKNS-uvw-tran}, we arrive at
\begin{align}
v=\dfrac{h}{f}=\dfrac{\eta^{-1}(-\eta|e^{\Ga}|^{-2})^{N+1}|T|g_{-1}}
{(-\eta|e^{\Ga}|^{-2})^{N+1}|T|f_{-1}}=\dfrac{g_{-1}}{\eta f_{-1}}=\eta u_{-1},
\end{align}	
which coincides with the reduction \eqref{real-Re} for the rndsG equation \eqref{rndsG}.
\end{proof}

Inserting \eqref{rndsG-Phsi-T} into \eqref{rndsG-solu} manifests that
exact solutions to equation \eqref{rndsG} can then be described as
$u=g/f$ and $w=\wt{f}\wh{f}/(f\wh{\wt{f}})$ with
\begin{align}
f=|e^{-N\Ga}\widehat{\Phi^{(N)}};e^{N\Ga}T\widehat{\Phi_{\sigma}^{(-N)}}|,\quad
g=(1/\epsilon)|e^{-N\Ga}\widehat{\Phi^{(N+1)}};e^{N\Ga}T\widehat{\Phi_{\sigma}^{(-N+1)}}|,
\end{align}
where $T$ and $\Ga$ satisfy matrix equations \eqref{rndsG-AT}.

\subsection{Some examples of solutions}

The aim of this part is to give some types of exact solutions to the nonlocal rndsG equation ($\sigma=-1$).
The key step is to solve the matrix equations \eqref{rndsG-AT}.
Without loss of generality, we just consider the case of $\eta=1$.
For simplicity, solutions $\Ga$ and $T$ are taken as block matrices
\begin{align}
\label{Ga-T-ex}
\Ga=\left(
\begin{array}{cc}
L & \bm 0  \\
\bm 0 & -L
\end{array}\right),\quad
T=\left(
\begin{array}{cc}
I & \bm 0  \\
\bm 0 & -I
\end{array}\right),
\end{align}	
where $L$ is Jordan canonical matrix.
For further analysis, we need to distinguish two forms of matrix $L$: diagonal form and Jordan-block form,
leading to multi-soliton solutions and Jordan-block solutions, respectively.
Notations
\begin{align}
\label{xi}
\xi_j=k_jn+\tau_j m, \quad e^{\tau_j}=\bigg(
\dfrac{4+\delta\epsilon\coth k_j}{4-\delta \epsilon \coth k_j}\bigg)^{\frac{1}{2}}, \quad j=1,2,\ldots,N+1
\end{align}	
are introduced, where we assume $(\delta\epsilon \coth k_j)^2<16$ to guarantee the real property of $\tau_j$.

Due to the block structure of matrix $T$, we note that $C^{+}$ can be gauged to be
$\bar{I}=(1,1,\dots,1;1,1,\dots,1)^{\st}$ or $\breve{I}=(1,0,\dots,0;1,0,\dots,0)^{\st}$.
It implies that the solutions for equation \eqref{rndsG} are independent of phase parameters $C^{+}$, i.e., the initial phase has always to be 0.

\vspace{.2cm}
\noindent{\it Soliton solutions}: Let $L$ be a diagonal matrix composed by distinct real nonzero eigenvalues $k_j$, given by
\begin{align}
\label{drL-Diag}
L=\mathrm{Diag}(k_{1},k_{2},\dots,k_{N+1}), \quad |L| \neq 0, \quad k_i \neq k_j, \quad (i \neq j).
\end{align}	
Here we take $C^{+}=\bar{I}$, so that $\Phi$ is obtained as
\begin{align*}
\Phi_{j}=\left\{
\begin{aligned}
& e^{\xi_j},\quad j=1, 2, \ldots, N+1, \\
& e^{-\xi_{s}},\quad j=N+1+s,\quad s=1,2,\ldots,N+1.
\end{aligned}	
\right.
\end{align*}

In the case of $N=0$, we have one-soliton solution
\begin{subequations}
\label{dr-1ss}
\begin{align}
\label{dr-1ss-u}
& u=(\sinh 2k_{1}\sech 2\xi_1)/\epsilon, \\
& w=\cosh2(k_1+\xi_1)\cosh2(\tau_1+\xi_1)\sech2\xi_1\sech2(k_1+\tau_1+\xi_1).
\end{align}
\end{subequations}
Solution $u$ in \eqref{dr-1ss-u} describes a stable wave with the travelling velocity $-\tau_1/k_1$,
which is single-peaked and unidirectional.
Its height relative to $u=0$ is approximately equals to $\sinh 2k_{1}/\epsilon$, and its width is proportional to $(2k_1)^{-1}$.
In view of the sign of the parameter $k_1\epsilon$, the ``amplitude'' of the wave can be positive or negative,
which corresponds to soliton or anti-soliton, as depicted in Fig. 1.
\vskip20pt
\begin{center}
\begin{picture}(120,80)
\put(-160,-23){\resizebox{!}{4.0cm}{\includegraphics{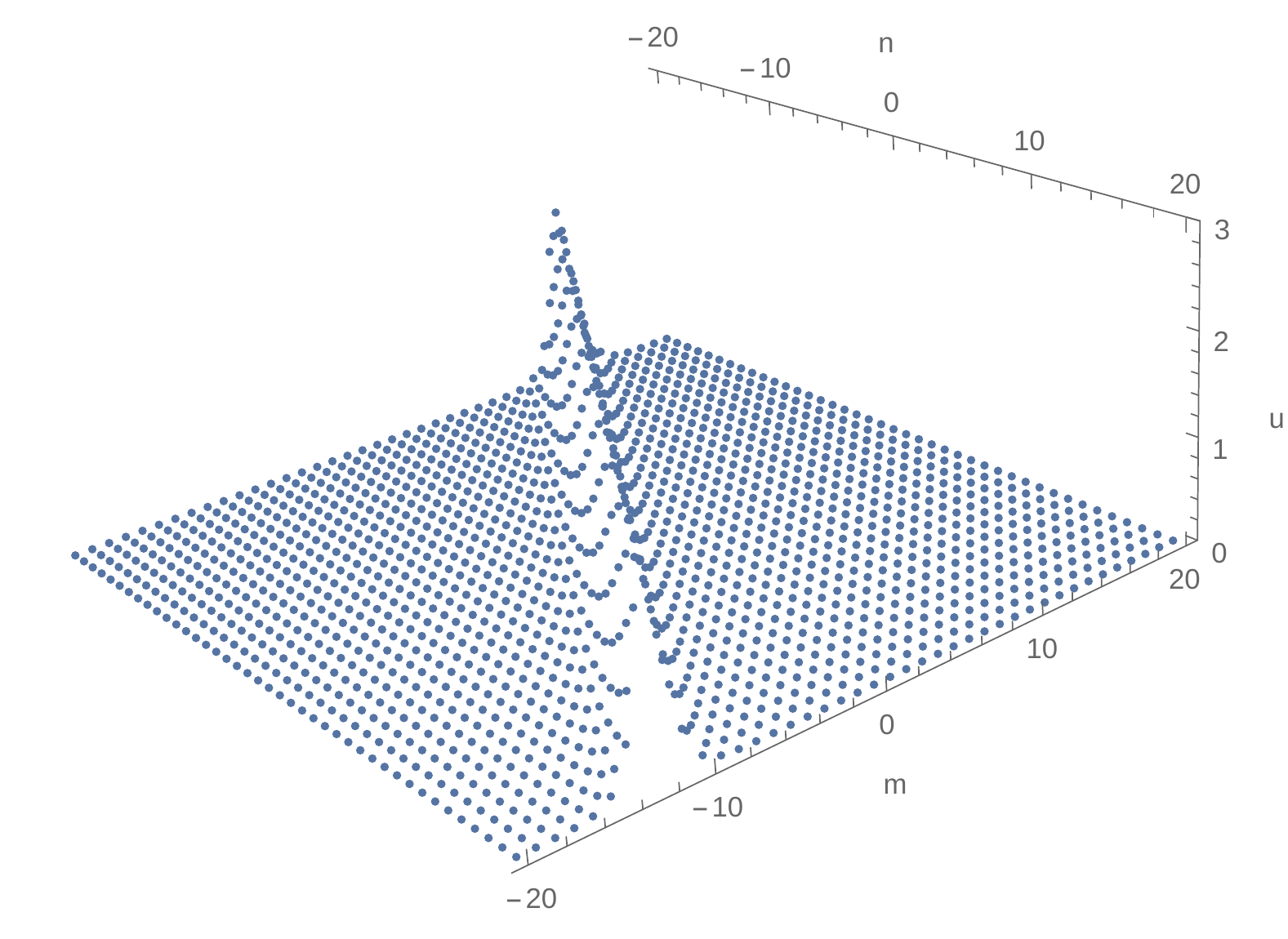}}}
\put(10,-23){\resizebox{!}{3.5cm}{\includegraphics{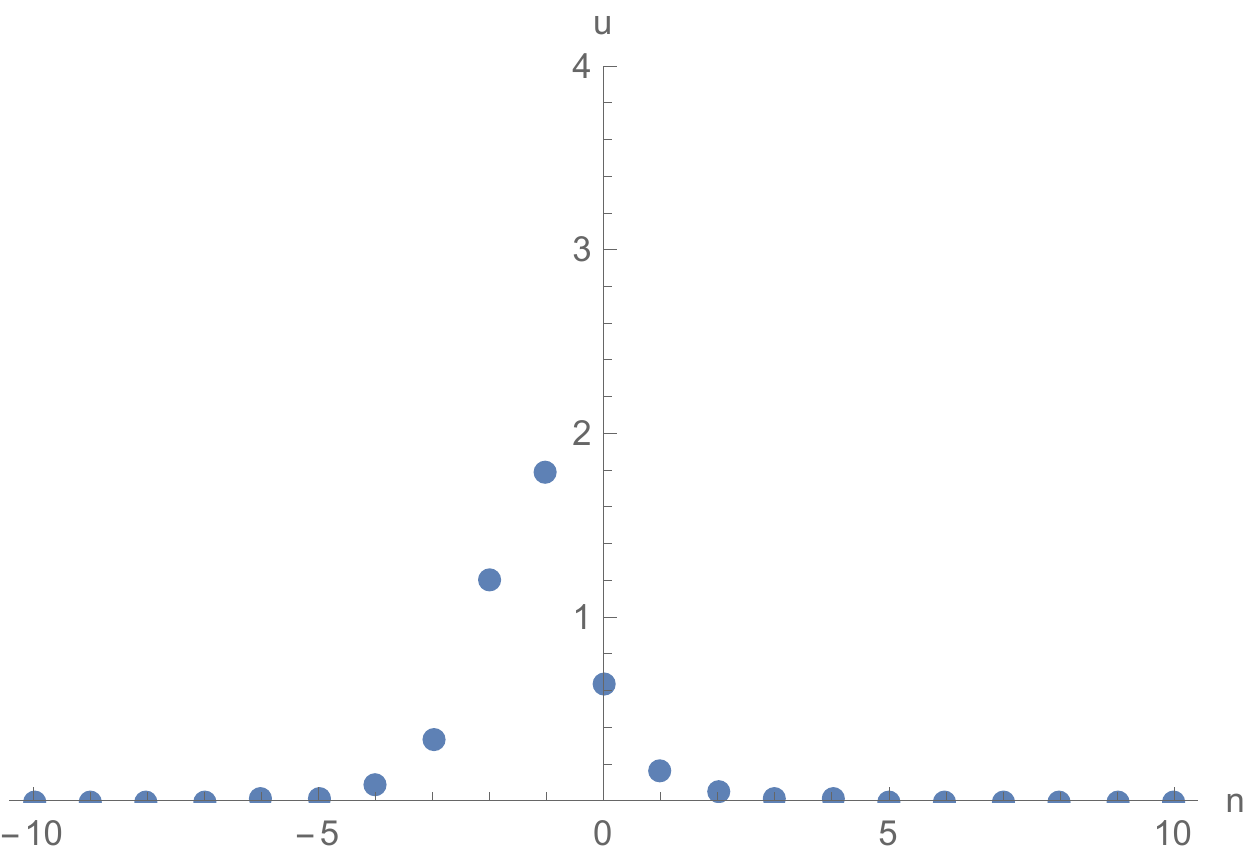}}}
\put(150,-23){\resizebox{!}{3.5cm}{\includegraphics{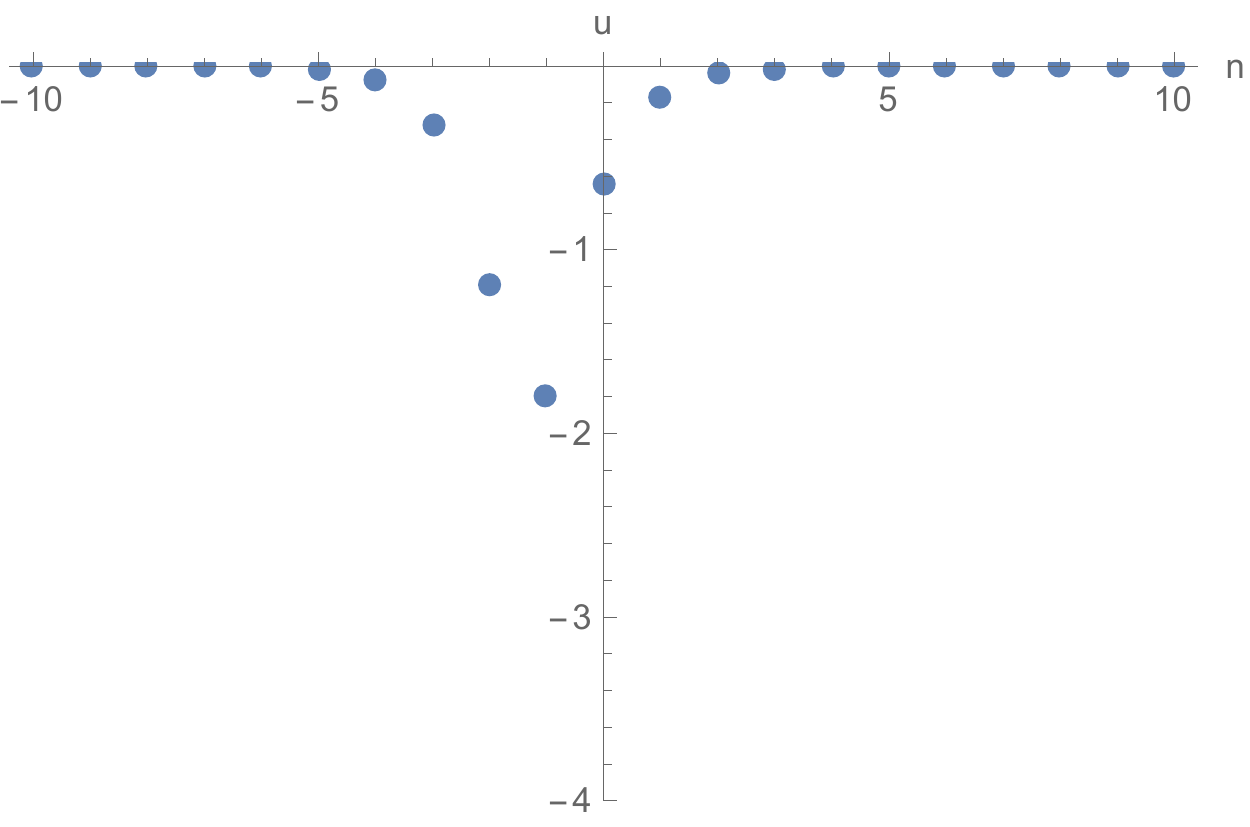}}}
\end{picture}
\end{center}
\vskip 20pt
\begin{center}
\begin{minipage}{15cm}{\footnotesize
\qquad\qquad\qquad\quad(a)\qquad\qquad\qquad\qquad\qquad\qquad\qquad\qquad (b) \qquad\qquad\qquad\qquad\qquad\qquad\quad (c)\\
{\bf Fig. 1.} One-soliton solution $u$ given by \eqref{dr-1ss-u} with $\epsilon=\delta=1$:
(a) shape and movement for $k_1=0.7$;
(b) soliton for $k_1=0.7$ at $m=2$;
(c) anti-soliton for $k_1=-0.7$ at $m=2$.}
\end{minipage}
\end{center}

In the case of $N=1$, we get two-soliton solutions
\begin{align}
\label{dr-2ss}
u=g/f, \quad w=\wt{f}\wh{f}/(f\wh{\wt{f}}),
\end{align}	
with
\begin{subequations}
\label{fg-2ss}
\begin{align}
f=& \cosh2(k_{1}+k_{2})\sinh^2(\xi_1-\xi_2)+\cosh2(k_{1}-k_{2})\cosh^2(\xi_1+\xi_2)-\cosh2\xi_1\cosh2\xi_2, \\
g=& (\cosh2k_{1}-\cosh2k_{2})(\sinh2k_{1}\cosh2\xi_{2}-\sinh2k_{2}\cosh 2\xi_{1})/\epsilon.
\end{align}	
\end{subequations}

We now examine the asymptotic form of the two-soliton solutions $u$ in \eqref{dr-2ss} as $m\rightarrow \pm \infty$,
which can be derived by the analysis of moving-coordinate expansions.
For convenience, we call the asymptotic solitons as $k_{1}$-soliton and $k_{2}$-soliton, respectively.
Without the loss of generality, we set
\begin{align}
0<\delta\epsilon<4, \quad \dfrac{1}{2}\ln\dfrac{4+\delta\epsilon}{4-\delta\epsilon}<k_{1}<k_{2},
\end{align}
to guarantee $\tau_1>\tau_2>0$.

To proceed, we consider $\xi_1=$const., and express $\xi_2=\frac{k_2}{k_1}\xi_1+\zeta_1$ in terms of $\xi_1$ and $\zeta_1=m\Delta c_1$, where $\Delta c_1=\tau_2-\frac{k_2}{k_1}\tau_1<0$ is the relative speed of the moving coordinates.
Note that $m\rightarrow \pm \infty$ corresponds to $\zeta_1\rightarrow \mp \infty$.
We then asymptotically expand $f$ and $g$ for large $\zeta_1$ with fixed $\xi_1$.
This yields, after neglecting subdominant exponential terms,
\begin{subequations}
\begin{align}
& f\simeq \frac{1}{2}(\sinh^2(k_2\pm k_1)e^{2\xi_1}+\sinh^2(k_2\mp k_1)e^{-2\xi_1})e^{\mp 2\xi_2}, \quad m\rightarrow \pm \infty, \\
& g\simeq \frac{1}{2\epsilon}\sinh 2k_1(\cosh 2k_1-\cosh 2k_2)e^{\mp 2\xi_2}, \quad m\rightarrow \pm \infty,
\end{align}	
\end{subequations}
and hence the $k_{1}$-soliton appears
\begin{align}
u\simeq \frac{\sinh 2k_1(\cosh 2k_1-\cosh 2k_2)}{\epsilon(\sinh^2(k_2\pm k_1)e^{2\xi_1}+\sinh^2(k_2\mp k_1)e^{-2\xi_1})},
\quad m\rightarrow \pm \infty,
\end{align}	
whose asymptotical behavior follows:
\begin{subequations}
\begin{align}
top \: point \: traces&: n(m)=\pm \dfrac{1}{2k_{1}}\ln\dfrac{\sinh(k_{2}-k_{1})}
{\sinh(k_{1}+k_{2})}-\frac{\tau_1}{k_1}m,\\
amplitude &: u=\dfrac{\sinh2k_{1}(\cosh2k_{1}-\cosh2k_{2})}
{2\epsilon\sinh(k_{1}+k_{2})\sinh(k_{2}-k_{1})},\\
speed &: -\frac{\tau_1}{k_1}, \\
phase \: shift &: \dfrac{1}{k_{1}}\ln\dfrac{\sinh(k_{2}-k_{1})}
{\sinh(k_{1}+k_{2})}.
\end{align}	
\end{subequations}

To continue, we next consider $\xi_2=$const., and express $\xi_1=\frac{k_1}{k_2}\xi_2+\zeta_2$ in terms of $\xi_2$ and $\zeta_2=m\Delta c_2$ with $\Delta c_2=\tau_1-\frac{k_1}{k_2}\tau_2>0$.
By asymptotically expanding $f$ and $g$ for large $\zeta_2$ with fixed $\xi_2$, and neglecting subdominant exponential terms, we obtain
\begin{subequations}
\begin{align}
& f\simeq \frac{1}{2}(\sinh^2(k_2\mp k_1)e^{2\xi_2}+\sinh^2(k_2\pm k_1)e^{-2\xi_2})e^{\pm 2\xi_1}, \quad m\rightarrow \pm \infty, \\
& g\simeq \frac{1}{2\epsilon}\sinh 2k_2(\cosh 2k_2-\cosh 2k_1)e^{\pm 2\xi_1}, \quad m\rightarrow \pm \infty,
\end{align}	
\end{subequations}
and thus the $k_{2}$-soliton reads
\begin{align}
u\simeq \frac{\sinh 2k_2(\cosh 2k_2-\cosh 2k_1)}{\epsilon(\sinh^2(k_2\mp k_1)e^{2\xi_2}+\sinh^2(k_2\pm k_1)e^{-2\xi_2})},
\quad m\rightarrow \pm \infty,
\end{align}	
whose asymptotical behavior follows:
\begin{subequations}
\begin{align}
top \: point \: traces&: n(m)=\mp \dfrac{1}{2k_{2}}\ln\dfrac{\sinh(k_{2}-k_{1})}
{\sinh(k_{1}+k_{2})}-\frac{\tau_2}{k_2}m,\\
amplitude&: u=\dfrac{\sinh2k_{2}(\cosh2k_{2}-\cosh2k_{1})}
{2\epsilon\sinh(k_{1}+k_{2})\sinh(k_{2}-k_{1})},\\
speed&: -\frac{\tau_2}{k_2}, \\
phase \: shift &: -\dfrac{1}{k_{2}}\ln\dfrac{\sinh(k_{2}-k_{1})}
{\sinh(k_{1}+k_{2})}.
\end{align}	
\end{subequations}

The solution $u$ in \eqref{dr-2ss} is illustrated in Fig. 2.
\vskip20pt
\begin{center}
\begin{picture}(120,80)
\put(-170,-23){\resizebox{!}{4.0cm}{\includegraphics{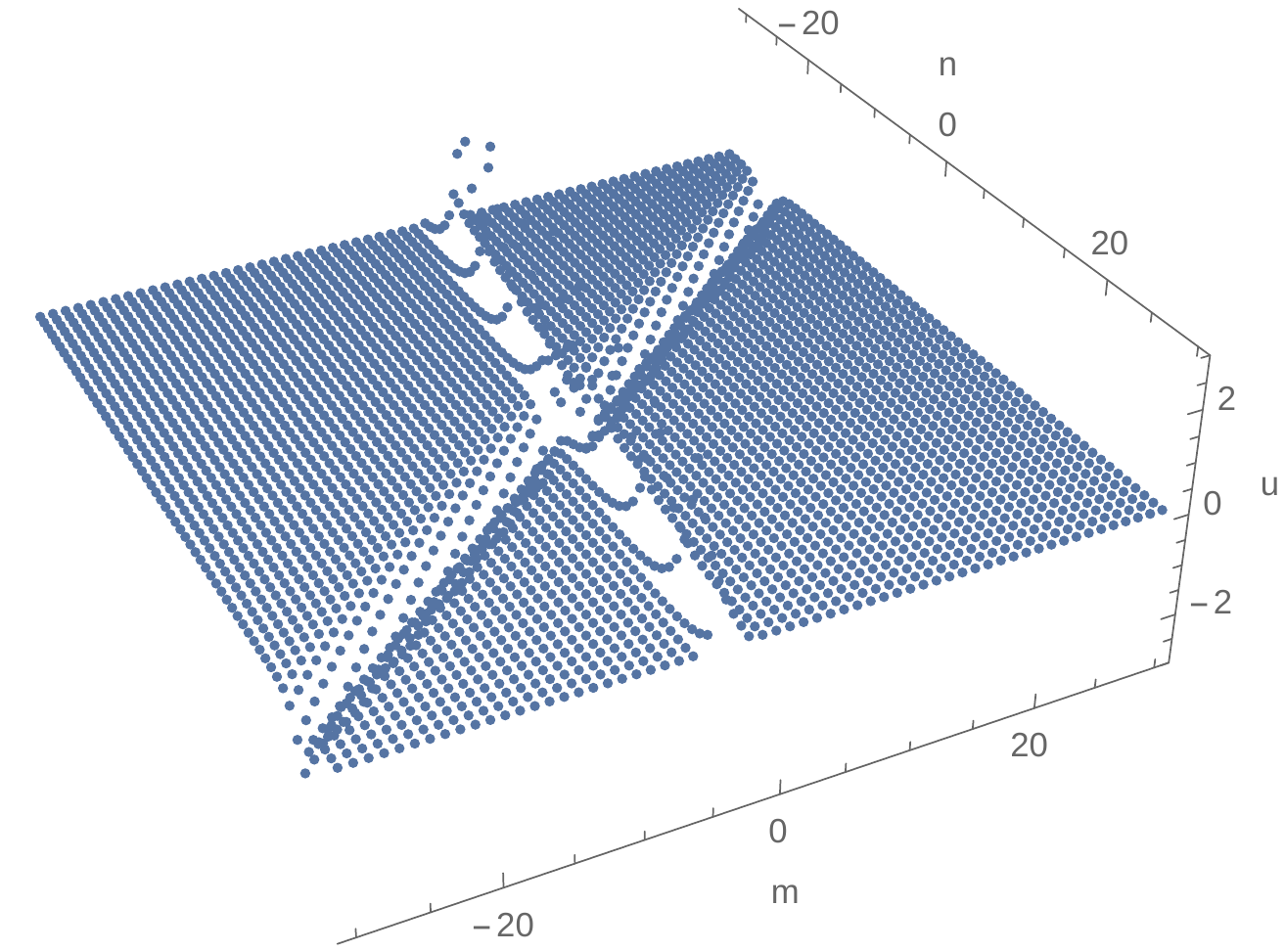}}}
\put(-10,-23){\resizebox{!}{3.5cm}{\includegraphics{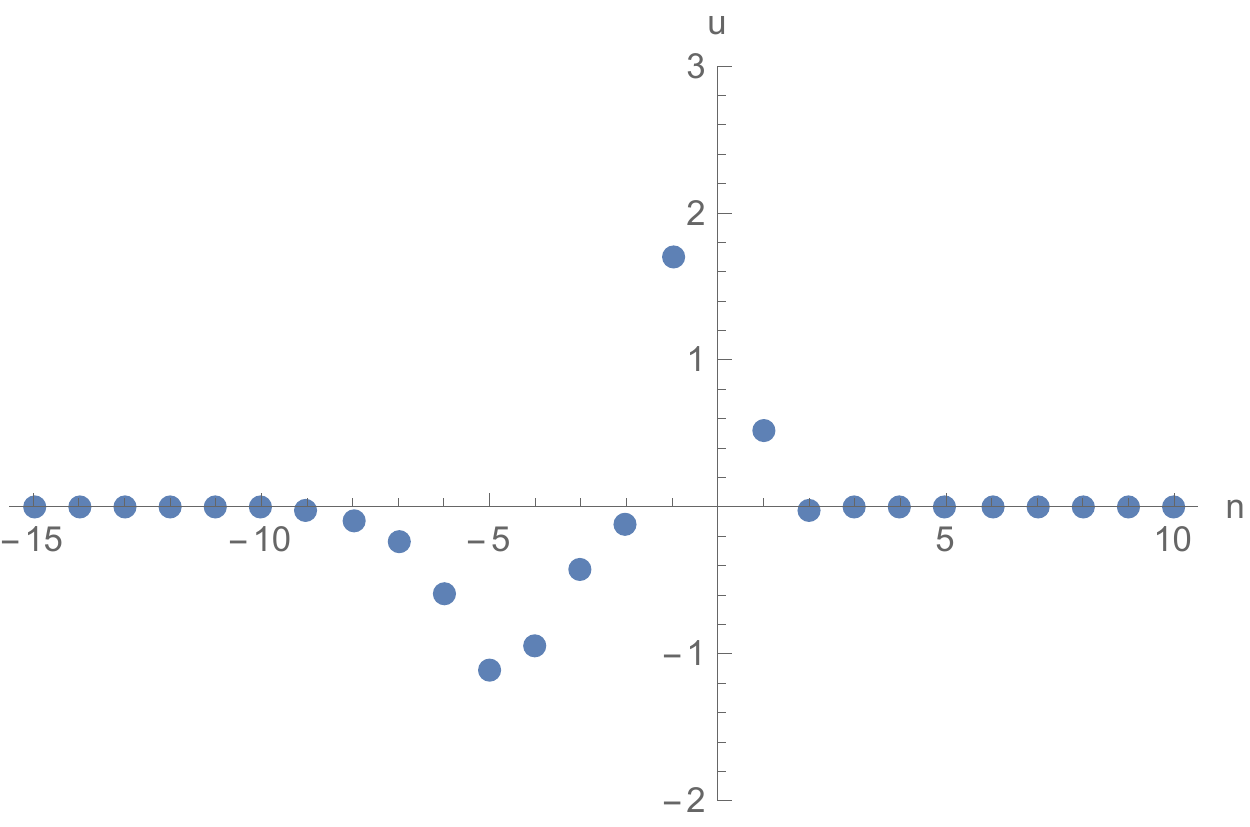}}}
\put(150,-23){\resizebox{!}{3.5cm}{\includegraphics{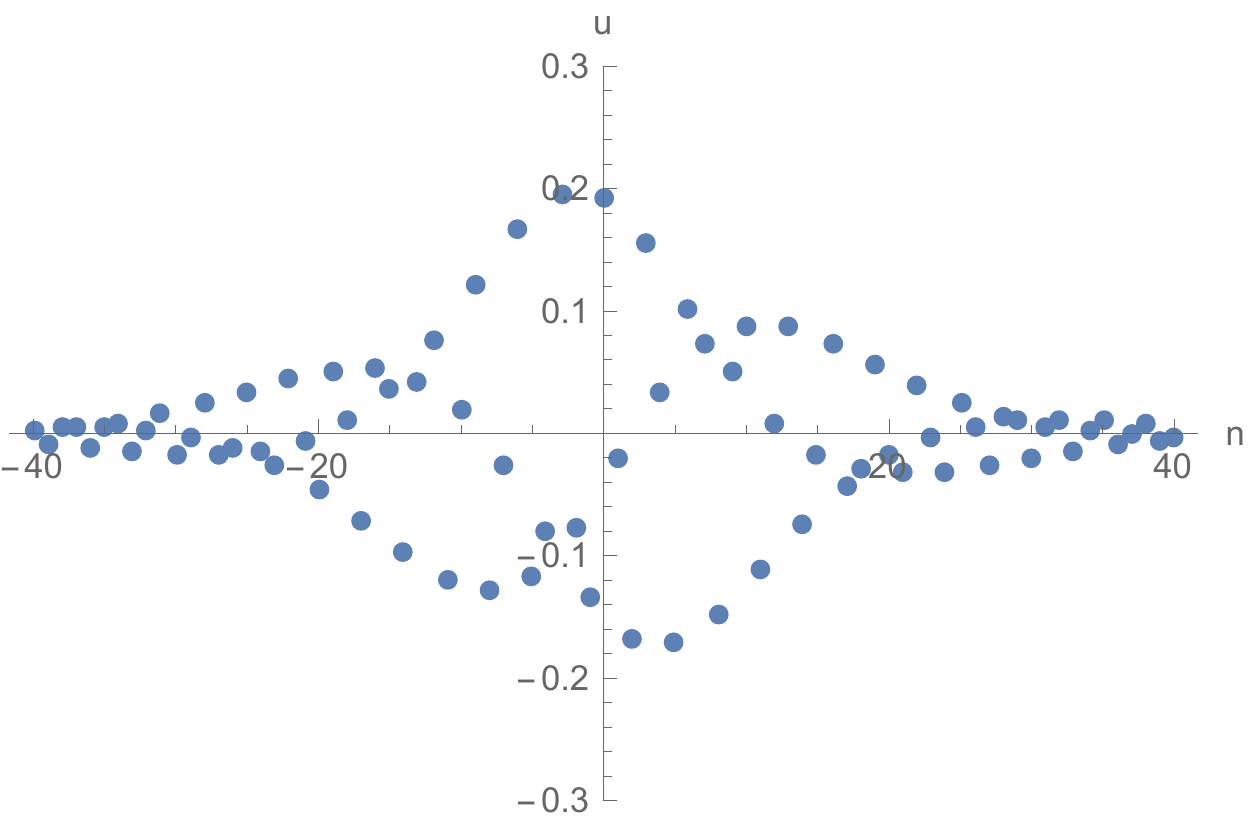}}}
\end{picture}
\end{center}
\vskip 20pt
\begin{center}
\begin{minipage}{15cm}{\footnotesize
\qquad\qquad\qquad\quad(a)\qquad\qquad\qquad\qquad\qquad\qquad\qquad\qquad (b) \qquad\qquad\qquad\qquad\qquad\qquad\quad (c)\\
{\bf Fig. 2.}
Solution $u$ given by \eqref{dr-2ss} with $\epsilon=\delta=1$:
(a) two-soliton solutions with $k_1=0.5$ and $k_2=2$; (b) 2D-plot of (a) at $m=3$; (c) breather solution with $k_1=0.05+i$ and $k_2=0.05-i$
at $m=1$.}
\end{minipage}
\end{center}

\vspace{.2cm}
\noindent{\it Remark 1: The eigenvalues in \eqref{drL-Diag} can be extended to complex numbers.
For example, if taking $k_{2j}=k^*_{2j-1}$ with $j=1,2,\ldots, N+1$, then we obtain the breather solutions (see Fig. 2 (c)).}

\vspace{.2cm}
\noindent{\it Jordan-block solution}:
Now let $L$ be a Jordan-block matrix
\begin{align}
\label{drL-Jor}
L=\left(
\begin{array}{cccc}
k_1 &   0 &  \cdots&  0\\
1 & k_1 &  \cdots&  0\\		
\vdots& \ddots&\ddots&\vdots\\
0&\cdots& 1 & k_1
\end{array}\right)_{(N+1)\times(N+1)},\quad k_1 \neq 0 \in \mathbb{R},
\end{align}	
and $C^{+}=\breve{I}$.
In principle, the Jordan-block solutions can be obtained from the multi-soliton solutions by taking limit \cite{ZZSZ}.
Here we omit this procedure and directly give the result.
For \eqref{drL-Jor},
solution $\Phi$ is composed by
\begin{align*}
\Phi_{j}=\left\{
\begin{aligned}
&\dfrac{\partial^{j-1}_{k_1}e^{\xi_1}}{(j-1)!}, \quad j=1, 2, \dots, N+1,\\
&\dfrac{\partial^{s-1}_{k_1}e^{-\xi_1}}{(s-1)!}, \quad j=N+1+s,\quad s=1,2,\dots,N+1.
\end{aligned}	
\right.
\end{align*}

When $N=1$, the simplest Jordan-block solution is
\begin{subequations}
\begin{align}
\label{dr-JB-solu-u}
& u=\dfrac{\sinh 2k_1[(1+\varsigma)\cosh 2(k_1-\xi_1)+(1-\varsigma)\cosh 2(k_1+\xi_1)]}
{\epsilon(\varsigma^{2}\sinh^2 2k_1+\cosh^2 2\xi_1)}, \\
& w=\wt{f}\wh{f}/(f\wh{\wt{f}}), \quad f=\varsigma^{2}\sinh^2 2k_1+\cosh^2 2\xi_1,
\end{align}	
\end{subequations}
where $\varsigma=n-\dfrac{4m\delta\epsilon}{16 \sinh^2k_1-(\delta\epsilon)^2 \cosh^2 k_1}$.
This solution is depicted in Fig. 3.
\vskip20pt
\begin{center}
\begin{picture}(120,80)
\put(-140,-23){\resizebox{!}{4.0cm}{\includegraphics{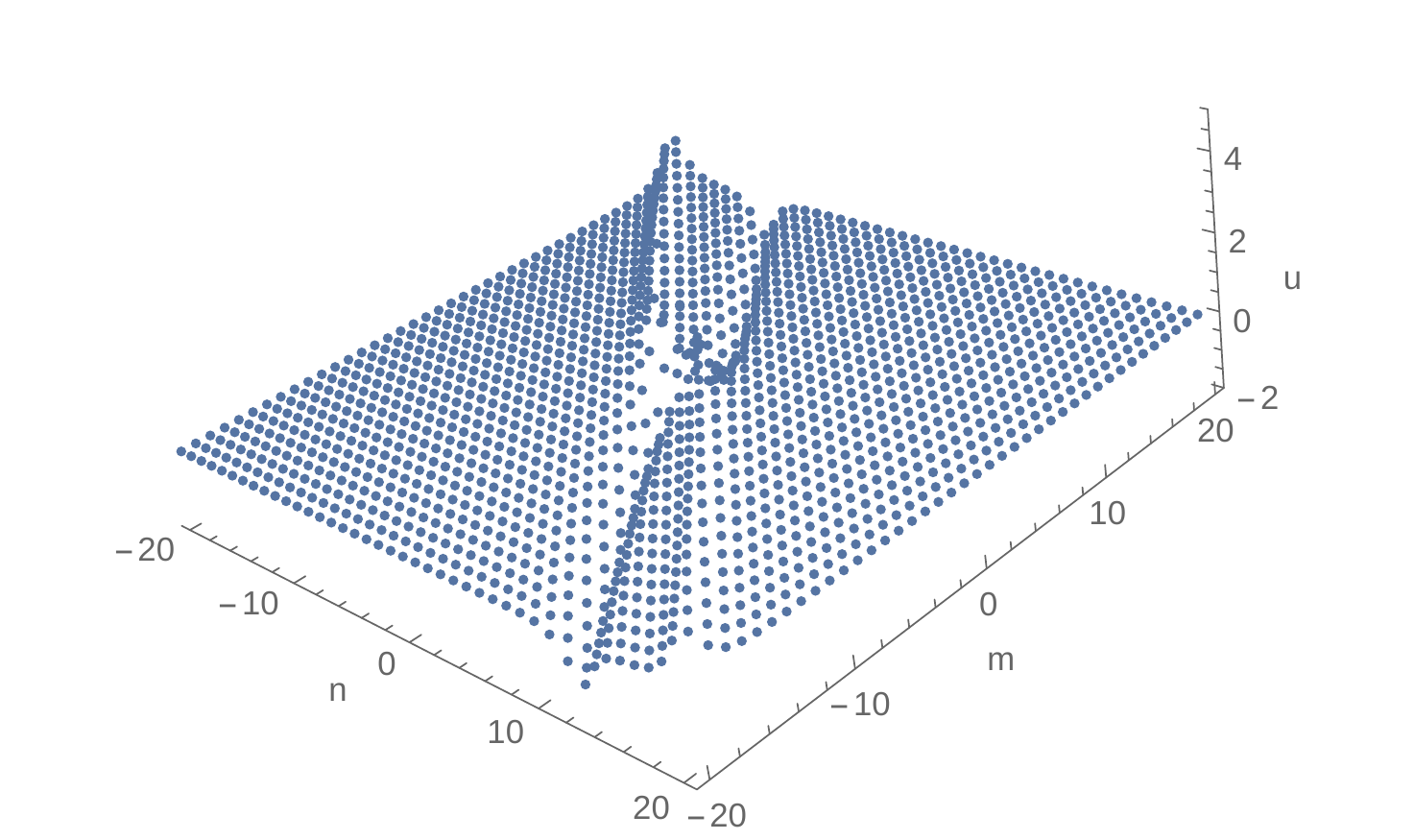}}}
\put(80,-23){\resizebox{!}{4.0cm}{\includegraphics{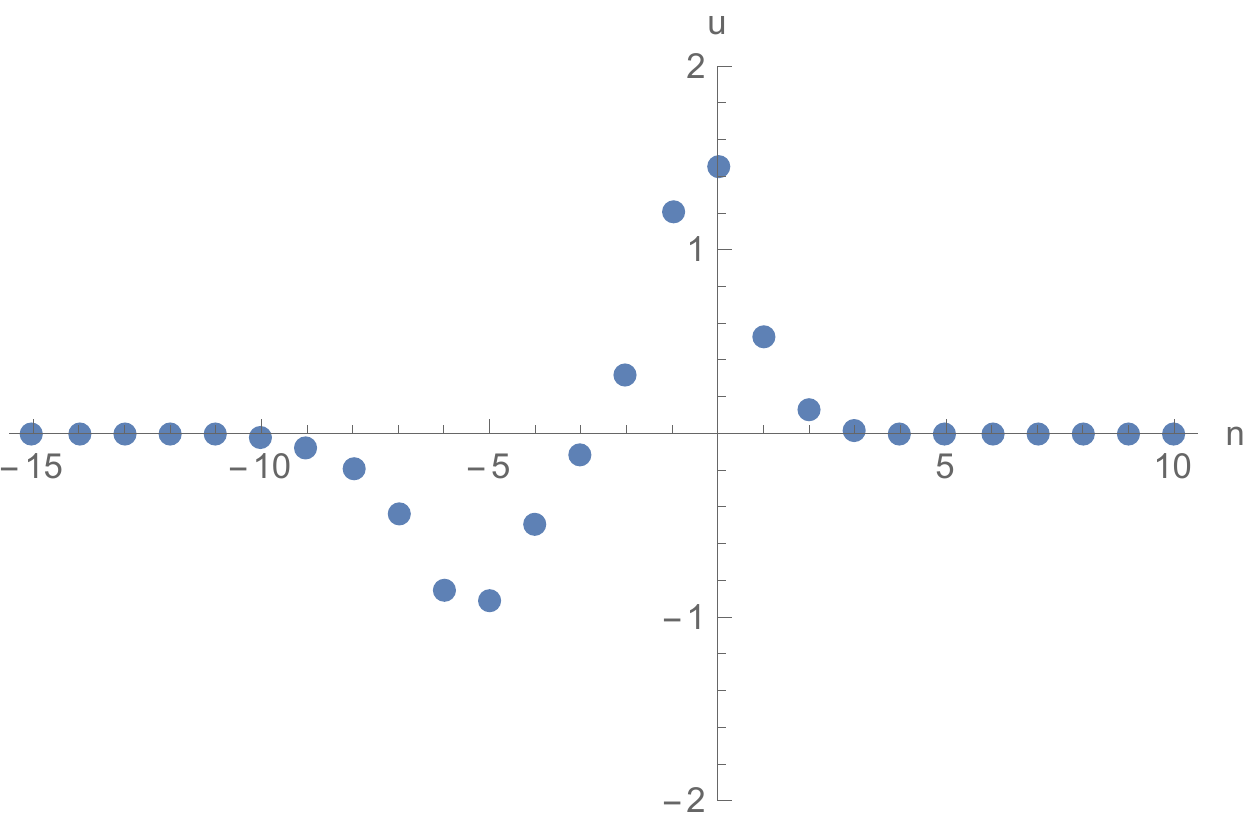}}}
\end{picture}
\end{center}
\vskip 20pt
\begin{center}
\begin{minipage}{15cm}{\footnotesize
\qquad\qquad\qquad\qquad\quad(a)\qquad\qquad\qquad\qquad\qquad\qquad\qquad\qquad\qquad\qquad\qquad\qquad\quad (b) \\
{\bf Fig. 3.} Jordan-block solution $u$ given by \eqref{dr-JB-solu-u} with $k_1=0.5$ and $\epsilon=\delta=1$:
(a) shape and motion; (b) 2D-plot of (a) at $m=2$.}
	\end{minipage}
\end{center}

\subsection{Continuum limits}
\label{re-CL}

We will now carry out (semi-)continuous limits of the rndsG equation \eqref{rndsG}.
Through the semi-continuous limits in the $m$-direction and the $n$-direction respectively,
we derive two real local and nonlocal semi-discrete sG equations.
Furthermore, by continuous limits we arrive at the real local and nonlocal continuous sG equation.

Together with the connections between the various parameters in the equation \eqref{rndsG} and the lattice spacing,
we use the formula
\begin{align}
\label{dc-re}
\lim\limits_{m\rightarrow\infty}(1+k/m)^m=e^k,
\end{align}
in the purpose to give the discrete-continuous relation, where the discrete exponential functions
\begin{align}
e^{\xi_j}:=e^{k_jn}\bigg(\dfrac{4+\delta\epsilon\coth k_j}{4-\delta \epsilon \coth k_j}\bigg)^{\frac{m}{2}}, \quad
j=1,2,\ldots,N+1
\end{align}
should be considered.

\vspace{.2cm}
\noindent{\bf Semi-continuous limit in $m$-direction:} The semi-continuous limit in $m$-direction used here is
the so-called straight continuum limit, namely, the discrete variable $m=0$ corresponds to the continuous variable $t=0$.
In order to use \eqref{dc-re} as a semi-continuous limit in the $m$-direction, we write
\begin{align}
\label{rstl-m}
\bigg(\dfrac{4+\delta\epsilon\coth k_j}{4-\delta \epsilon \coth k_j}\bigg)^{\frac{m}{2}}
=\bigg(1+\dfrac{2\delta\epsilon\coth k_j}{4-\delta \epsilon \coth k_j}\bigg)^{\frac{m}{2}},
\quad j=1,2,\ldots,N+1,
\end{align}
and therefore
\begin{align}
\label{rcl-nt}
\delta \quad \text{must approach zero as} \quad m\rightarrow\infty, \quad \text{i.e.}, \quad m\delta=t.
\end{align}
As a consequence, the discrete exponential functions become
\begin{align}
\label{lambdaj-def}
e^{\xi_j} \rightarrow e^{\lambda_j}, \quad \text{with} \quad \lambda_j:=k_{j}n+\frac{\epsilon\coth k_{j}}{4}t,
\quad j=1,2,\ldots,N+1.
\end{align}

Interpreting the dependent variables $u(n,m):=\mu(n,t)$, $w(n,m):=\omega(n,t)$ and substituting the Taylor expansions\footnote{
Noticing
transformation $w=\wt{f}\wh{f}/(f\wh{\wt{f}})$ and limit \eqref{rcl-nt}, we know
$w=\dfrac{(f+\delta f'+\cdots)\wt{f}}
{f(\wt{f}+\delta\wt{f}'+\cdots)}
\rightarrow\dfrac{(f+\delta f')\wt{f}}
{f(\wt{f}+\delta\wt{f}')}
=1-\delta\dfrac{\wt{f}'{f}-f'\wt{f}}
{f\wt{f}}=1-\delta \epsilon \omega'$, where $\omega=\epsilon^{-1}\ln(\wt{f}/f)$.}
\begin{subequations}
\label{Tay-exp-dsd}
\begin{align}
& \wh{u}=\mu(t+\delta)=\mu+\delta \mu'+\ldots, \\
& \wh{\wt{u}}=\wt{\mu}(t+\delta)=\wt{\mu}+\delta\wt{\mu}'+\ldots, \\
& \wh{u}_{\sigma}=\mu(\sigma t+\sigma\delta)=\mu_{\sigma}+\delta (\mu_{\sigma})'+\ldots, \\
\label{w-omega-rt}
& w\rightarrow 1-\delta \epsilon \omega'
\end{align}
\end{subequations}
into the rndsG equation \eqref{rndsG}, we obtain as coefficient of the leading term of order $\mathcal{O}(\delta)$ the
real local and nonlocal semi-discrete sG (rnsdsG-$t$) equation
\begin{subequations}
\label{rnsdsG-t}
\begin{align}
& 2(\wt{\mu}-\mu)'=\epsilon(\mu+\wt{\mu})(1-2\omega'),\\
& e^{\epsilon(\omega-\undertilde{\omega})}-1=\eta\epsilon^{2} \mu\mu_{\sigma},
\end{align}
\end{subequations}
where the prime denotes the derivative with respect to $t$ and $\omega=\omega_{\sigma}$.
Similar to the discrete case, equation \eqref{rnsdsG-t} is preserved under transformation $\mu\rightarrow -\mu$
and equation \eqref{rnsdsG-t} with $(\sigma,\eta)=(\pm 1,1)$ and $(\sigma,\eta)=(\pm 1,-1)$ can be
transformed into each other by taking $\mu\rightarrow i\mu$.

Next, we show double Casoratian solutions to the rnsdsG-$t$ equation \eqref{rnsdsG-t}.
\begin{Thm}
The functions $\mu=g/f$ and $\omega=\epsilon^{-1}\ln(\wt{f}/f)$ with
\begin{align}
\label{rnsdsG-t-solu}
f=|e^{-N\Ga}\wh{\Phi^{(N)}};e^{N\Ga}T\wh{\Phi_{\sigma}^{(-N)}}|,\quad
g=(1/\epsilon)|e^{-N\Ga}\widehat{\Phi^{(N+1)}};e^{N\Ga}T\wh{\Phi_{\sigma}^{(-N+1)}}|,
\end{align}
solve the rnsdsG-$t$ equation \eqref{rnsdsG-t}, where $\Phi=e^{\Ga n+\epsilon\coth \Ga t/4}C^{+}$ and $T$ is a constant matrix of order $2(N+1)$
satisfying
\begin{align}
\label{rnsdsG-t-Ga-T}
\Ga T+\sigma T\Ga=\bm 0,\quad T^{2}=\left\{
\begin{array}{l}
-\eta I, \quad \mbox{with} \quad \sigma=1,\\
\eta|e^{\Ga}|^{2}I, \quad \mbox{with} \quad \sigma=-1.
\end{array}\right.
\end{align}
\end{Thm}

In what follows, taking $\Ga$ and $T$ as \eqref{Ga-T-ex}, we discuss explicit solutions
for the equation \eqref{rnsdsG-t} with $(\eta,\sigma)=(1,-1)$.

\vspace{.2cm}
\noindent{\it Soliton solutions}: Let $L$ be the diagonal matrix \eqref{drL-Diag}.
The basic column vector $\Phi$ is thus given by
\begin{align}
\Phi_{j}=\left\{
\begin{aligned}
& e^{\lambda_j},\quad j=1, 2, \ldots, N+1,\\
& e^{-\lambda_s},\quad j=N+1+s, \quad s=1,2,\ldots,N+1.
\end{aligned}	
\right.
\end{align}

In the case of $N=0$, equation \eqref{rnsdsG-t} has one-soliton solution
\begin{subequations}
\label{sdr-t-1ss}
\begin{align}
\label{sdr-1ss-mu}
& \mu=(\sinh 2k_1\sech 2\lambda_1)/\epsilon, \\
& \omega=[\ln (\cosh2(k_1+\lambda_1)\sech 2\lambda_1)]/\epsilon.
\end{align}
\end{subequations}
The solution $\mu$ in \eqref{sdr-1ss-mu} still describes a stable travelling wave with
velocity $-\epsilon\coth k_{1}/(4k_1)$ and amplitude $\sinh 2k_{1}/\epsilon$, where
the width is proportional to $(2k_1)^{-1}$. We depict soliton and anti-soliton in Fig. 4
in terms of the sign of the parameter $k_1\epsilon$.
\vskip20pt
\begin{center}
\begin{picture}(120,80)
\put(-160,-23){\resizebox{!}{4.0cm}{\includegraphics{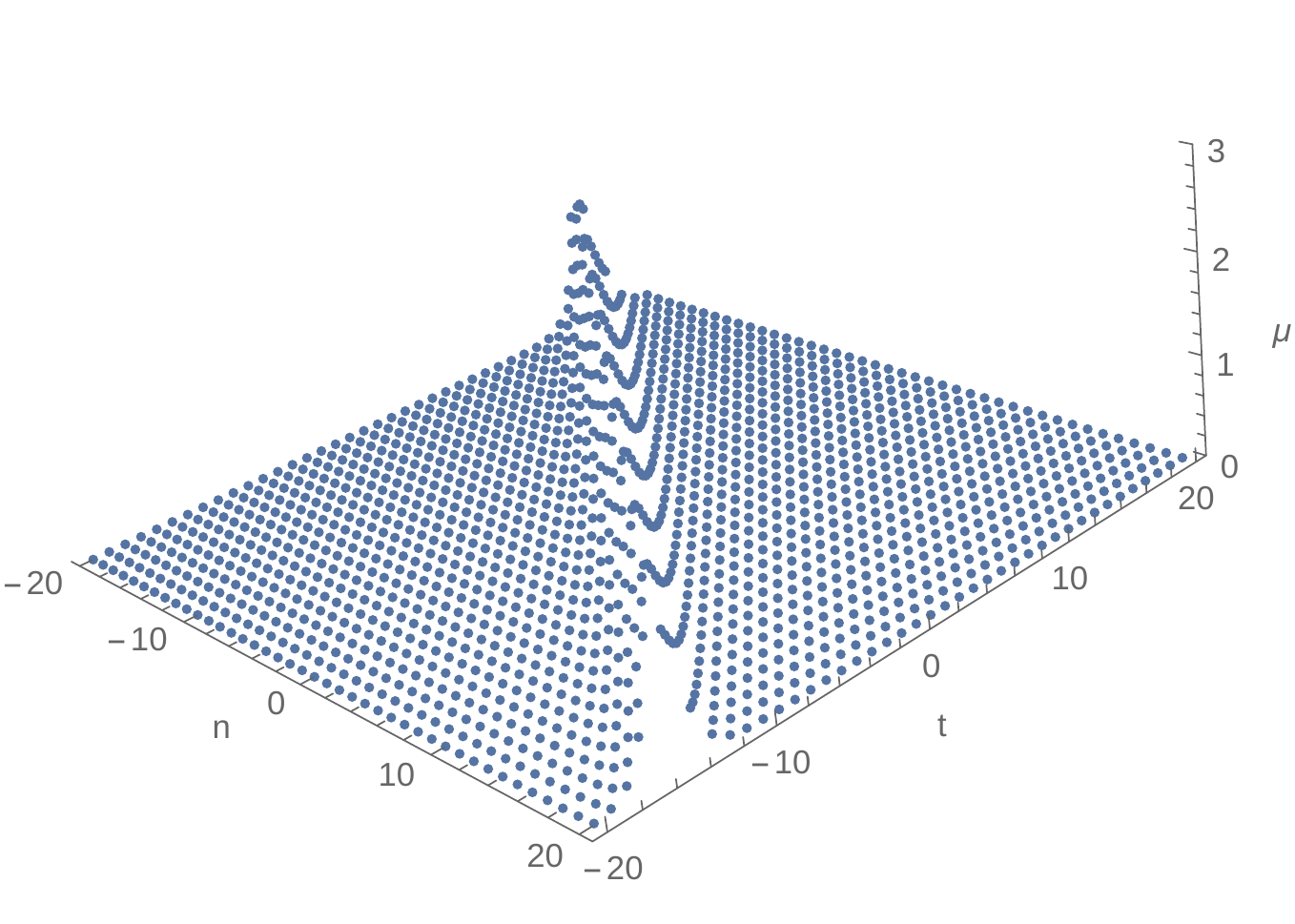}}}
\put(10,-23){\resizebox{!}{3.5cm}{\includegraphics{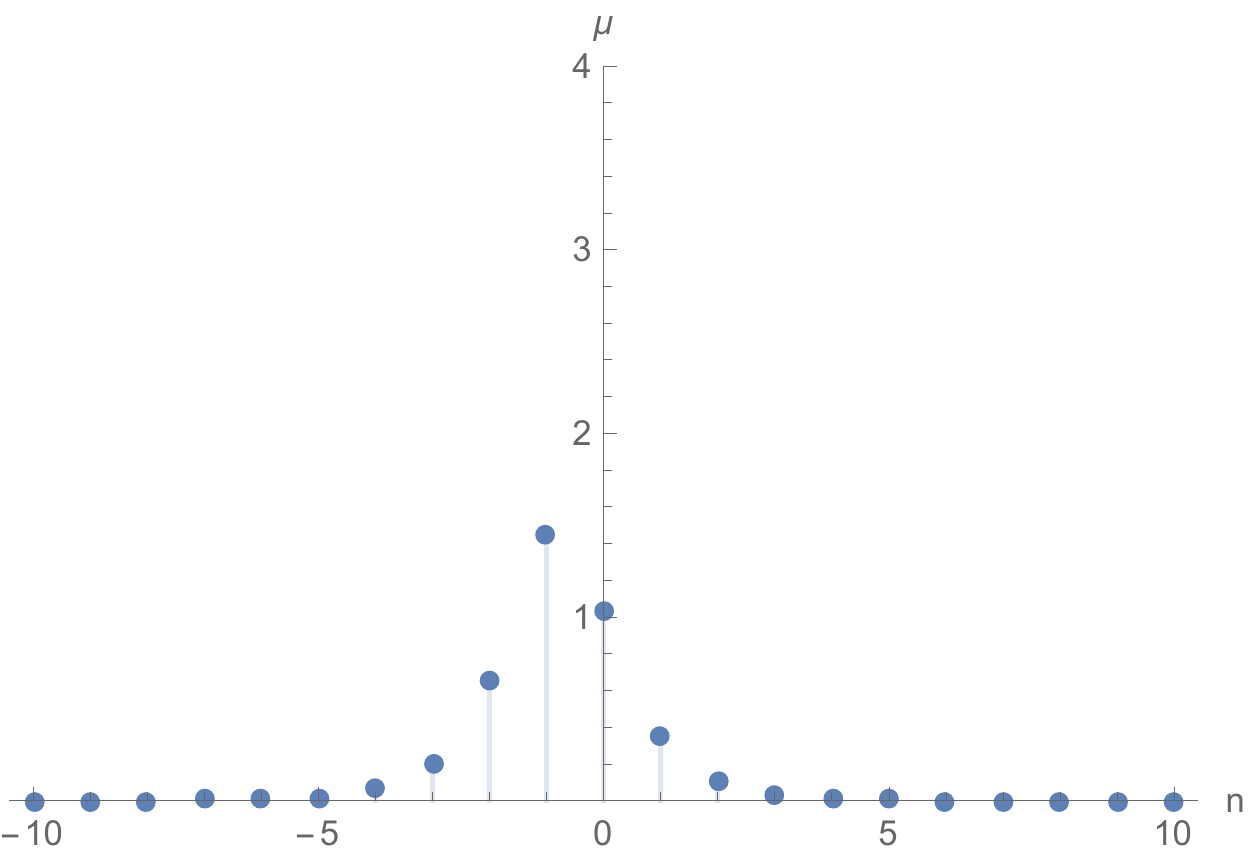}}}
\put(150,-23){\resizebox{!}{3.5cm}{\includegraphics{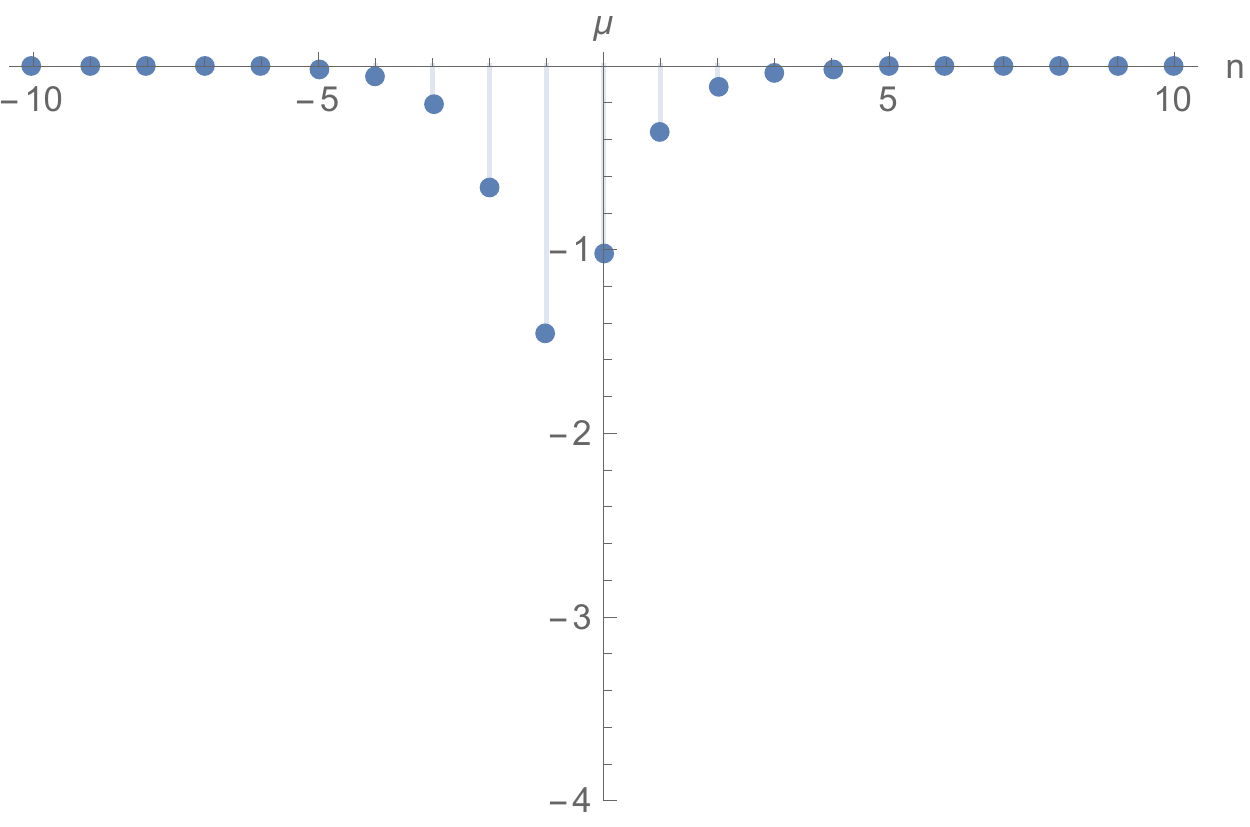}}}
\end{picture}
\end{center}
\vskip 20pt
\begin{center}
\begin{minipage}{15cm}{\footnotesize
\qquad\qquad\qquad\quad(a)\qquad\qquad\qquad\qquad\qquad\qquad\qquad\qquad (b) \qquad\qquad\qquad\qquad\qquad\qquad\quad (c)\\
{\bf Fig. 4}. One-soliton solution $\mu$ given by \eqref{sdr-1ss-mu} with $\epsilon=1$:
(a) shape and movement with $k_1=0.6$;
(b) soliton for $k_1=0.6$ at $t=1$;
(c) anti-soliton for $k_1=-0.6$ at $t=1$.}
\end{minipage}
\end{center}

In the case of $N=1$, equation \eqref{rnsdsG-t} has two-soliton solutions
\begin{align}
\label{sdr-t-2ss}
\mu=g/f, \quad \omega=\epsilon^{-1}\ln(\wt{f}/f),
\end{align}	
where $f$ and $g$ are given by \eqref{fg-2ss} up to a replacement of $\xi_i$ by $\lambda_i,~(i=1,2)$.
Solution $\mu$ in \eqref{sdr-t-2ss} has a similar asymptotical behavior as $u$ in \eqref{dr-2ss}, which is sketched in Fig. 5.
\begin{Thm}
Suppose that $0<k_{1}<k_{2}$ and $\epsilon>0$. Then, when $t\rightarrow\pm\infty$, the $k_{1}$-soliton asymptotically follows:
\begin{subequations}
\begin{align}
top \: point \: traces
&: n(t)=\pm\dfrac{1}{2k_{1}}\ln\dfrac{\sinh(k_{2}-k_{1})}
{\sinh(k_{1}+k_{2})}-\dfrac{\epsilon\coth k_{1}}{4k_1}t,\\
amplitude
&: \mu=\dfrac{\sinh 2k_{1}(\cosh 2k_{1}-\cosh 2k_{2})}
{2\epsilon\sinh(k_{2}-k_{1})\sinh(k_{1}+k_{2})},\\
speed
&: -\dfrac{\epsilon}{4k_{1}}\coth k_{1}, \\
phase \: shift &: \dfrac{1}{k_{1}}
\ln\dfrac{\sinh{(k_{2}-k_{1})}}
{\sinh(k_{1}+k_{2})},
\end{align}	
\end{subequations}
and the $k_{2}$-soliton asymptotically follows:
\begin{subequations}
\begin{align}
top \: point \: traces
&:n(t)=\mp\dfrac{1}{2k_{2}}\ln\dfrac{\sinh(k_2-k_{1})}
{\sinh(k_{1}+k_{2})}-\dfrac{\epsilon\coth k_{2}}{4k_2}t,\\
amplitude
&: \mu=\dfrac{\sinh 2k_{2}(\cosh 2k_{2}-\cosh 2k_{1})}
{2\epsilon\sinh(k_{2}-k_{1})\sinh(k_{1}+k_{2})},\\
speed
&: -\dfrac{\epsilon}{4k_{2}}\coth k_{2}, \\
phase \: shift &: -\dfrac{1}{k_{2}}
\ln\dfrac{\sinh{(k_{2}-k_{1})}}
{\sinh(k_{1}+k_{2})}.
\end{align}	
\end{subequations}
\end{Thm}
\vskip20pt
\begin{center}
\begin{picture}(120,80)
\put(-170,-23){\resizebox{!}{4.0cm}{\includegraphics{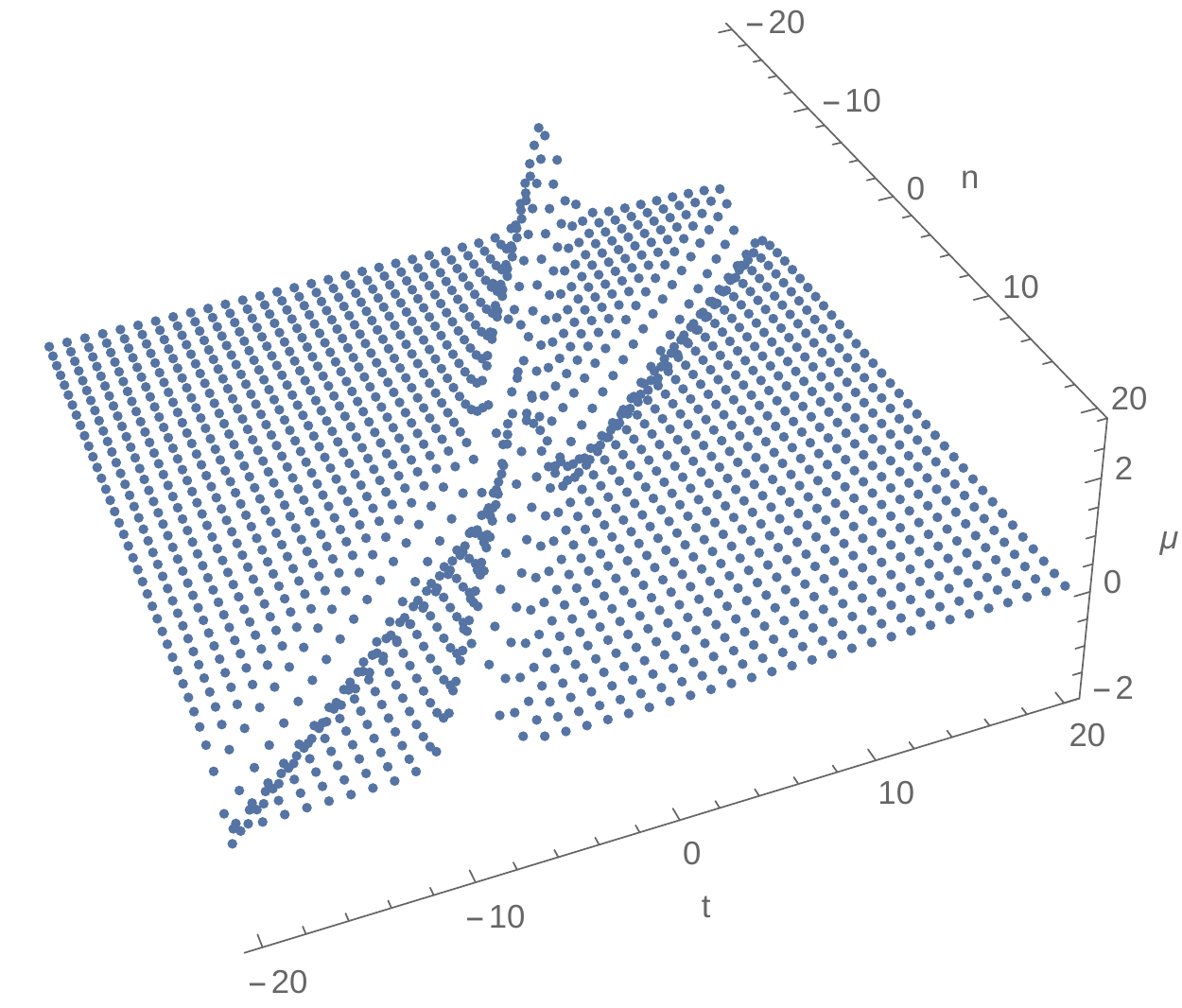}}}
\put(-10,-23){\resizebox{!}{3.5cm}{\includegraphics{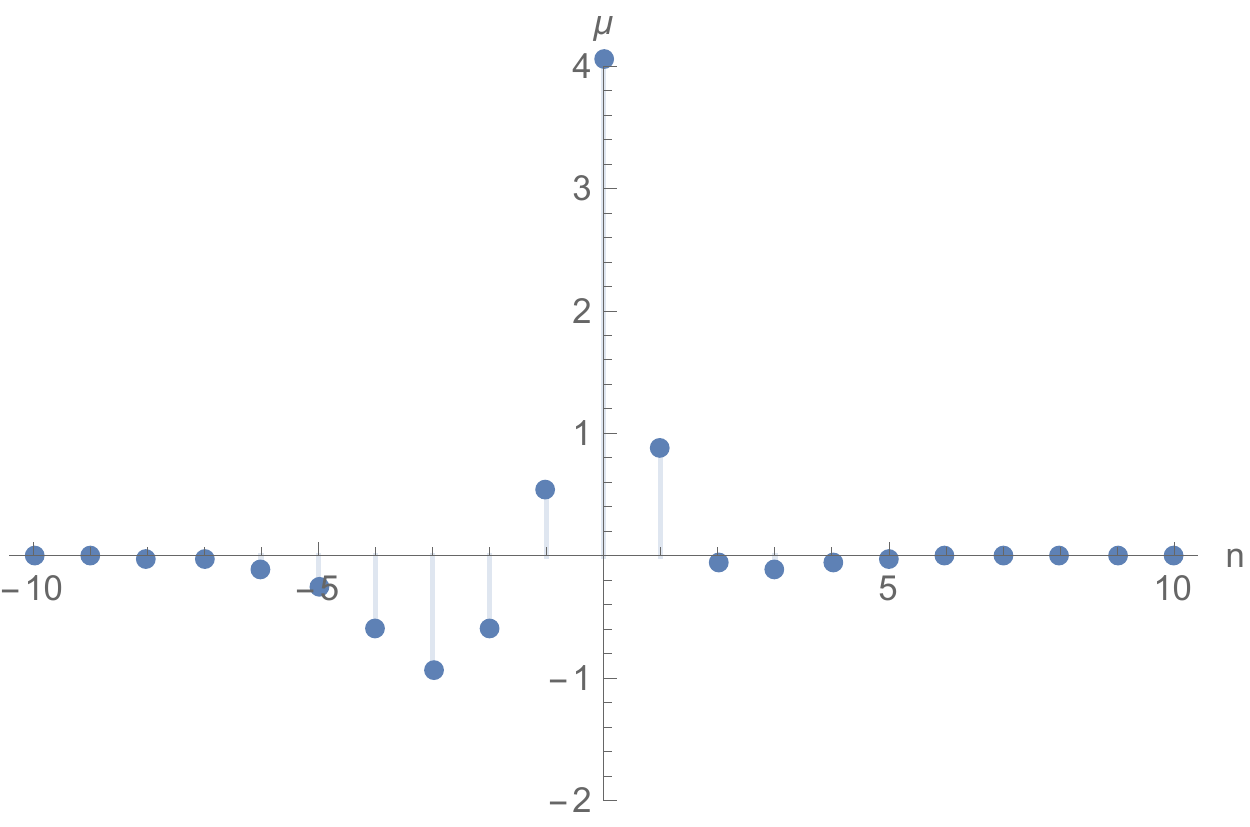}}}
\put(150,-23){\resizebox{!}{3.5cm}{\includegraphics{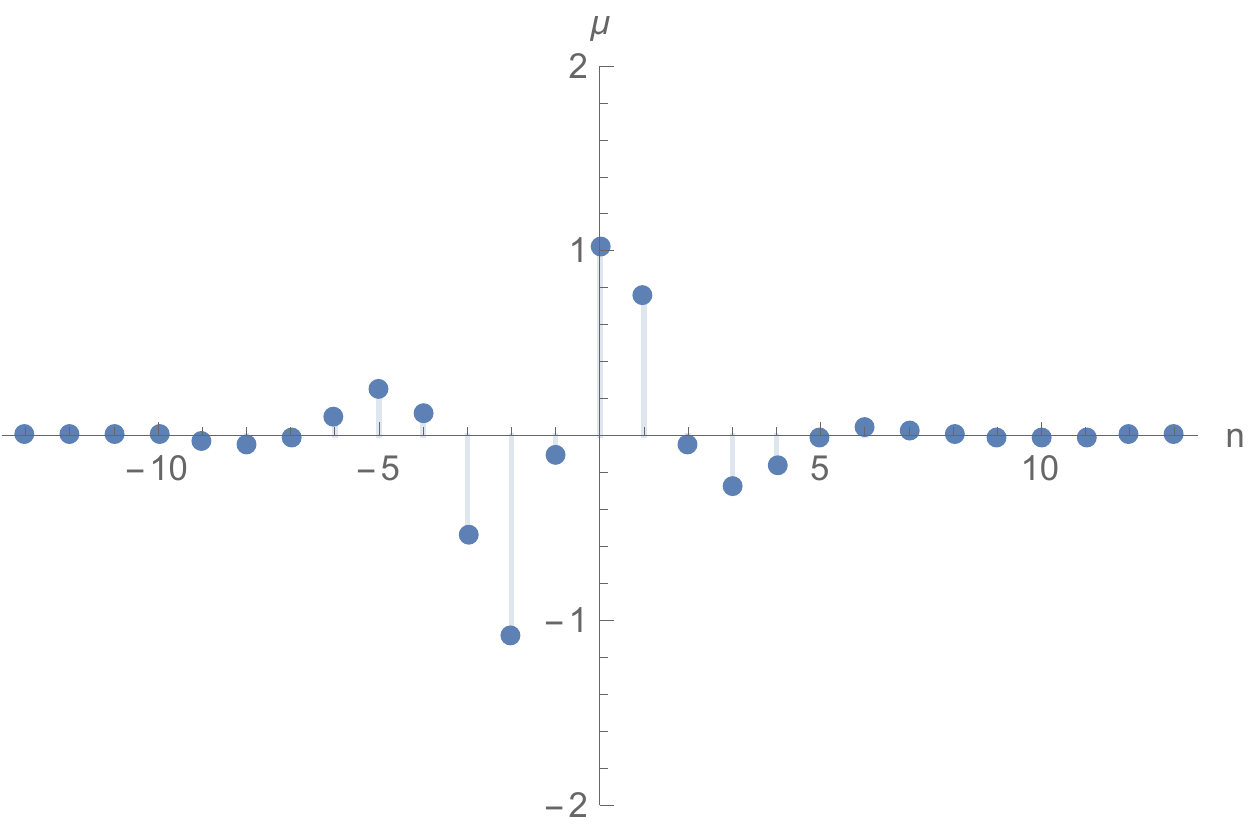}}}
\end{picture}
\end{center}
\vskip 20pt
\begin{center}
\begin{minipage}{15cm}{\footnotesize
\qquad\qquad\qquad\quad(a)\qquad\qquad\qquad\qquad\qquad\qquad\qquad (b) \qquad\qquad\qquad\qquad\qquad\qquad\qquad\quad (c)\\
{\bf Fig. 5.} Solution $\mu$ given by \eqref{sdr-t-2ss} with $\epsilon=1$:
(a) two-soliton solutions with $k_1=1$ and $k_2=1.5$; (b) 2D-plot of (a) at $t=1$;
(c) breather solution with $k_1=0.6+i$ and $k_2=0.6-i$ at $t=1$.}
\end{minipage}
\end{center}

\vspace{.2cm}
\noindent{\it Jordan-block solutions}: Let $L$ be the Jordan-block matrix \eqref{drL-Jor}, the basic entries $\{\Phi_j\}$ are thereby of the form
\begin{align}
\Phi_{j}=\left\{
\begin{aligned}
&\dfrac{\partial^{j-1}_{k_1}e^{\lambda_1}}{(j-1)!},\quad j=1, 2, \ldots, N+1,\\
&\dfrac{\partial^{s-1}_{k_1}e^{-\lambda_1}}{(s-1)!}, \quad j=N+1+s, \quad s=1,2,\ldots,N+1.
\end{aligned}	
\right.
\end{align}

In the case of $N=1$, the simplest Jordan-block solution is
\begin{subequations}
\begin{align}
\label{sdr-t-JB-solu-mu}
& \mu=\dfrac{4[\sinh{4k_1}\cosh2\lambda_1+2\cosh^{2}{k_1}(t\epsilon-4n\sinh^2k_1)\sinh{2\lambda_1}]}
{\epsilon[4\cosh^2{2\lambda_1}+(t\epsilon\coth k_1-2n\sinh{2k_1})^2]}, \\
& \omega=\epsilon^{-1}\ln(\wt{f}/f), \quad f=4\cosh^2{2\lambda_1}+(t\epsilon\coth k_1-2n\sinh{2k_1})^2,
\end{align}	
\end{subequations}
where the behavior of $\mu$ is depicted in Fig. 6.
\vskip20pt
\begin{center}
\begin{picture}(120,80)
\put(-140,-23){\resizebox{!}{4.0cm}{\includegraphics{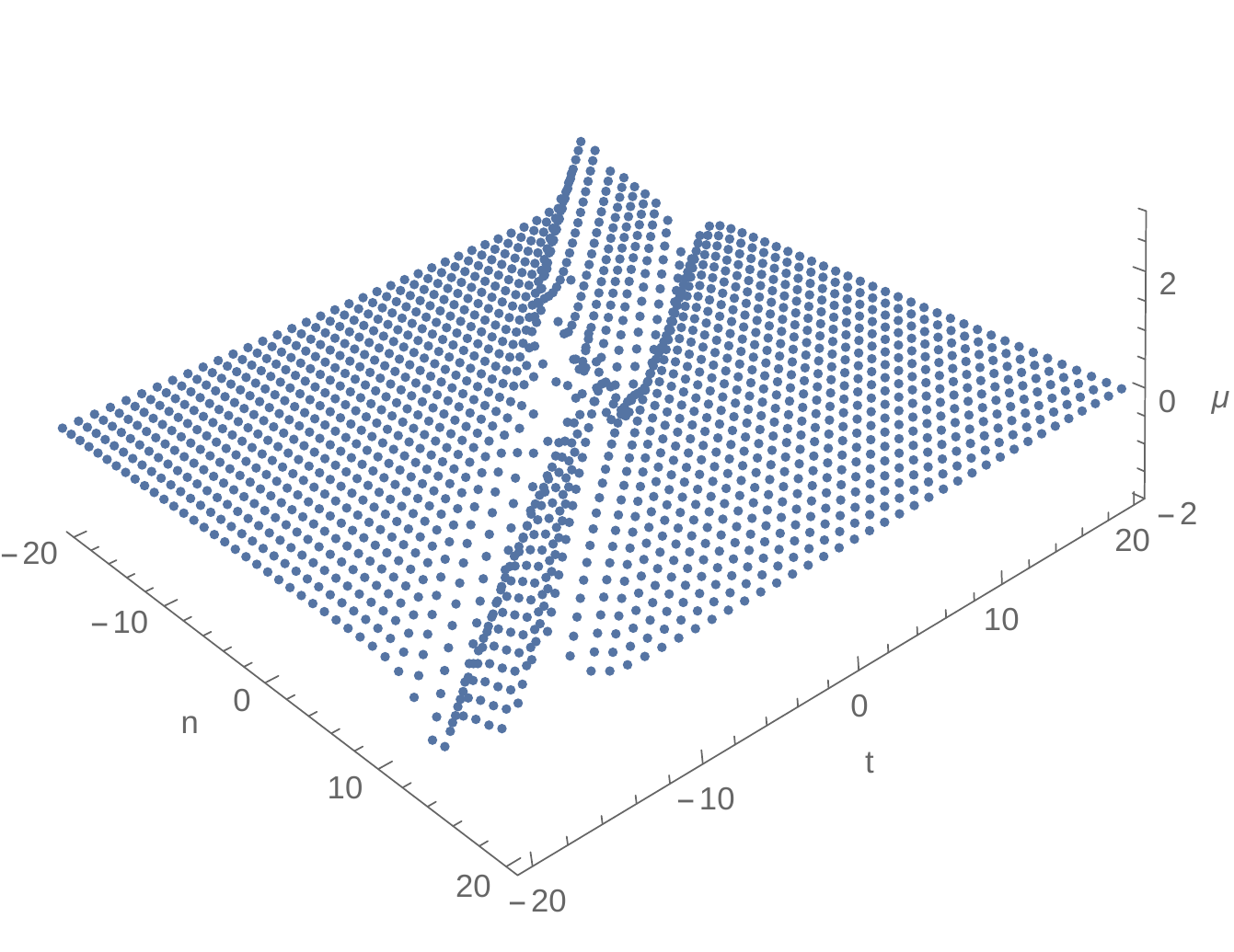}}}
\put(80,-23){\resizebox{!}{4.0cm}{\includegraphics{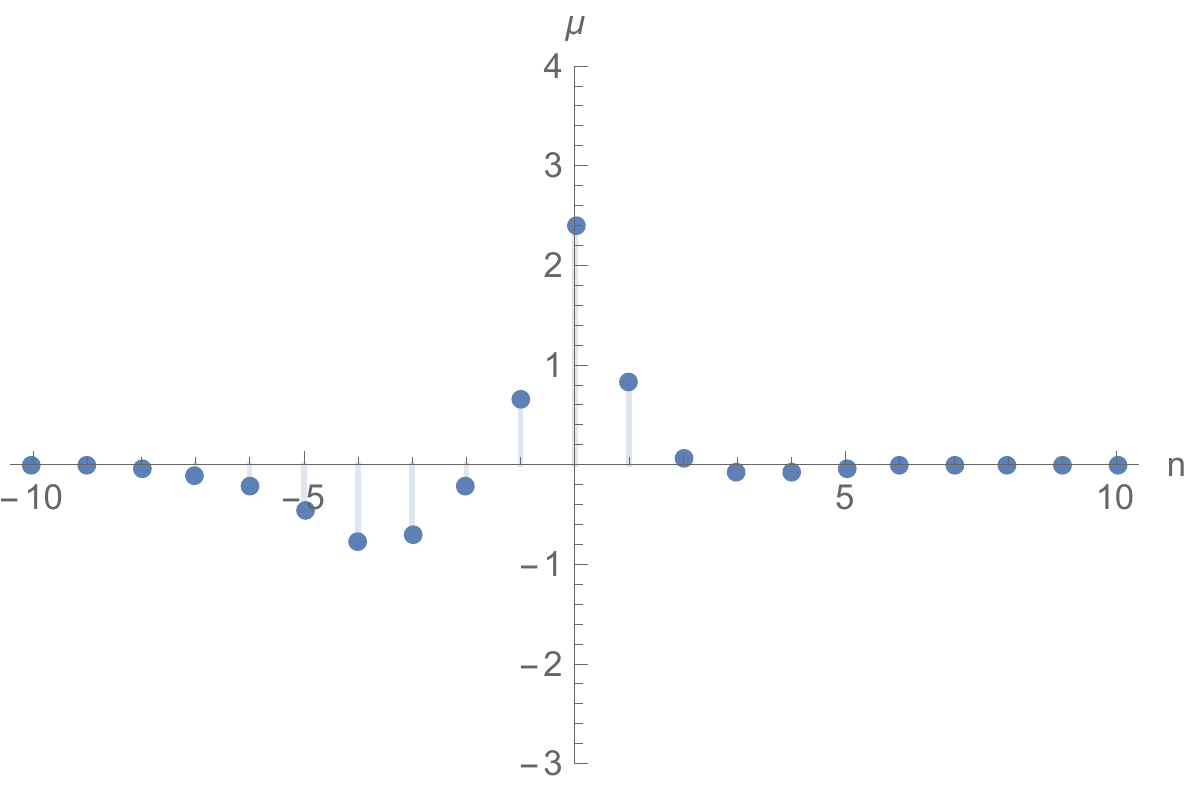}}}
\end{picture}
\end{center}
\vskip 20pt
\begin{center}
\begin{minipage}{15cm}{\footnotesize
\qquad\qquad\qquad\qquad\quad(a)\qquad\qquad\qquad\qquad\qquad\qquad\qquad\qquad\qquad\qquad\qquad\qquad\quad (b) \\
{\bf Fig. 6.} Jordan-block solution $\mu$ given by \eqref{sdr-t-JB-solu-mu} with $k_1=0.5$ and $\epsilon=1$:
(a) shape and motion; (b) 2D-plot of (a) at $t=1$.}
\end{minipage}
\end{center}

\vspace{.2cm}
\noindent{\bf Semi-continuous limit in $n$-direction:} The straight continuum limit can't be
applied to determine the semi-continuous limit in $n$-direction for the discrete exponential
function \eqref{dc-re}. This difficulty can be by-passed, if we introduce
\begin{align}
e^{2l_j}:=\dfrac{4+\delta\epsilon\coth k_j}{4-\delta \epsilon \coth k_j}, \quad
j=1,2,\ldots,N+1,
\end{align}
and rewrite the discrete exponential functions $\{e^{\xi_j}\}$ as
\begin{align}
e^{\xi_j}=e^{\theta_j}:=
e^{l_j m}\bigg(\dfrac{4+\delta\epsilon\coth l_j}{4-\delta \epsilon \coth l_j}\bigg)^{\frac{n}{2}}, \quad
j=1,2,\ldots,N+1.
\end{align}
For the functions $\{e^{\theta_j}\}$, we set
\begin{align}
\label{rcl-nx}
n\rightarrow\infty, \quad \epsilon\rightarrow0, \quad x=n\epsilon\sim O(1),
\end{align}
and then get the limit
\begin{align}
\label{eth-vath}
e^{\theta_j}\rightarrow e^{\vartheta_j}:= e^{l_jm+\frac{\delta\coth l_j}{4}x},\quad
j=1,2,\ldots,N+1.
\end{align}

To proceed, we reinterpret the variables $u$ and $w$ as $u(n,m):=\chi(x,m)$
and $w(n,m)=\varpi(x,m)$. Then, by substituting the Taylor expansions\footnote{Under the limit \eqref{rcl-nx} we have
$w=\dfrac{(f+\epsilon \cdd{f}+\cdots)\wh{f}}
{f(\wh{f}+\epsilon\wh{\cdd{f}}+\cdots)}
\rightarrow\dfrac{(f+\epsilon \cdd{f})\wh{f}}
{f(\wh{f}+\epsilon\wh{\cdd{f}})}
=1+\epsilon\dfrac{\cdd{f}\wh{f}-f\wh{\cdd{f}}}{f\wh{f}}=1+\epsilon(\varpi-\wh{\varpi})$, where $\varpi=\cdd{f}/f$.}
\begin{subequations}
\begin{align}	
& \wt{u}=\chi(x+\epsilon)=\chi+\epsilon \cdd{\chi}+\ldots, \\
& \wh{\wt{u}}=\wh{\chi}(x+\epsilon)=\wh{\chi}+\epsilon\wh{\cdd{\chi}}+\ldots, \\
& w \rightarrow 1+\epsilon(\varpi-\wh{\varpi})
\end{align}	
\end{subequations}
into the rndsG equation \eqref{rndsG}, we immediately obtain another real local and nonlocal semi-discrete sG (rnsdsG-$x$) equation
\begin{subequations}
\label{rnsdsG-x}
\begin{align}
& 2(\wh{\cdd{\chi}}-\cdd{\chi})+(\wh{\chi}+\chi)[2(\wh\varpi-\varpi)-\delta]=0,\\
& \cdd{\varpi}=\eta \chi\chi_{\sigma},
\end{align}
\end{subequations}
where $\varpi=\varpi_{\sigma}$ and $\cdd{f}$ denotes the derivative of $f$ with respect to $x$. Unsurprisingly,
equation \eqref{rnsdsG-x} is preserved under transformation $\chi\rightarrow -\chi$,
and the equation \eqref{rnsdsG-x} with $(\sigma,\eta)=(\pm 1,1)$ and $(\sigma,\eta)=(\pm 1,-1)$ can be
transformed into each other by taking $\chi\rightarrow i\chi$.

Now we absorb $e^{\Gamma N}$ in $\Phi(n,m)$ and focus on the basic column vector $\Phi(n,m):=\phi(x,m)$.
Under the continuum limit \eqref{rcl-nx}, the Taylor expansion of $E^{2j}\Phi(n,m)$ is given by
\begin{align*}
E^{2j}\Phi(n,m)=\Phi(n+2j,m)=\phi(x+2j\epsilon,m)=\phi+2j\epsilon\partial_{x}\phi+
\dfrac{(2j\epsilon)^{2}}{2!}\partial_{x}^{2}\phi+\ldots
\end{align*}	
with $j=1,2,\ldots,N+1$.
Simultaneously, we have
\begin{align}	
f\rightarrow |\Delta_{N}^{2}||\wh{\phi^{(N)}};T\wh{\phi_{\sigma}^{(N)}}|, \quad
g\rightarrow\frac{1}{\epsilon}|\Delta_{N+1}\Delta_{N-1}||\wh{\phi^{(N+1)}};T\wh{\phi_{\sigma}^{(N-1)}}|,
\end{align}	
where
\begin{align}
	\Delta_{N}=
	\left(
	\begin{array}{cccc}
		2\epsilon &  \dfrac{(2\epsilon)^{2}}{2!} &  \cdots& \dfrac{(2\epsilon)^{N}}{N!}\\
		4\epsilon &  \dfrac{(4\epsilon)^{2}}{2!} &  \cdots& \dfrac{(4\epsilon)^{N}}{N!}\\		
		\vdots& \vdots&\cdots&\vdots\\
		2N\epsilon&\dfrac{(2N\epsilon)^{2}}{2!}& \cdots & \dfrac{(2N\epsilon)^{N}}{N!}
	\end{array}\right).
\end{align}
Consequently, for the dependent variables $u$ and $w$ we have
\begin{subequations}
\begin{align}
& u=\dfrac{g}{f}\rightarrow
\chi=\dfrac{|\Delta_{N+1}\Delta_{N-1}||\wh{\phi^{(N+1)}};T\wh{\phi_{\sigma}^{(N-1)}}|}
{\epsilon|\Delta_{N}^{2}||\wh{\phi^{(N)}};T\wh{\phi_{\sigma}^{(N)}}|}
=\dfrac{2|\wh{\phi^{(N+1)}};T\wh{\phi_{\sigma}^{(N-1)}}|}
{|\wh{\phi^{(N)}};T\wh{\phi_{\sigma}^{(N)}}|}, \\
& w=\dfrac{\wt{f}\wh{f}}{f\wh{\wt{f}}}\rightarrow \varpi=
\dfrac{\partial_x|\wh{\phi^{(N)}};T\wh{\phi_{\sigma}^{(N)}}|}{|\wh{\phi^{(N)}};T\wh{\phi_{\sigma}^{(N)}}|}.
\end{align}
\end{subequations}

Double Casoratian solutions to the rnsdsG-$x$ equation \eqref{rnsdsG-x} can be summarized by the following theorem.
\begin{Thm}
\label{Thm-rnsdsG-x-solu}
The functions $\chi=g/f$ and $\varpi=\cdd{f}/f$ with
\begin{align}
\label{rnsdsG-x-solu}
f=|\wh{\phi^{(N)}};T\wh{\phi_{\sigma}^{(N)}}|,\quad
g=2|\wh{\phi^{(N+1)}};T\wh{\phi_{\sigma}^{(N-1)}}|,
\end{align}
where $\phi=e^{\Omega m+\frac{\delta\coth \Omega}{4}x}C^{+}$,
solve the rnsdsG-$x$ equation \eqref{rnsdsG-x}, in which $T$ is a constant matrix of order $2(N+1)$,
satisfying matrix equations
\begin{align}
\label{rnsdsG-x-L-T}
\Omega T+\sigma T\Omega=\bm 0,\quad T^{2}=-\sigma\eta I.
\end{align}
\end{Thm}

We take $(\sigma,\eta)=(-1,1)$ and set
\begin{align}
\label{csd-OmT-form}
\Omega=\left(
\begin{array}{cc}
S & \bm 0  \\
\bm 0 & -S
\end{array}\right)	,\quad
T=\left(
\begin{array}{cc}
I & \bm 0  \\
\bm 0 & -I
\end{array}\right).
\end{align}	

\vspace{.2cm}
\noindent{\it Soliton solutions}: Let $S$ be a diagonal matrix
\begin{align}
\label{rsdx-S-diag}
S=\mathrm{Diag}(l_{1},l_{2},\dots,l_{N+1}),
\end{align}	
and $C^{+}=\bar{I}$. The one-soliton solution is thus given by
\begin{subequations}
\begin{align}
\label{sdr-x-1ss}
& \chi=\delta\coth{l_{1}}\sech 2\vartheta_1/2, \\
& \varpi=\delta\coth{l_{1}}\tanh 2\vartheta_1/2.
\end{align}	
\end{subequations}
Solution $\chi$ appears as a stable travelling wave with amplitude $\delta\coth{l_{1}}/2$ and travelling velocity $-4l_1/(\delta\coth{l_{1}})$.
The width is proportional to $2/(\delta\coth{l_{1}})$. Soliton and anti-soliton
determined by the sign of the parameter $l_1\delta$ are depicted in Fig. 7.
\vskip20pt
\begin{center}
\begin{picture}(120,80)
\put(-160,-23){\resizebox{!}{4.0cm}{\includegraphics{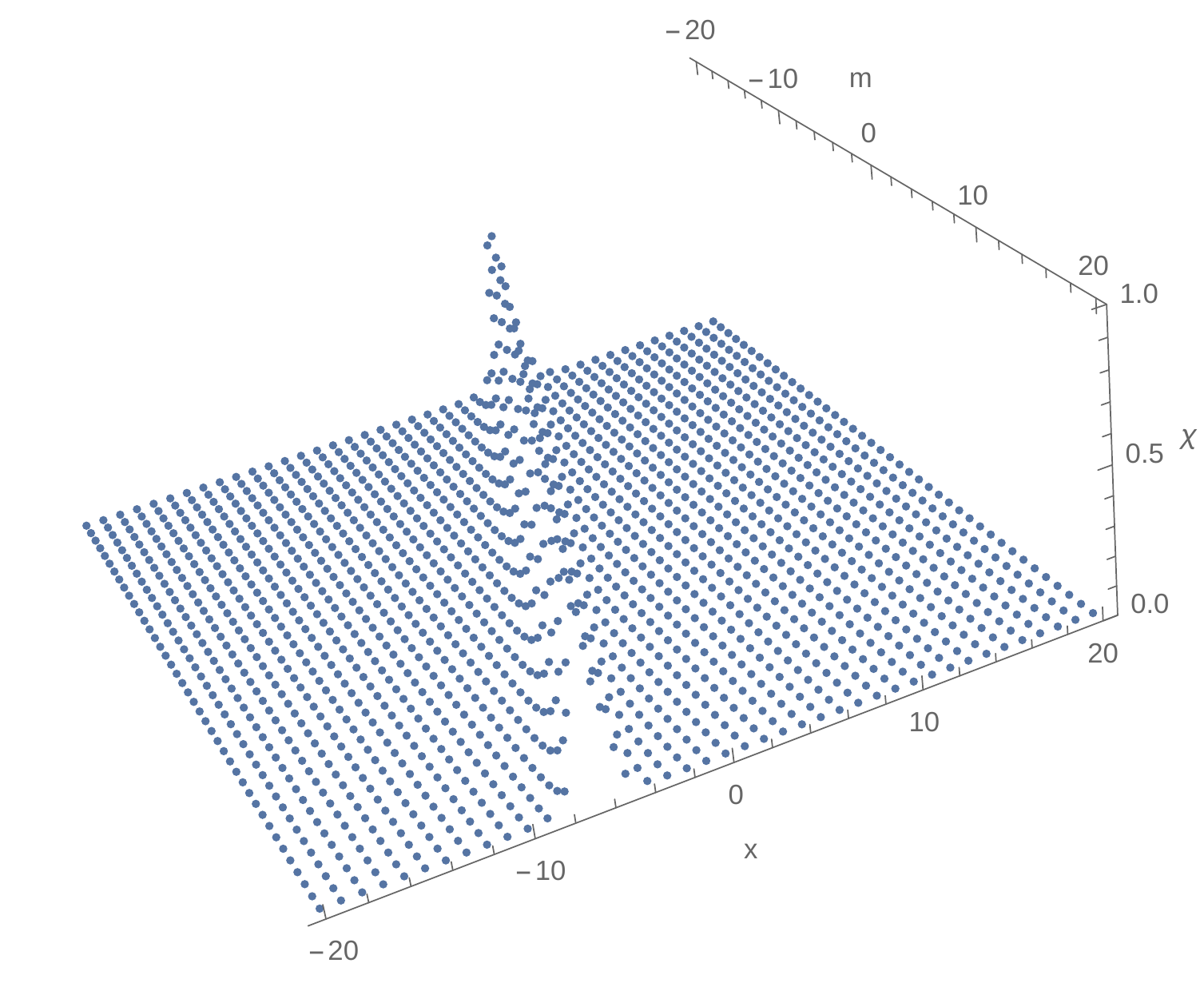}}}
\put(10,-23){\resizebox{!}{3.5cm}{\includegraphics{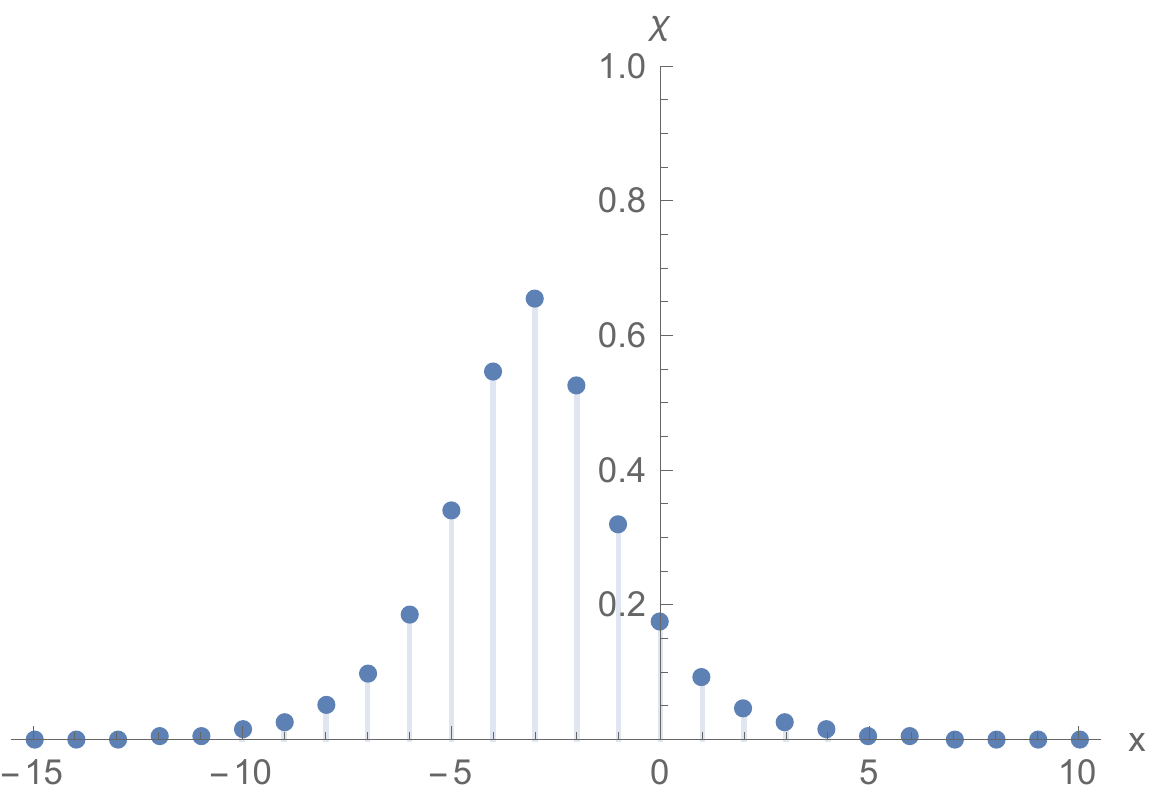}}}
\put(150,-23){\resizebox{!}{3.5cm}{\includegraphics{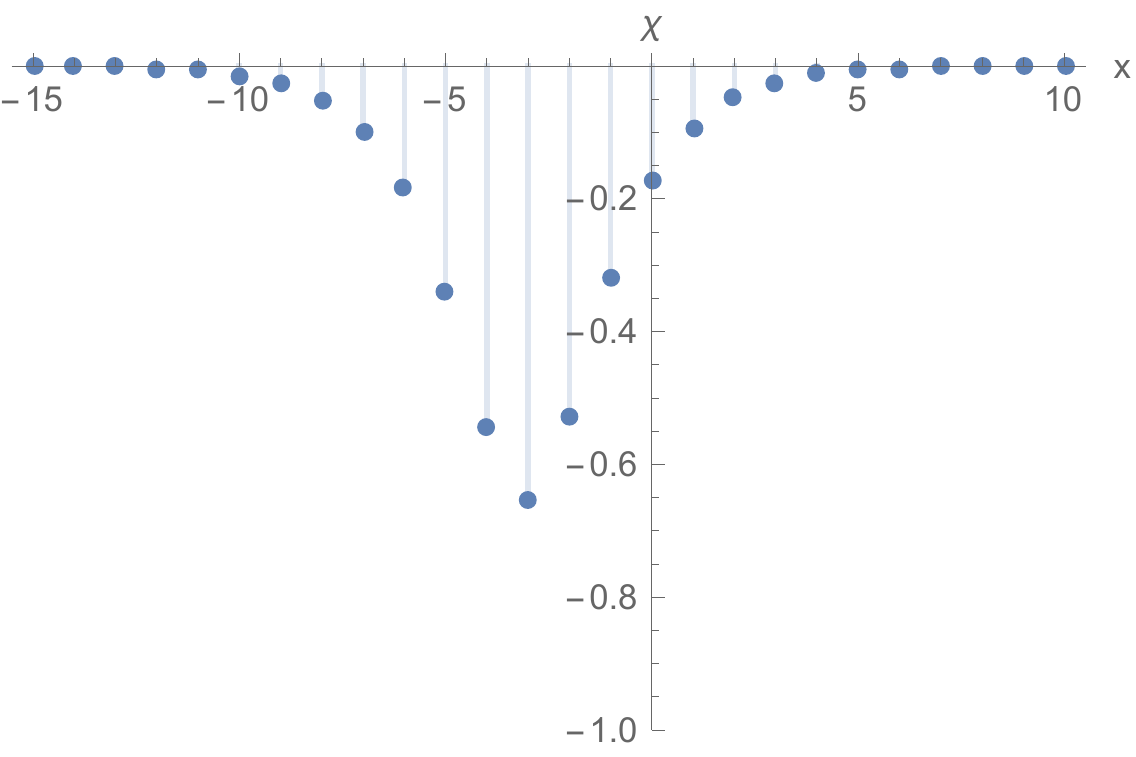}}}
\end{picture}
\end{center}
\vskip 20pt
\begin{center}
\begin{minipage}{15cm}{\footnotesize
\qquad\qquad\qquad\quad(a)\qquad\qquad\qquad\qquad\qquad\qquad\qquad\qquad (b) \qquad\qquad\qquad\qquad\qquad\qquad\quad (c)\\
{\bf Fig. 7}. One-soliton solution $\chi$ given by \eqref{sdr-x-1ss} with $\delta=1$:
(a) shape and movement with $k_1=1$;
(b) soliton for $k_1=1$ at $m=1$;
(c) anti-soliton for $k_1=-1$ at $m=1$.}
\end{minipage}
\end{center}

The two-soliton solutions are of form
\begin{align}
\label{sdr-x-2ss}
\chi=g/f, \quad \varpi=\cdd{f}/f,
\end{align}
in which
\begin{subequations}
\begin{align}
f= & (\csch{l_{1}}\csch{l_{2}})^{2}
[(\cosh2l_1\cosh2l_2-1)\cosh2\vartheta_1\cosh2\vartheta_2 \nn \\
& -\sinh2l_1\sinh2l_2(1+\sinh2\vartheta_1\sinh2\vartheta_2)], \\
g=& \delta (\coth^{2}{l_{1}}-\coth^{2}{l_{2}})(\coth l_1\cosh2\vartheta_2-\coth l_2\cosh2\vartheta_1).
\end{align}	
\end{subequations}
Fig. 8 displays the movement of solution $\chi$, whose dynamic behavior is listed in Theorem 6.
\vskip20pt
\begin{center}
\begin{picture}(120,80)
\put(-170,-23){\resizebox{!}{4.0cm}{\includegraphics{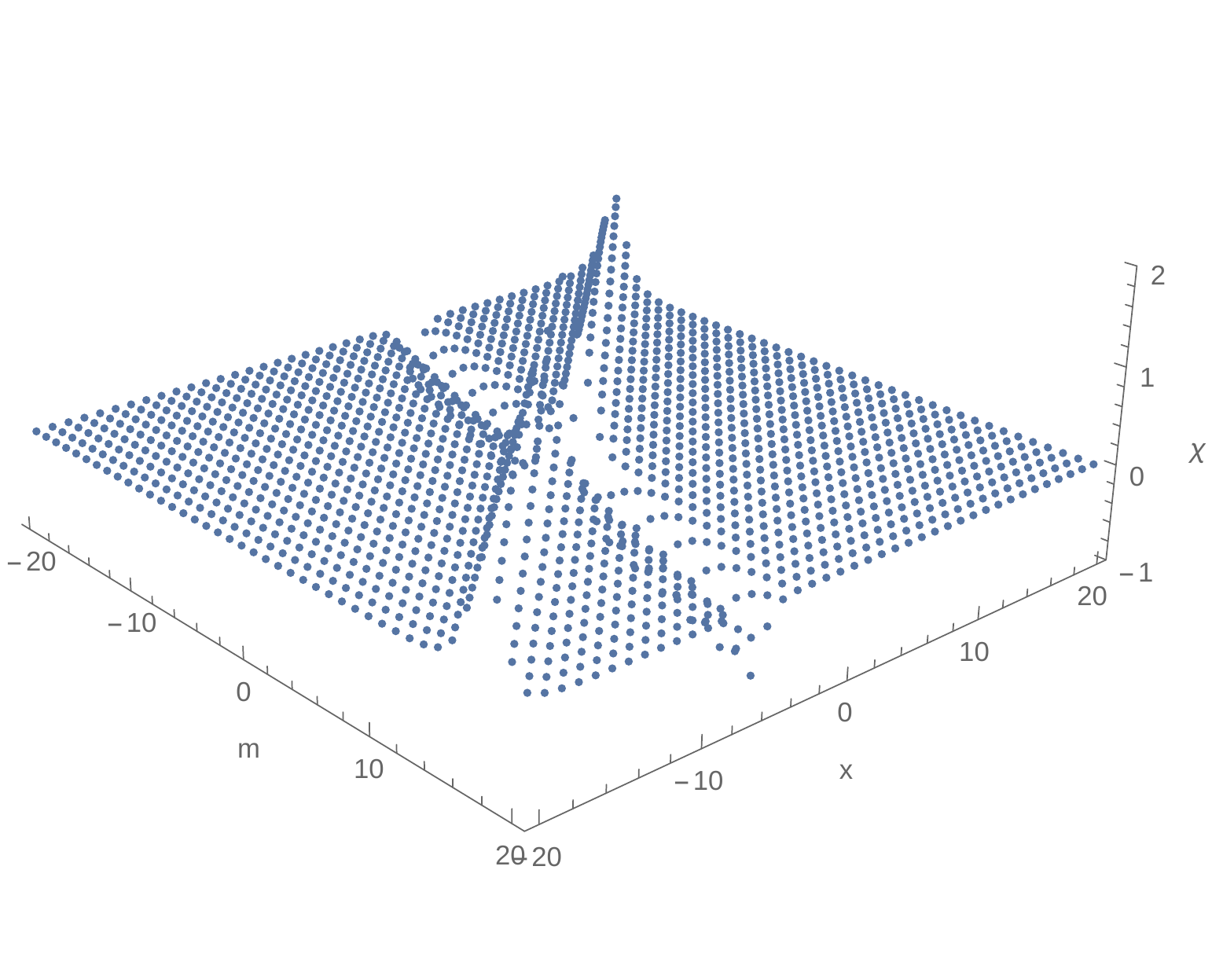}}}
\put(-10,-23){\resizebox{!}{3.5cm}{\includegraphics{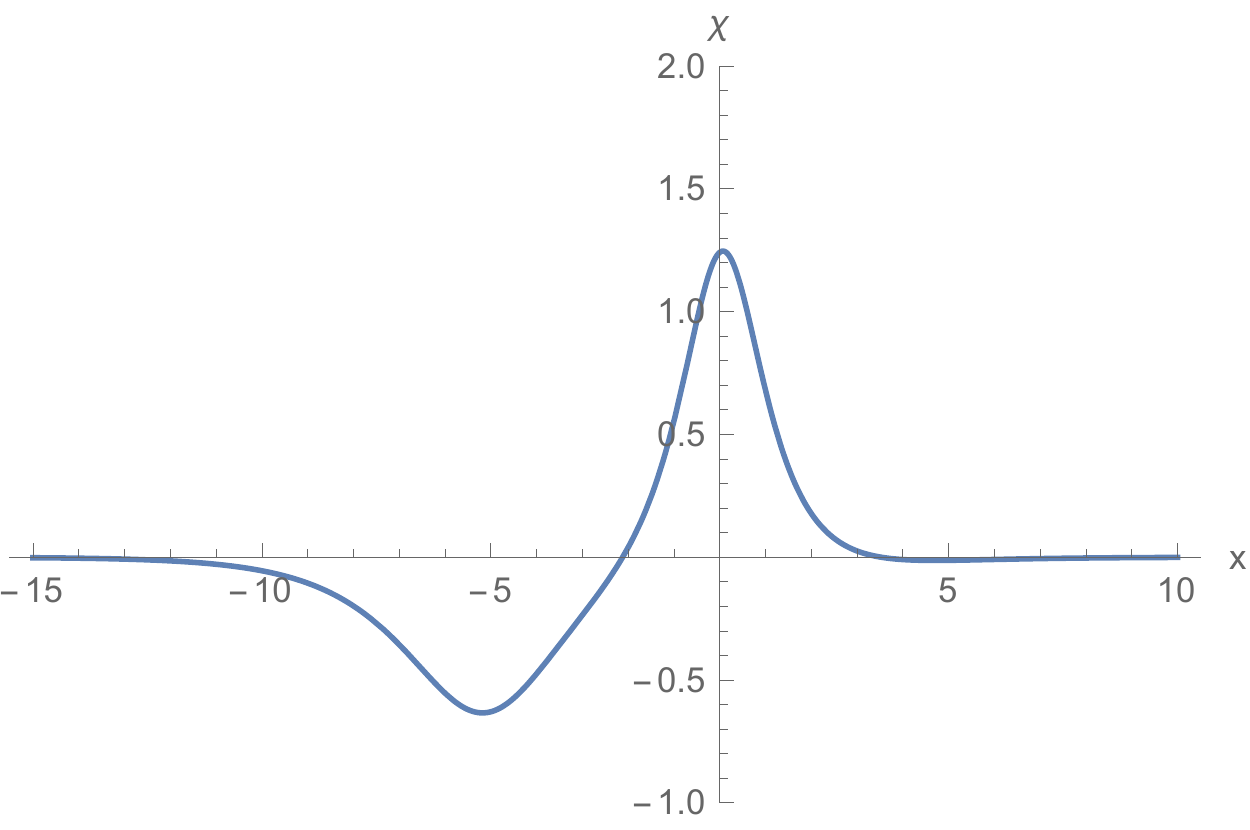}}}
\put(150,-23){\resizebox{!}{3.5cm}{\includegraphics{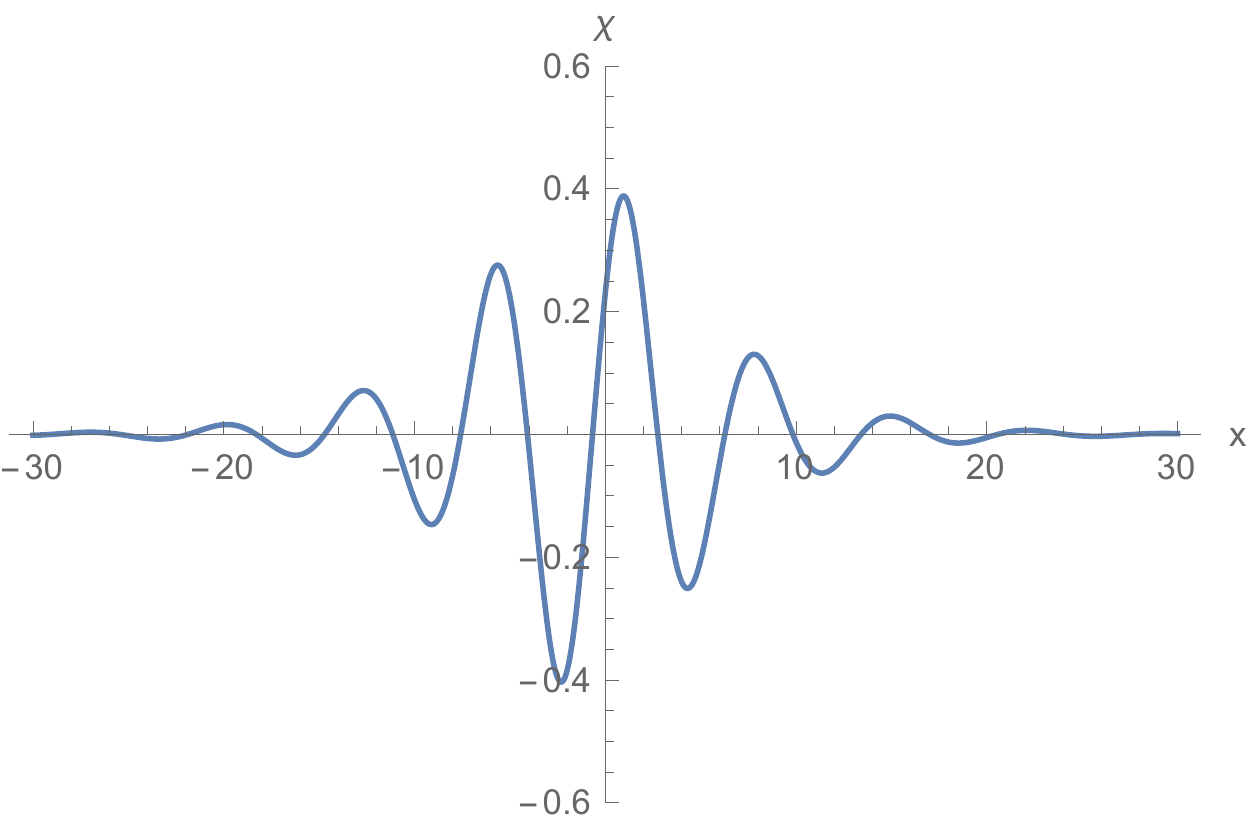}}}
\end{picture}
\end{center}
\vskip 20pt
\begin{center}
\begin{minipage}{15cm}{\footnotesize
\qquad\qquad\quad(a)\qquad\qquad\qquad\qquad\qquad\qquad\qquad\qquad\quad (b) \qquad\qquad\qquad\qquad\qquad\qquad\qquad\quad (c)\\
{\bf Fig. 8.} Solution $\chi$ given by \eqref{sdr-x-2ss} with $\epsilon=1$:
(a) two-soliton solutions with $k_1=1$ and $k_2=2$; (b) 2D-plot of (a) at $m=1$;
(c) breather solution with $k_1=0.2+i$ and $k_2=0.2-i$ at $m=1$.}
\end{minipage}
\end{center}
\begin{Thm}
Suppose that $0<l_{1}<l_{2}$ and $\delta>0$.
Then, when $m\rightarrow\pm\infty$, the $l_{1}$-soliton asymptotically follows:
\begin{subequations}
\begin{align}
top \: point \: traces
&: x(m)=\dfrac{2}{\delta \coth l_{1}}\bigg(\mp\ln\dfrac{\sinh(l_{2}-l_{1})}
{\sinh(l_{1}+l_{2})}-2l_{1}m\bigg),\\
amplitude
&: \chi=\dfrac{\coth{l_{1}}(\coth^{2}{l_{1}}-\coth^{2}{l_{2}})}
{2(\csch{l_{1}}\csch{l_{2}})^{2}\sinh({l_{2}-l_{1}})
	\sinh(l_{1}+l_{2})},\\
speed
&: -\dfrac{4l_{1}}{\delta\coth{l_{1}}}, \\
phase \: shift &: -\dfrac{4}{\delta \coth l_{1}}\ln\dfrac{\sinh(l_{2}-l_{1})}
{\sinh(l_{1}+l_{2})},
\end{align}	
\end{subequations}
and the $l_{2}$-soliton asymptotically follows:
\begin{subequations}
\begin{align}
top \: point \: traces
&: x(m)=\dfrac{2}{\delta \coth l_{2}}\bigg(\pm\ln\dfrac{\sinh(l_{2}-l_{1})}
{\sinh(l_{1}+l_{2})}-2l_{2}m\bigg),\\
amplitude
&: \chi=\dfrac{\coth{l_{2}}(\coth^{2}{l_{2}}-\coth^{2}{l_{1}})}
{2(\csch{l_{1}}\csch{l_{2}})^{2}\sinh({l_{2}-l_{1}})
	\sinh({l_{1}+l_{2}})}, \\
speed
&: -\dfrac{4l_{2}}{\delta\coth{l_{2}}}, \\
phase \: shift &: \dfrac{4}{\delta \coth l_{2}}\ln\dfrac{\sinh(l_{2}-l_{1})}
{\sinh(l_{1}+l_{2})}.
\end{align}	
\end{subequations}
\end{Thm}

\vspace{.2cm}
\noindent{\it Jordan-block solutions}: Let $S$ be the Jordan-block matrix of order $N+1$ given by
\begin{align}
\label{rsdx-S-Jor}
S=\left(
\begin{array}{cccc}
l_1 & 0 & \cdots& 0\\
1 & l_1 & \cdots & 0 \\		
\vdots& \ddots&\ddots&\vdots\\
0 &\cdots & 1 & l_1
\end{array}\right),
\end{align}	
and $C^{+}=\breve{I}$. Then $\phi$ is composed by
\begin{align}
\label{rsdx-S-Jor}
\phi_{j}=\left\{
\begin{aligned}
	&\dfrac{\partial_{l_1}^{j-1}e^{\vartheta_1}}{(j-1)!},\quad j=1,2,\dots,N+1, \\
	&\dfrac{\partial_{l_1}^{s-1}e^{-\vartheta_1}}{(s-1)!},
	\quad j=N+1+s,\quad s=1,2,\dots,N+1.
\end{aligned}	
\right.
\end{align}
The simplest Jordan-block solution is expressed as
\begin{subequations}
\begin{align}
\label{sdr-t-JB-solu-chi}
& \chi=\dfrac{\delta\coth{l_1}\csch^{2}{l_1}[4\csch^{2}{l_1}\cosh2\vartheta_1
+2(4m-\delta x\csch^{2}{l_1})\coth{l_1}\sinh2\vartheta_1]}
{4\csch^4{l_1}\cosh^22\vartheta_1+(4m-\delta x\csch^{2}{l_1})^{2}\coth^2l_1}, \\
& \varpi=\dfrac{2\delta\coth l_1\csch^4{l_1}(x\delta\coth l_1-4m\cosh l_1\sinh l_1+\sinh4\vartheta_1)}
{4\csch^4{l_1}\cosh^22\vartheta_1+(4m-\delta x\csch^{2}{l_1})^{2}\coth^2l_1}.
\end{align}	
\end{subequations}
\vskip20pt
\begin{center}
\begin{picture}(120,80)
\put(-140,-23){\resizebox{!}{4.0cm}{\includegraphics{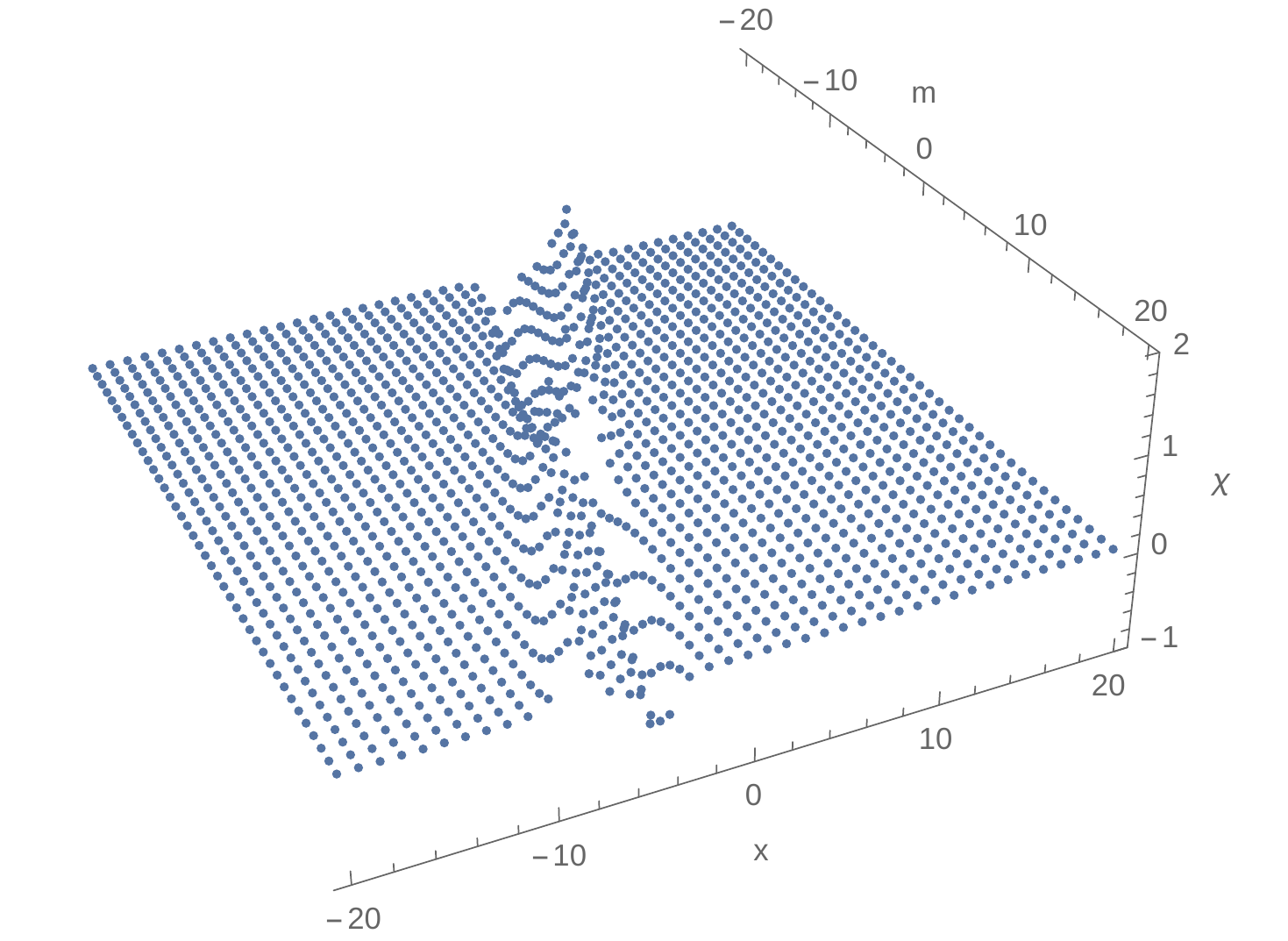}}}
\put(80,-23){\resizebox{!}{4.0cm}{\includegraphics{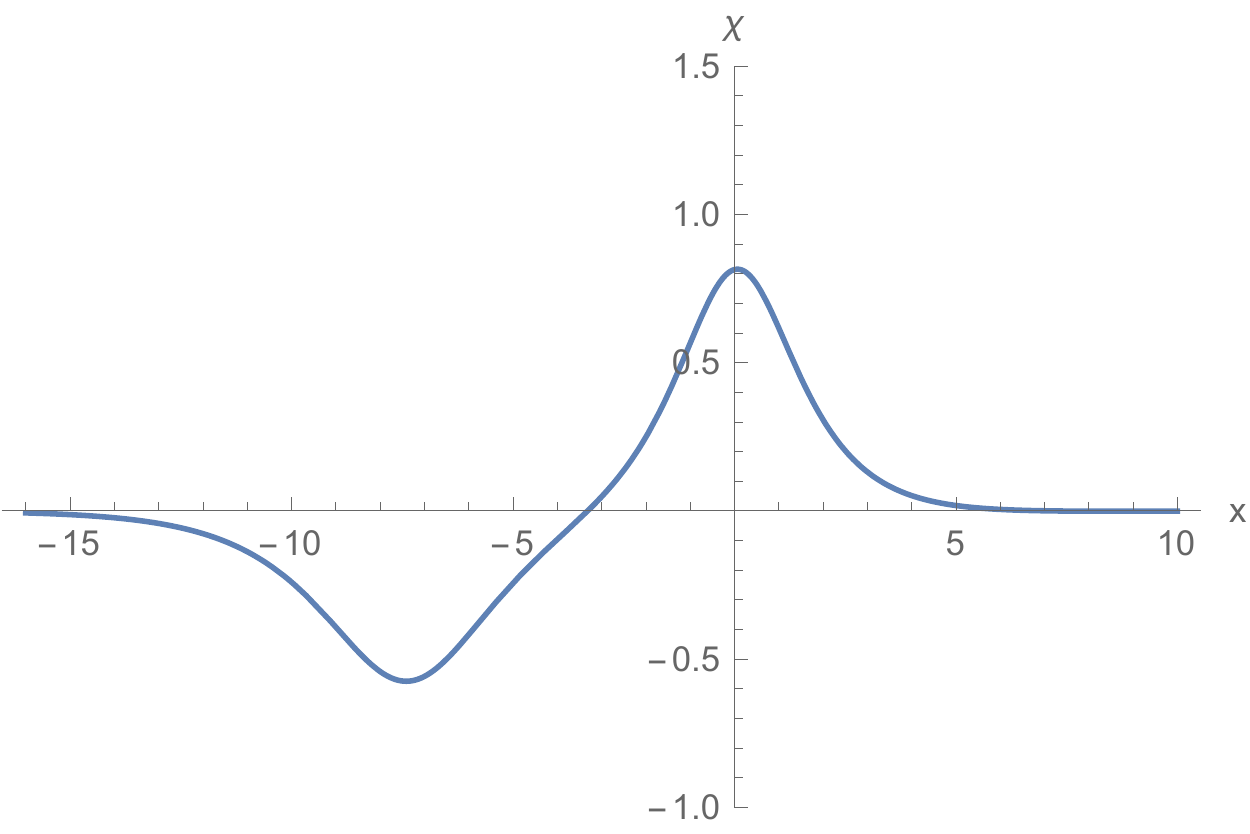}}}
\end{picture}
\end{center}
\vskip 20pt
\begin{center}
\begin{minipage}{15cm}{\footnotesize
\qquad\qquad\qquad\qquad\quad(a)\qquad\qquad\qquad\qquad\qquad\qquad\qquad\qquad\qquad\qquad\qquad\qquad\quad (b) \\
{\bf Fig. 9.} Jordan-block solution $\chi$ given by \eqref{sdr-t-JB-solu-chi} with $k_1=1$ and $\delta=1$:
(a) shape and motion; (b) 2D-plot of (a) at $m=1$.}
\end{minipage}
\end{center}

\vspace{.2cm}
\noindent{\bf Continuous limits:} Through continuous limits one can recover the rnsdsG-$t$ equation
\eqref{rnsdsG-t} or the rnsdsG-$x$ equation \eqref{rnsdsG-x} to the real local and nonlocal continuous sG (rncsG) equation.

To derive the rncsG equation from the rnsdsG-$t$ equation \eqref{rnsdsG-t}, we start from its semi-discrete exponential functions $\{e^{\lambda_j}\}$.
With new spectral parameters
\begin{align}
a_{j}=(\epsilon\coth k_{j})^{-1}, \quad
j=1,2,\ldots,N+1
\end{align}
and limit \eqref{rcl-nx}, we find $\{e^{\lambda_j}\rightarrow e^{\kappa_j}=e^{a_{j}x+\frac{t}{4a_{j}}}\}$.
Proceeding as before, we reinterpret the variables $\mu$ and $\omega$ as $\mu(n,t):=\al(x,t)$ and $\omega(n,t):=\beta(x,t)$.
Inserting the Taylor expansions
\begin{subequations}
\begin{align}	
& \wt{\mu}=\al(x+\epsilon)=\al+\epsilon\cdd{\al}+\ldots, \\
& \wt{\mu}'=\al'(x+\epsilon)=\al'+\epsilon \cdd{\al}'+\ldots, \\
& e^{\epsilon(\omega-\undertilde{\omega})}-1=\epsilon(\omega-\dt{\omega})+\epsilon^2(\omega-\dt{\omega})^2/2!+
\ldots=\epsilon^2\cdd{\be}+\ldots
\end{align}	
\end{subequations}
into the equation \eqref{rnsdsG-t} and then to leading order we obtain the rncsG equation
\begin{align}	
\label{rncsG}
\cdd{\al}'+2\al \be'=\al, \quad \cdd{\be}=\eta \al\al_{\sigma},
\end{align}	
where $\be=\be_{\sigma}$.
The equation \eqref{rncsG} is preserved under transformation $\al\rightarrow -\al$, and
the equation \eqref{rncsG} with $(\sigma,\eta)=(\pm 1,1)$ and $(\sigma,\eta)=(\pm 1,-1)$ can be
transformed into each other by taking $\al\rightarrow i\al$.
The rncsG equation \eqref{rncsG}, firstly proposed in \cite{AbMu-2016}, has been studied by many
researchers from various methods, such as inverse scattering transform \cite{AFLM-SAPM} and bilinearization reduction method \cite{CDLZ}.

Similar to earlier case, for the rnsdsG-$x$ equation \eqref{rnsdsG-x}, we introduce
\begin{align}
b_{j}=\dfrac{\delta}{4}\coth l_{j}, \quad
j=1,2,\ldots,N+1.
\end{align}
It is readily to see that by limit \eqref{rcl-nt} the semi-discrete exponential functions $\{e^{\vartheta_j}\}$ satisfy
$\{e^{\vartheta_j}\rightarrow e^{\iota_j}:=e^{b_{j}x+\frac{t}{4b_{j}}}\}$.
We reinterpret the variables $\chi$ and $\varpi$ as $\chi(x,m):=\al(x,t)$ and $\varpi(x,m):=\be(x,t)$. Then, taking the Taylor expansions
\begin{subequations}
\begin{align}	
& \wh{\chi}=\al(t+\delta)=\al+\delta\al'+\ldots, \\
& \wh{\cdd{\chi}}=\cdd{\al}(t+\delta)=\cdd{\al}+\delta\cdd{\al}'+\ldots, \\
& \wh{\varpi}=\be(t+\delta)=\be+\delta\be'+\ldots
\end{align}	
\end{subequations}
into equation \eqref{rnsdsG-x}, we get the rncsG equation \eqref{rncsG}.

As indicated in \cite{CDLZ} as well as the Theorem \ref{Thm-rnsdsG-x-solu}, we can make a clear description
of exact solutions for the rncsG equation \eqref{rncsG}.
\begin{Thm}
\label{Thm-rncsG-solu}
The functions $\al=g/f$ and $\be=\cdd{f}/f$ with
\begin{align}
\label{rncsG-solu}
f=|\wh{\phi^{(N)}};T\wh{\phi_{\sigma}^{(N)}}|,\quad
g=2|\wh{\phi^{(N+1)}};T\wh{\phi_{\sigma}^{(N-1)}}|,
\end{align}
solve the rncsG equation \eqref{rncsG}, where $\phi=e^{B x+(4B)^{-1}t}C^{+}$ and $T$ is a constant matrix of order $2(N+1)$ satisfying
\begin{align}
\label{rncsG-B-T}
B T+\sigma TB=\bm 0,\quad T^{2}=-\sigma\eta I.
\end{align}
\end{Thm}

\section{Complex reduction: solutions and continuum limits}

In the previous section, we have applied the bilinearization reduction technique to construct exact solutions
for the real local and nonlocal sG type equations.
In this section, we consider complex reduction of the dnAKNS equation \eqref{dnAKNS},
which yields the complex local and nonlocal discrete sG equation.
By solving the matrix equation set, one-soliton solution and dynamical behaviors are presented.
We then inspect continuum limits of the resulting complex local and nonlocal discrete sG equation,
as well as exact solutions and dynamics.
For convenience, we call the resulting complex local and nonlocal sG type equations as cndsG, cnsdsG-$t$, cnsdsG-$x$ and cncsG,
respectively.

\subsection{Complex reduction}

Imposing complex reduction
\begin{align}
\label{com-Re}
v=\eta u^*_\sigma, \quad \eta, \sigma=\pm 1,
\end{align}	
where asterisk denotes the complex conjugate, into the dnAKNS equation \eqref{dnAKNS}, we arrive at the cndsG equation
\begin{subequations}
\label{cndsG}
\begin{align}
& 4\big(u+\wh{\wt{u}}-(\wt{u}+\wh{u})w\big)
=\delta\epsilon\big(u+\wh{\wt{u}}+(\wt{u}+\wh{u})w \big), \\
\label{cndsG-rb}
& (1+\eta\epsilon^{2}\wh{u}\wh{u}^*_{\sigma})w=(1+\eta\epsilon^{2}uu^*_{\sigma})\dt{w},
\end{align}
\end{subequations}
where $w=w^*_{\sigma}$.
Equation \eqref{cndsG} is a local equation as $\sigma=1$ and a nonlocal equation as
$\sigma=-1$. This equation is preserved under transformations $u\rightarrow -u$ and $u\rightarrow \pm iu$.

To implement reduction on the level of double Casoratian solutions, we take $M=N$ and then have the following result.
\begin{Thm}
The functions $u=g/f$ and $w=\wt{f}\wh{f}/(f\wh{\wt{f}})$ with
\begin{align}
\label{cndsG-solu}
f=|e^{-N\Ga}\wh{\Phi^{(N)}};
e^{N\Ga}T\wh{\Phi_{\sigma}^{*(-N)}}|,\quad
g=(1/\epsilon)|e^{-N\Ga}\wh{\Phi^{(N+1)}};
e^{N\Ga}T\wh{\Phi_{\sigma}^{*(-N+1)}}|,
\end{align}
solve the cndsG equation \eqref{cndsG}, if the $(2N+2)$-th order column vector $\Phi$ is given by \eqref{dnKANS-Phsi-Ga},
where $T\in\mathbb{C}^{(2N+2)\times(2N+2)}$ in \eqref{cndsG-solu} is a constant matrix, satisfying
\begin{align}
\label{cndsG-AT}
\Ga T+\sigma T\Ga^*=\bm 0,\quad TT^*=\left\{
\begin{array}{l}
-\eta I, \quad \mbox{with} \quad \sigma=1,\\
\eta|e^{\Ga^*}|^{2}I, \quad \mbox{with} \quad \sigma=-1.
\end{array}\right.
\end{align}
\end{Thm}

\subsection{One-soliton solution}

We note that, compared with the double Casoratian solutions of the rndsG equation \eqref{rndsG},
the solutions for the cndsG equation \eqref{cndsG} are more complicated. Hence,
we just construct one-soliton solution of equation \eqref{cndsG} in the case of
$(\eta,\sigma)=(1,-1)$. Let matrices $\Ga$ and $T$ be of form
\begin{align}
\label{cd-LT-form}
\Ga=\left(
\begin{array}{cc}
L & \bm 0  \\
\bm 0 & L^*
\end{array}\right),
\quad
T=\left(
\begin{array}{cc}
\bm 0 & I  \\
I & \bm0
\end{array}\right)|e^{\Ga^{*}}|.
\end{align}	
When the discrete spectral parameters $\{k_j\}$ in diagonal matrix \eqref{drL-Diag} are different nonzero complex numbers,
the basic column vector $\Phi$ is made up of
\begin{align}
\label{no}
\Phi_{j}=\left\{
\begin{aligned}
& c_{j}e^{\xi_j}, \quad j=1, 2, \ldots, N+1, \\
& d_{s}e^{\xi_s^*}, \quad j=N+1+s,\quad s=1,2,\dots,N+1,
\end{aligned}	
\right.
\end{align}
where and whereafter $\{c_{j},d_{s}\}$ are complex constants.

The one-soliton solution of equation \eqref{cndsG} is
\begin{subequations}
\label{cndsG-1ss}
\begin{align}
& \label{cndsG-1ss-u}
u=\dfrac{2c_{1}d_{1}\sinh(k_{1}^{*}-k_{1})}
{\epsilon(|c_{1}|^{2}e^{-2\xi_{1}^{*}}-|d_{1}|^{2}e^{-2\xi_{1}})}, \\
& w=\wt{f}\wh{f}/(f\wh{\wt{f}}), \quad f=|c_{1}|^{2}e^{\xi_1-\xi_1^{*}}-|d_{1}|^{2}e^{\xi_1^*-\xi_1},
\end{align}
\end{subequations}
where $\xi_1$ is defined as \eqref{xi}.
To demonstrate the dynamics of solution $u$ in an analytic way, we take $k_{1}=k_{11}+ik_{12}$, with which \eqref{cndsG-1ss-u} becomes
\begin{subequations}
\begin{align}
\label{dc-1ss}
u=& \dfrac{-2ic_{1}d_{1}\varepsilon_1^{m/2}e^{2nk_{11}}\sin2k_{12}}
{\epsilon [|c_{1}|^{2} e^{i(2nk_{12}+m\varepsilon_2)}
-|d_{1}|^{2}e^{-i(2nk_{12}+m\varepsilon_2)}]}, \\
\varepsilon_1=& \dfrac{[((\delta\epsilon)^{2}-16)(1+e^{4k_{11}})+2((\delta\epsilon)^{2}+16)e^{2k_{11}}\cos2k_{12}]^{2}
+(16\delta\epsilon e^{2k_{11}}\sin2k_{12})^{2}}{
(4+\delta\epsilon+(\delta\epsilon-4)e^{2k_{11}}\cos2k_{12})^{2}+((\delta\epsilon-4)e^{2k_{11}}\sin2k_{12})^{2}}, \\
\varepsilon_2=& \arctan\dfrac{16\delta\epsilon e^{2k_{11}}\sin2k_{12}}{((\delta\epsilon)^{2}-16)(1+e^{4k_{11}})
+2((\delta\epsilon)^{2}+16)e^{2k_{11}}\cos2k_{12}},
\end{align}	
\end{subequations}
and the corresponding envelop reads
\begin{align}
\label{dc-1ss-mo}
|u|^2=\dfrac{4|c_1d_1|^2\varepsilon_1^{m}e^{4nk_{11}}\sin^22k_{12}} {\epsilon^2[|c_1|^{4}+|d_{1}|^{4} -2|c_1d_1|^2\cos(4nk_{12}+2m\varepsilon_2)]}.
\end{align}
The carrier wave \eqref{dc-1ss-mo} is oscillatory due to the cosine in the denominator.
When $|c_{1}|=|d_{1}|$, wave \eqref{dc-1ss-mo} has singularities along points
\begin{align}
n(m)=\dfrac{\ell\pi-m\varepsilon_2}{2k_{12}},\quad \ell\in\mathbb{Z}.
\end{align}
While when $|c_{1}|\neq|d_{1}|$, wave \eqref{dc-1ss-mo} is nonsingular, and reaches its extrema along points
\begin{align}
n(m)=-\dfrac{m\varepsilon_2}{2k_{12}}+\frac{1}{4k_{12}}\bigg(\arcsin\frac{(|c_1|^4+|d_1|^4)k_{11}}
{2|c_1d_1|^2\sqrt{k^2_{11}+k^2_{12}}}-\jmath+2\ell\pi\bigg),\quad \ell\in\mathbb{Z},
\end{align}
where $\sin\jmath=k_{11}/\sqrt{k^2_{11}+k^2_{12}}$. The velocity is $-\varepsilon_2/(2k_{12})$.
It is easy to see that for any $m$, $|u|^2$ approaches to $0$ as either $(k_{11}<0,n\rightarrow +\infty)$ or $(k_{11}>0,n\rightarrow -\infty)$.
\vskip20pt
\begin{center}
\begin{picture}(120,80)
\put(-160,-23){\resizebox{!}{4.0cm}{\includegraphics{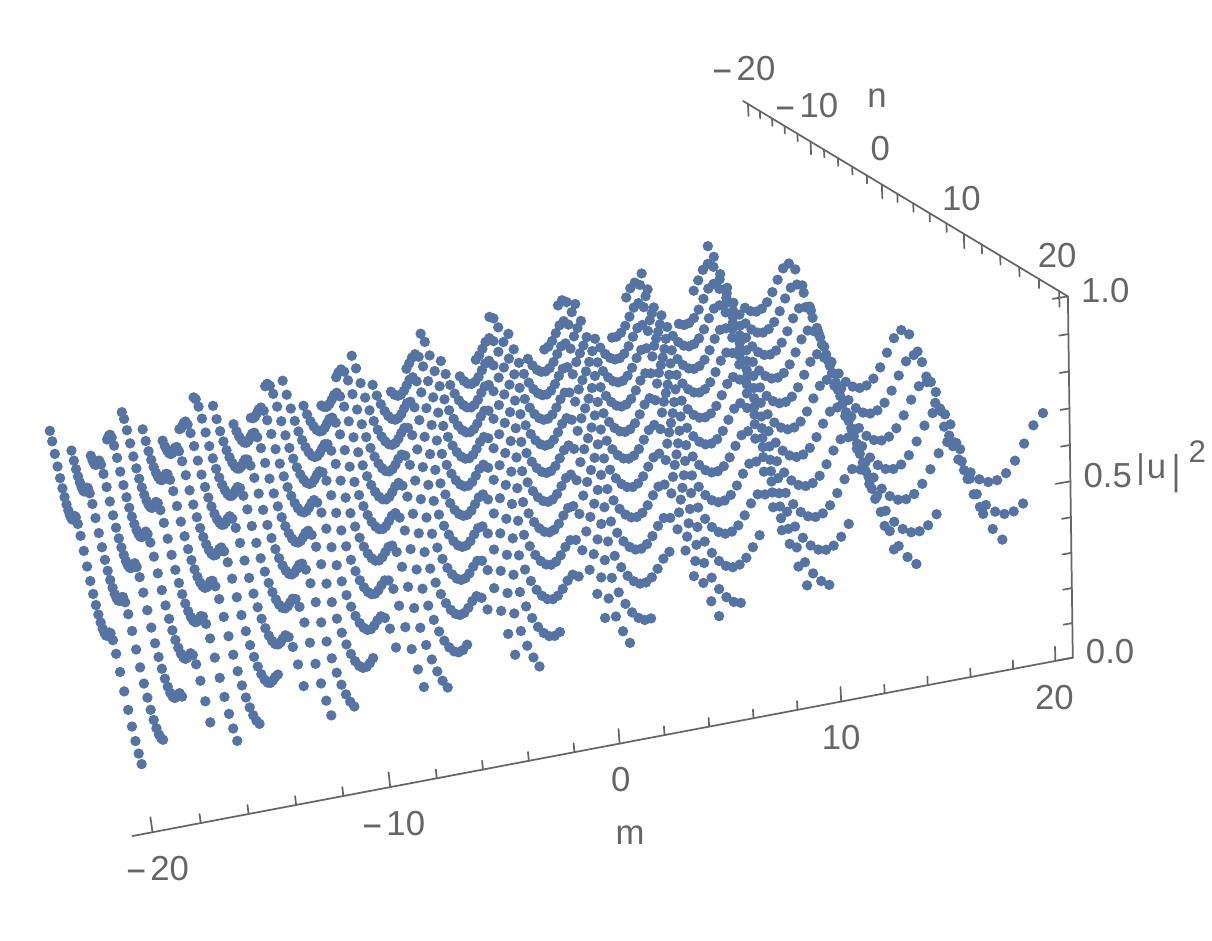}}}
\put(0,-23){\resizebox{!}{3.5cm}{\includegraphics{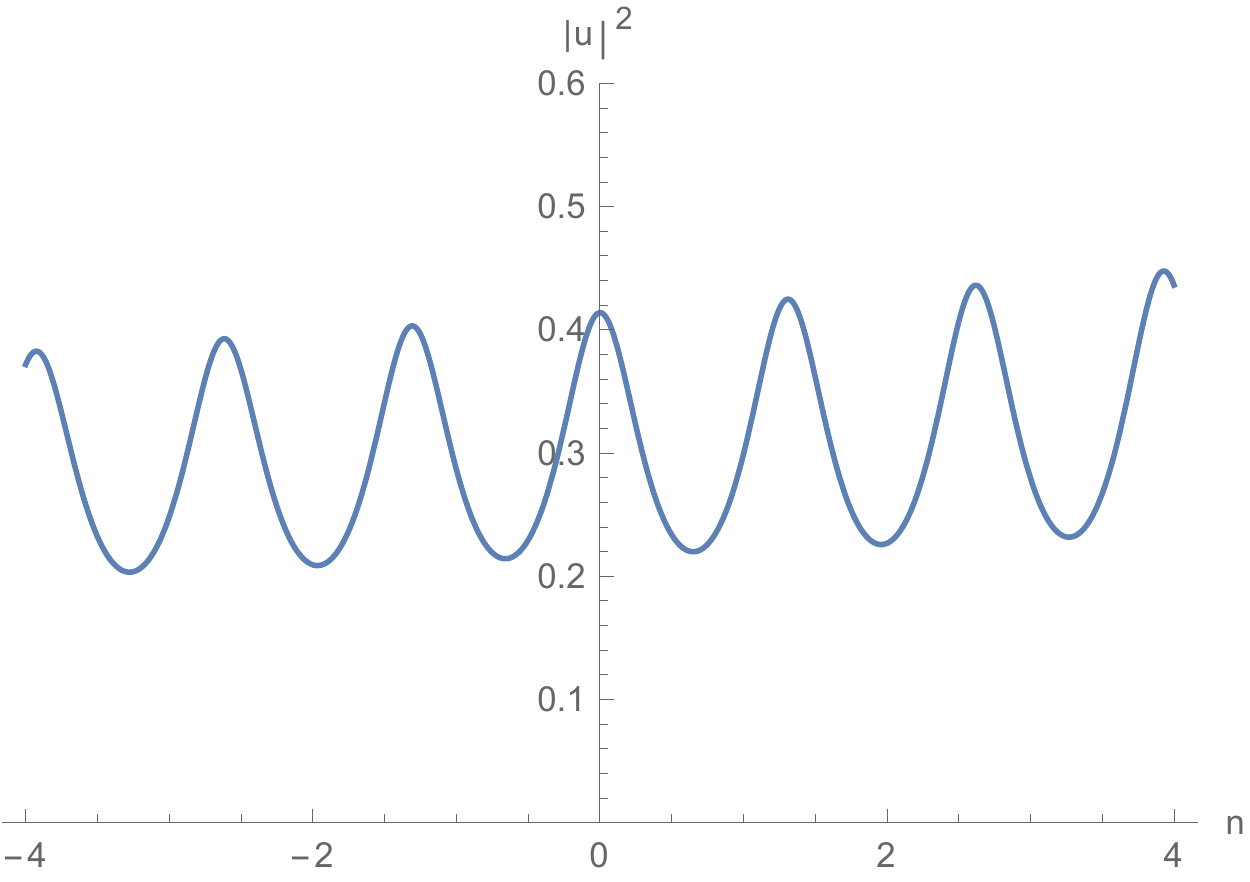}}}
\put(150,-23){\resizebox{!}{3.5cm}{\includegraphics{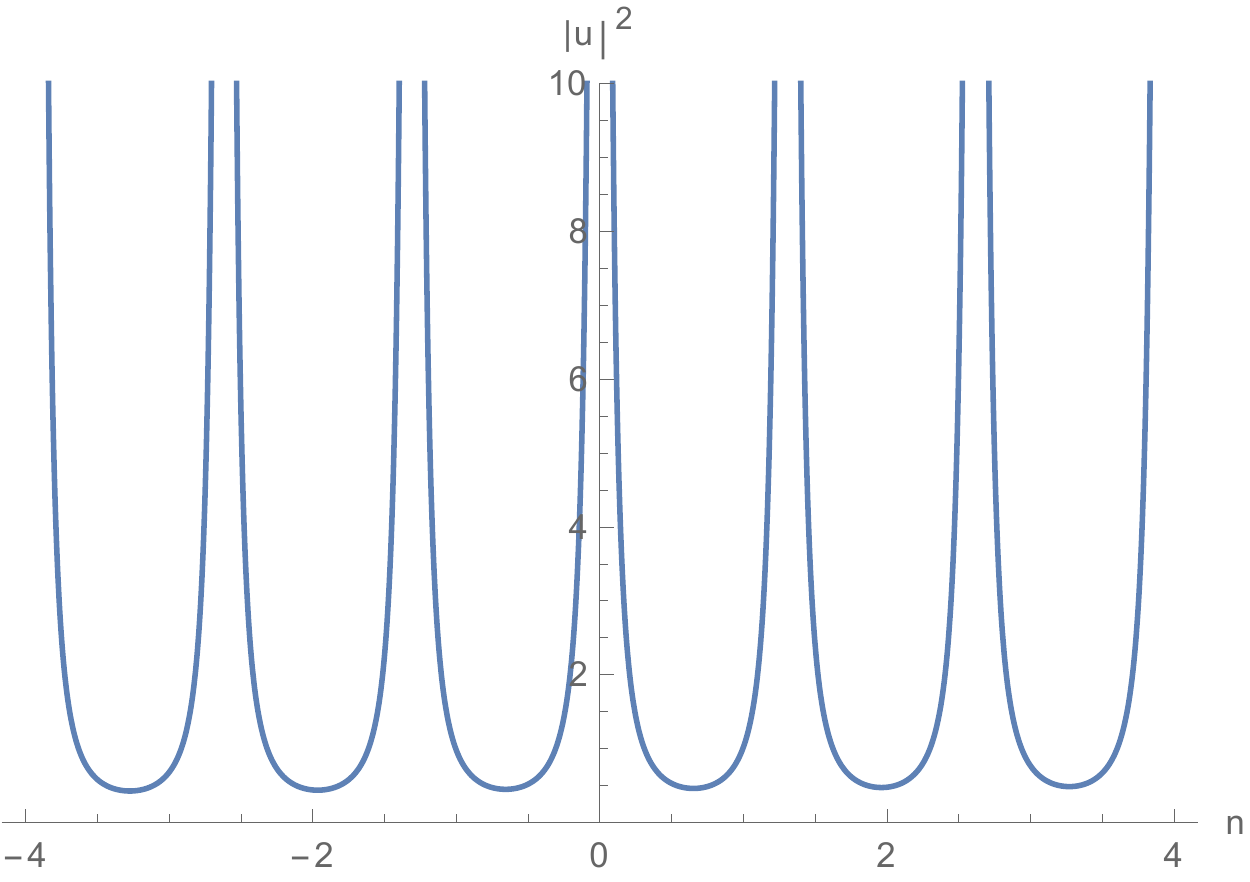}}}
\end{picture}
\end{center}
\vskip 20pt
\begin{center}
\begin{minipage}{15cm}{\footnotesize
\qquad\qquad\qquad\quad(a)\quad\qquad\qquad\qquad\qquad\qquad\qquad\qquad (b) \quad\qquad\qquad\qquad\qquad\qquad\qquad\quad (c)\\
{\bf Fig. 10}. One-soliton solution $|u|^2$ given by \eqref{dc-1ss-mo} with $\delta=\epsilon=1$ and $k_1=0.005+1.2i$:
(a) shape and movement with $c_1=0.5+0.5i$ and $d_1=0.2+0.2i$;
(b) 2D-plot of (a) at $m=0$;
(c) 2D-plot with $c_1=d_1=1+i$ at $m=0$.}
\end{minipage}
\end{center}

\subsection{Continuum limits} Applying the two semi-continuous limits introduced in subsection \ref{re-CL} to the cndsG equation \eqref{cndsG},
we can obtain the cnsdsG-$t$ and cnsdsG-$x$ equations, respectively. Moreover, the cncsG equation can be derived
by performing the continuous limits.

\vspace{.2cm}
\noindent{\bf Semi-continuous limit in $m$-direction:} Under the semi-continuous limit in $m$-direction
\eqref{rcl-nt}, inserting the Taylor expansions \eqref{Tay-exp-dsd} into the equation \eqref{cndsG}, we
realize that the coefficient of the leading order $\mathcal{O}(\delta)$ is exactly the cnsdsG-$t$ equation
\begin{subequations}
\label{cnsdsG-t}
\begin{align}
& 2(\wt{\mu}-\mu)'=\epsilon(\mu+\wt{\mu})(1-2\omega'), \\
& e^{\epsilon(\omega-\undertilde{\omega})}-1=\eta\epsilon^{2} \mu\mu^*_{\sigma}.
\end{align}
\end{subequations}
It can be checked that this equation is preserved under transformations $\mu\rightarrow -\mu$ and $\mu\rightarrow \pm i\mu$.

Double Casoratian solutions of the cnsdsG-$t$ equation \eqref{cnsdsG-t} are presented in the following theorem.
\begin{Thm}
The functions $\mu=g/f$ and $\omega=\epsilon^{-1}\ln(\wt{f}/f)$ with
\begin{align}
\label{cnsdsG-t-solu}
f=|e^{-N\Ga}\wh{\Phi^{(N)}};e^{N\Ga}T\wh{\Phi_{\sigma}^{*(-N)}}|,\quad
g=(1/\epsilon)|e^{-N\Ga}\widehat{\Phi^{(N+1)}};e^{N\Ga}T\wh{\Phi_{\sigma}^{*(-N+1)}}|,
\end{align}
solve the cnsdsG-$t$ equation \eqref{cnsdsG-t}, where $\Phi=e^{\Ga n+\epsilon\coth \Ga t/4}C^{+}$
and $T$ is a constant matrix of order $2(N+1)$, satisfying
\begin{align}
\label{cnsdsG-t-Ga-T}
\Ga T+\sigma T\Ga^*=\bm 0,\quad TT^*=\left\{
\begin{array}{l}
-\eta I, \quad \mbox{with} \quad \sigma=1,\\
\eta|e^{\Ga^*}|^{2}I, \quad \mbox{with} \quad \sigma=-1.
\end{array}\right.
\end{align}
\end{Thm}

For $(\sigma,\eta)=(-1,1)$, we take matrices $\Ga$ and $T$ as block form \eqref{cd-LT-form}.
To the diagonal matrix $L$ defined as \eqref{drL-Diag} with distinct complex nonzero
eigenvalues $\{k_j\}$, the column vector $\Phi$ with entries
\begin{align}
\Phi_{j}=\left\{
\begin{aligned}
& c_{j}e^{\lambda_j},\quad j=1, 2, \dots, N+1,\\
& d_{s}e^{\lambda^*_s},\quad j=N+1+s,\quad s=1,2,\dots,N+1,
\end{aligned}	
\right.
\end{align}
yields the multi-soliton solutions, where $\{\lambda_j\}$ are defined as \eqref{lambdaj-def}.

In particular, the one-soliton solution is described as
\begin{subequations}
\begin{align}
& \mu=\dfrac{2c_{1}d_{1}\sinh(k_{1}-k^*_{1})}
{\epsilon(|d_{1}|^{2}e^{-2\lambda_1}-|c_{1}|^{2}e^{-2\lambda_{1}^{*}})},\\
& \omega=\frac{1}{\epsilon}\ln\dfrac{|c_{1}|^{2}e^{2(k_{1}+\lambda_{1})}-|d_{1}|^{2}e^{2(k_{1}^{*}+\lambda_{1}^{*})}}
{e^{k_{1}+k_{1}^{*}}(|c_{1}|^{2}e^{2\lambda_{1}}-|d_{1}|^{2}e^{2\lambda_{1}^{*}})}.
\end{align}	
\end{subequations}
For $k_{1}=k_{11}+ik_{12}$, we have
\begin{align}
\mu &=\dfrac{2ic_{1}d_{1}e^{2k_{11}n+\Upsilon_1 t}\sin2k_{12}}
{\epsilon
(|d_{1}|^{2}e^{-i(2nk_{12}-\Upsilon_2 t)}-|c_{1}|^{2}e^{i(2nk_{12}-\Upsilon_2 t)})},
\end{align}	
where $\Upsilon_1=\frac{\epsilon \sinh2k_{11}}{2(\cosh2k_{11}-\cos2k_{12})}$,
 $\Upsilon_2=\frac{\epsilon \sin 2k_{12}}{2(\cosh2k_{11}-\cos2k_{12})}$. The wave package
\begin{align}
\label{sdc-t-1ss-mo}
|\mu|^2=\dfrac{4|c_{1}d_1|^{2}
e^{4k_{11}n+2\Upsilon_1t}\sin^22k_{12}}
{\epsilon^2(|c_{1}|^4+|d_1|^{4}-2|c_{1}d_1|^{2}\cos(4nk_{12}-2\Upsilon_2 t))},
\end{align}	
is still quasi-periodic. When $|c_{1}|=|d_{1}|$, wave \eqref{sdc-t-1ss-mo} has singularities along points
\begin{align}
n(t)=\dfrac{\ell\pi+\Upsilon_2t}{2k_{12}},\quad \ell\in\mathbb{Z}.
\end{align}
While when $|c_{1}|\neq|d_{1}|$, this wave is nonsingular,
and reaches its extrema along points
\begin{align}
n(t)=\dfrac{t\Upsilon_2}{2k_{12}}+\frac{1}{4k_{12}}\bigg(\arcsin\frac{(|c_1|^4
+|d_1|^4)k_{11}}{2|c_1d_1|^2\sqrt{k^2_{11}+k^2_{12}}}-\jmath+2\ell\pi\bigg),\quad \ell\in\mathbb{Z}.
\end{align}
The travelling speed is $\Upsilon_2/(2k_{12})$.
For any $t$, $|\mu|^2$ approaches to zero as either $(k_{11}<0,n\rightarrow +\infty)$ or $(k_{11}>0,n\rightarrow -\infty)$.
The dynamic of \eqref{sdc-t-1ss-mo} is depicted in Fig. 11.
\vskip20pt
\begin{center}
\begin{picture}(120,80)
\put(-160,-23){\resizebox{!}{4.0cm}{\includegraphics{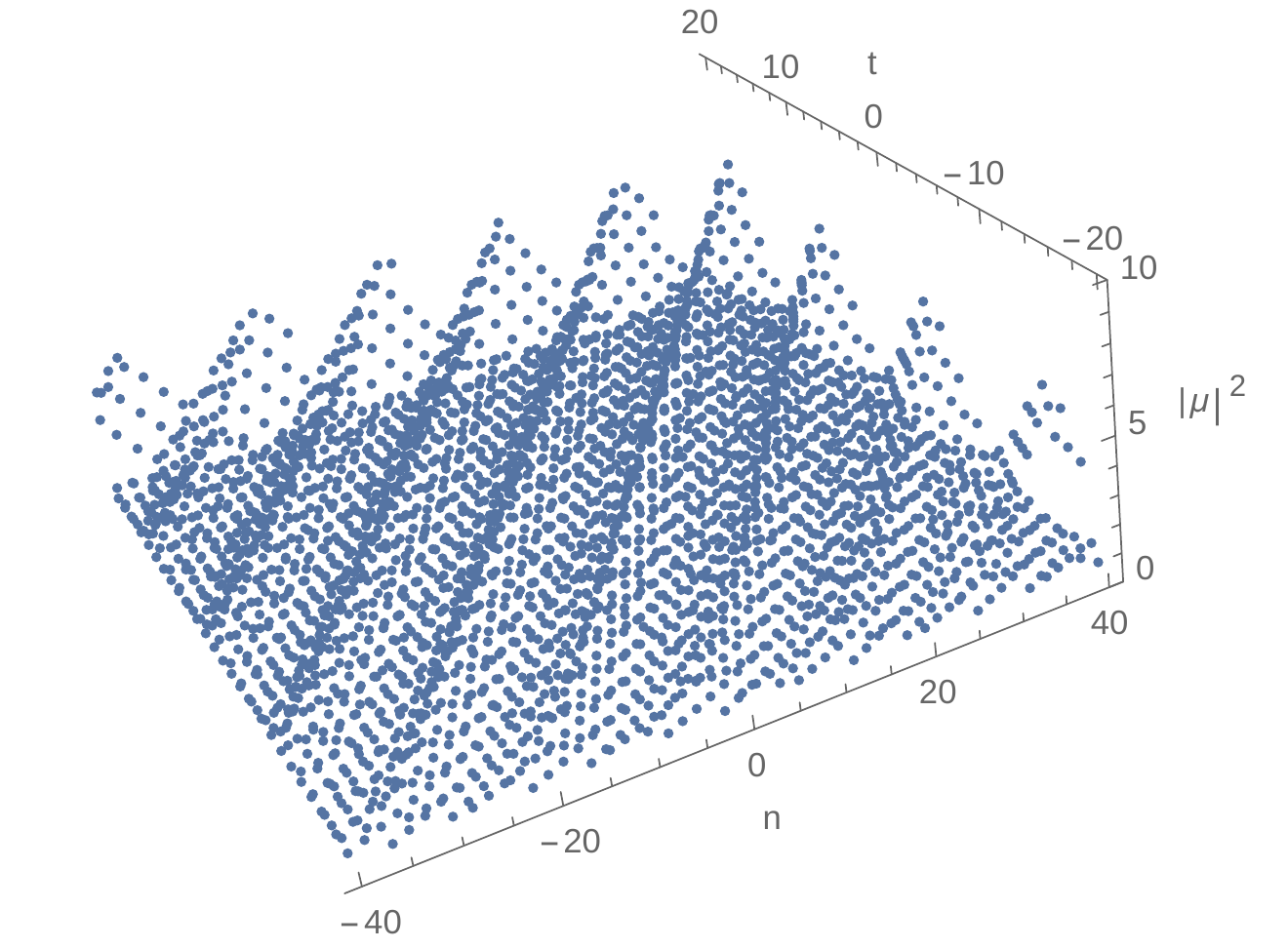}}}
\put(0,-23){\resizebox{!}{3.5cm}{\includegraphics{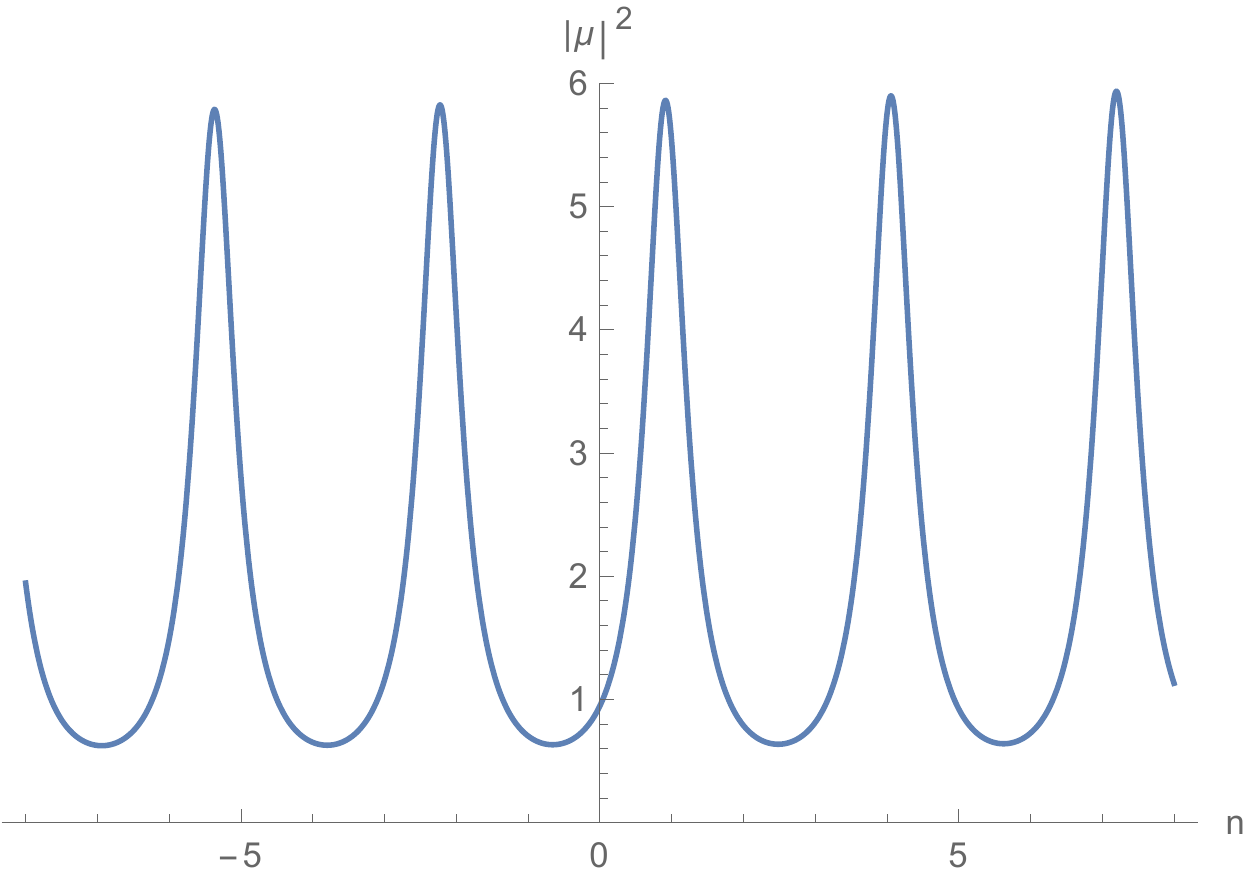}}}
\put(150,-23){\resizebox{!}{3.5cm}{\includegraphics{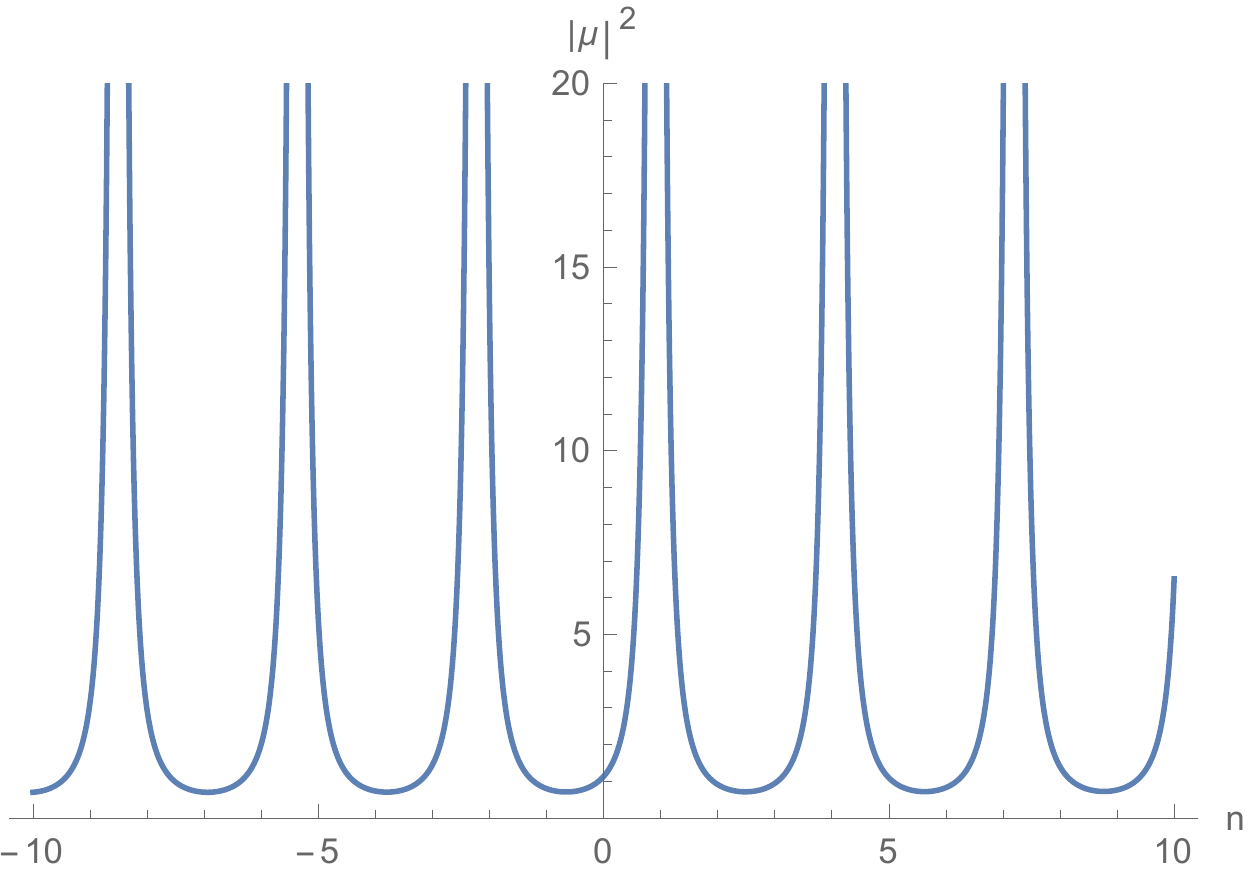}}}
\end{picture}
\end{center}
\vskip 20pt
\begin{center}
\begin{minipage}{15cm}{\footnotesize
\qquad\qquad\qquad\quad(a)\quad\qquad\qquad\qquad\qquad\qquad\qquad\qquad (b) \qquad\qquad\qquad\qquad\qquad\qquad\qquad\quad (c)\\
{\bf Fig. 11}. One-soliton solution $|\mu|^2$ given by \eqref{sdc-t-1ss-mo} with $\epsilon=1$ and $k_1=0.001+i$:
(a) shape and movement with $c_1=1+0.1i$ and $d_1=1+i$;
(b) 2D-plot of (a) at $t=1$;
(c) 2D-plot with $c_1=d_1=1+i$ at $t=1$.}
\end{minipage}
\end{center}

\vspace{.2cm}
\noindent{\bf Semi-continuous limit in $n$-direction:} Employing the semi-continuous limit in $n$-direction, i.e., \eqref{rcl-nx},
we find that the leading order of Taylor expansion for equation \eqref{cndsG} gives rise to the cnsdsG-$x$ equation
\begin{subequations}
\label{cnsdsG-x}
\begin{align}
& 2(\wh{\cdd{\chi}}-\cdd{\chi})+(\wh{\chi}+\chi)[2(\wh\varpi-\varpi)-\delta]=0,\\
& \cdd{\varpi}=\eta \chi\chi^*_{\sigma},
\end{align}
\end{subequations}
where $\varpi=\varpi^*_{\sigma}$.
This equation is preserved under transformations $\chi\rightarrow -\chi$ and $\chi\rightarrow \pm i\chi$.
\begin{Thm}
\label{Thm-cnsdsG-x-solu}
The functions $\chi=g/f$ and $\varpi=\cdd{f}/f$ with
\begin{align}
\label{cnsdsG-x-solu}
f=|\wh{\phi^{(N)}};T\wh{\phi_{\sigma}^{*(N)}}|,\quad
g=2|\wh{\phi^{(N+1)}};T\wh{\phi_{\sigma}^{*(N-1)}}|,
\end{align}
solve the cnsdsG-$x$ equation \eqref{cnsdsG-x}, where $\phi=e^{\Omega m+\frac{\delta\coth \Omega}{4}x}C^{+}$ and
$T$ is a constant matrix of order $2(N+1)$, satisfying
\begin{align}
\label{cnsdsG-x-L-T}
\Omega T+\sigma T\Omega^*=\bm 0,\quad TT^{*}=-\sigma\eta I.
\end{align}
\end{Thm}

We still take $(\sigma,\eta)=(-1,1)$. In the case of block matrices \eqref{csd-OmT-form}
with diagonal matrix $S$ in \eqref{rsdx-S-diag}, the column vector $\phi$ composed by
\begin{align}
\phi_{j}=\left\{
\begin{aligned}
& c_{j}e^{\vartheta_j},\quad j=1, 2, \dots, N+1,\\
& d_{s}e^{\vartheta^*_s},\quad j=N+1+s,\quad s=1,2,\dots,N+1,
\end{aligned}	
\right.
\end{align}
produces the multi-soliton solutions, where $\{\vartheta_j\}$ are defined as \eqref{eth-vath}.

In the present case, the one-soliton solution is of form
\begin{subequations}
\begin{align}
& \chi=\dfrac{\delta c_{1}d_{1}(\coth{l_{1}^{*}}-\coth{l_{1}})}
{2(|c_{1}|^{2}e^{-2\vartheta^*_1}-|d_{1}|^{2}e^{-2\vartheta_1})},\\
& \varpi=\dfrac{\delta(\coth l_{1}-\coth l_{1}^{*})(|c_{1}|^{2}e^{2\vartheta_{1}}+|d_{1}|^{2}e^{2\vartheta_{1}^{*}})}
{4(|c_{1}|^{2}e^{2\vartheta_{1}}-|d_{1}|^{2}e^{2\vartheta_{1}^{*}})}.
\end{align}
\end{subequations}
Setting $l_{1}=l_{11}+il_{12}$ leads to the modulus of $\chi$, given by
\begin{align}
\label{sdc-x-1ss-mo}
|\chi|^2=\dfrac{|\delta c_{1}d_1|^{2}
e^{4k_{11}m+2\hbar_1x}(\cosh2l_{11}-\cos 2l_{12})^{-2}\sin^22l_{12}}
{|c_{1}|^4+|d_1|^{4}-2|c_{1}d_1|^{2}\cos(4ml_{12}-2\hbar_2 x)},
\end{align}	
where $\hbar_1=\frac{\delta\sinh2l_{11}}{2(\cosh 2l_{11}-\cos2l_{12})}$ and $\hbar_2=\frac{\delta\sin2l_{12}}{2(\cosh 2l_{11}-\cos2l_{12})}$.

Wave $|\chi|^2$ still has a quasi-periodic phenomenon. When $|c_{1}|=|d_{1}|$, it has singularities along points
\begin{align}
x(m)=\dfrac{2ml_{12}+\ell\pi}{\hbar_2},\quad \ell\in\mathbb{Z}.
\end{align}
While when $|c_{1}|\neq|d_{1}|$, this wave is nonsingular,
and reaches its extrema along points
\begin{align}
x(m)=\dfrac{2ml_{12}}{\hbar_2}+\dfrac{1}{2\hbar_2}\bigg(
\arcsin\dfrac{(|c_{1}|^4+|d_1|^{4})\hbar_1}{2|c_{1}d_1|^{2}\sqrt{\hbar^2_{1}+\hbar^2_{2}}}-\wp+2\ell\pi\bigg), \quad \ell\in\mathbb{Z}.
\end{align}
where $\sin\wp=\hbar_{1}/\sqrt{\hbar^2_{1}+\hbar^2_{2}}$.
This wave travels with speed $2l_{12}/\hbar_2$.
For any $m$, $|\chi|^2$ approaches to zero as either $(\hbar_{1}<0,x\rightarrow +\infty)$ or $(\hbar_{1}>0,x\rightarrow -\infty)$.

\vskip20pt
\begin{center}
\begin{picture}(120,80)
\put(-160,-23){\resizebox{!}{4cm}{\includegraphics{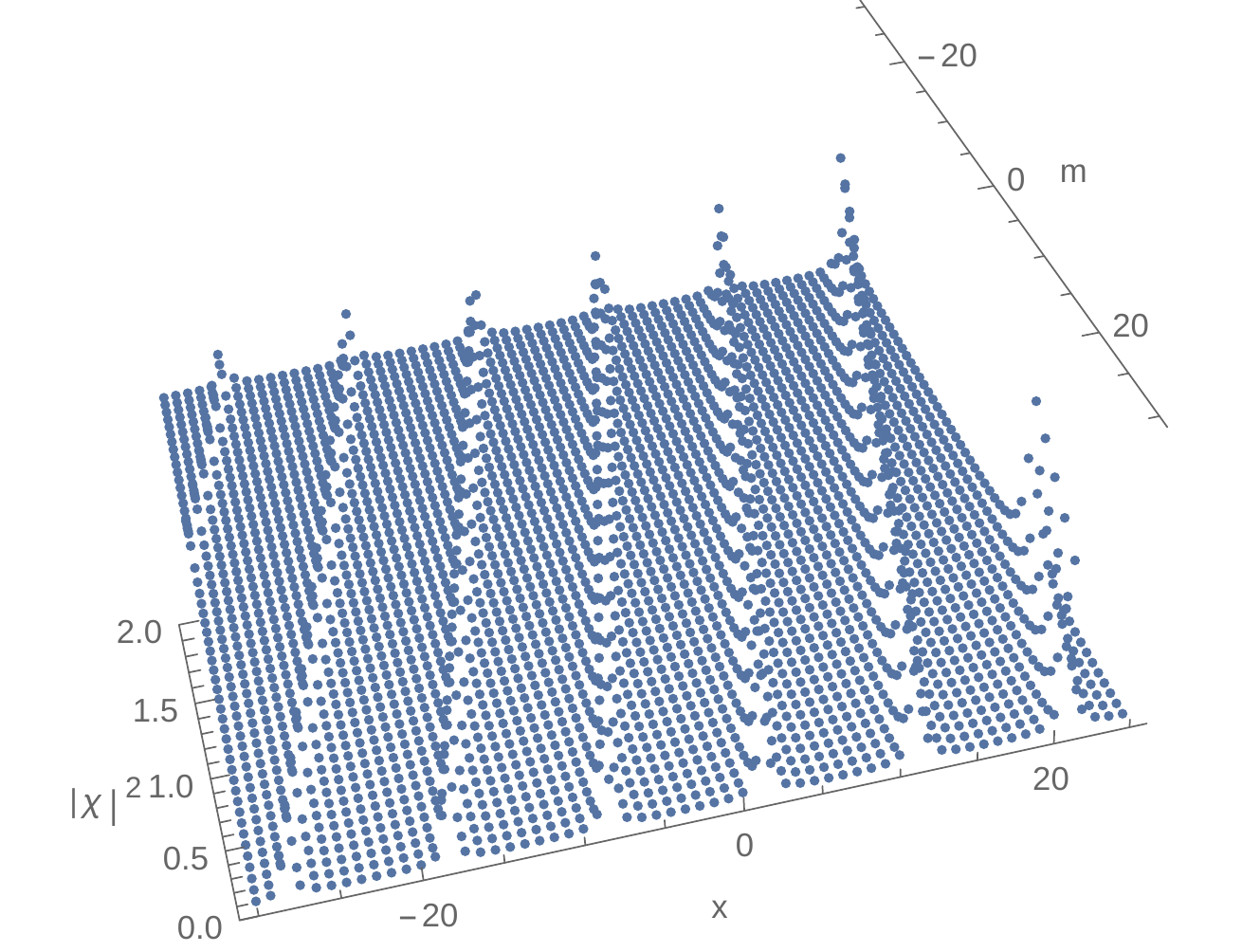}}}
\put(0,-23){\resizebox{!}{3.5cm}{\includegraphics{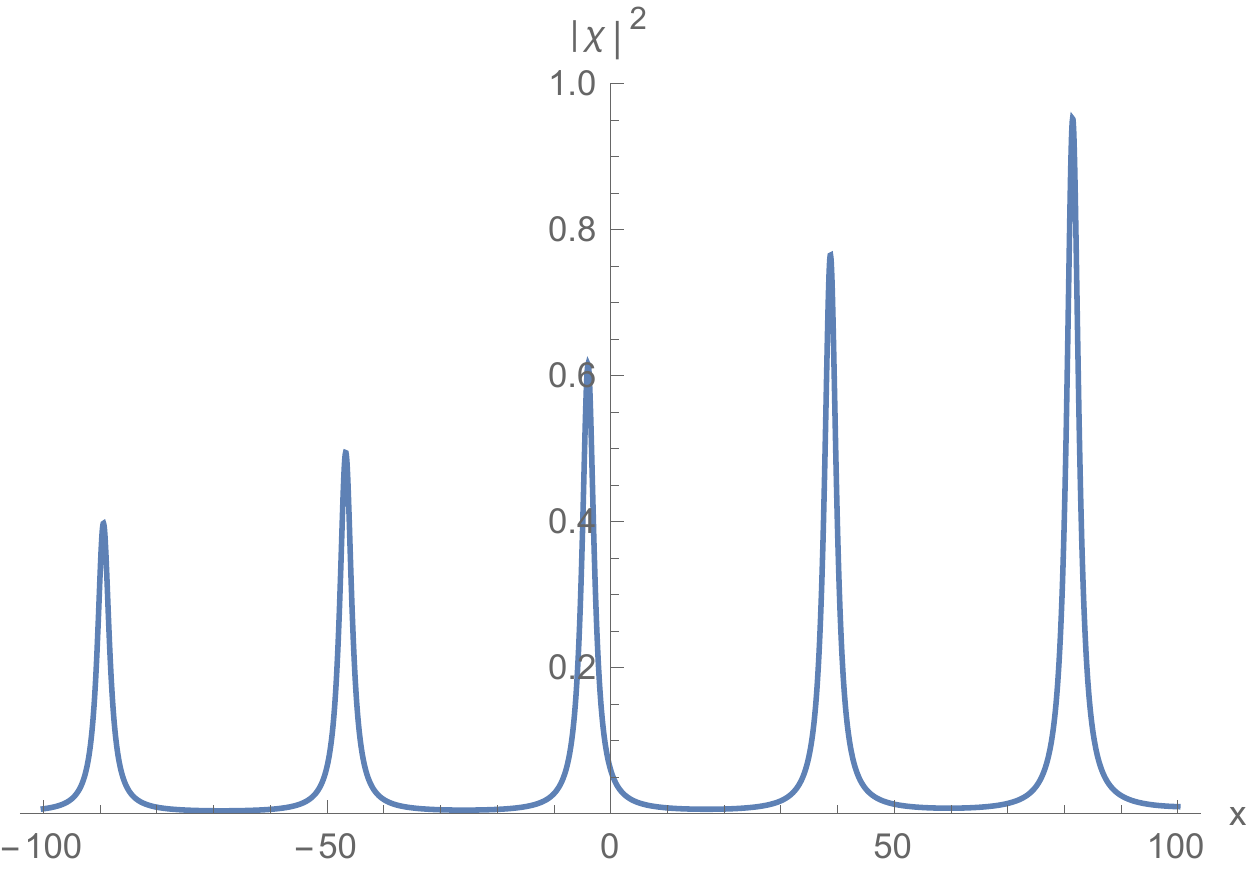}}}
\put(150,-23){\resizebox{!}{3.5cm}{\includegraphics{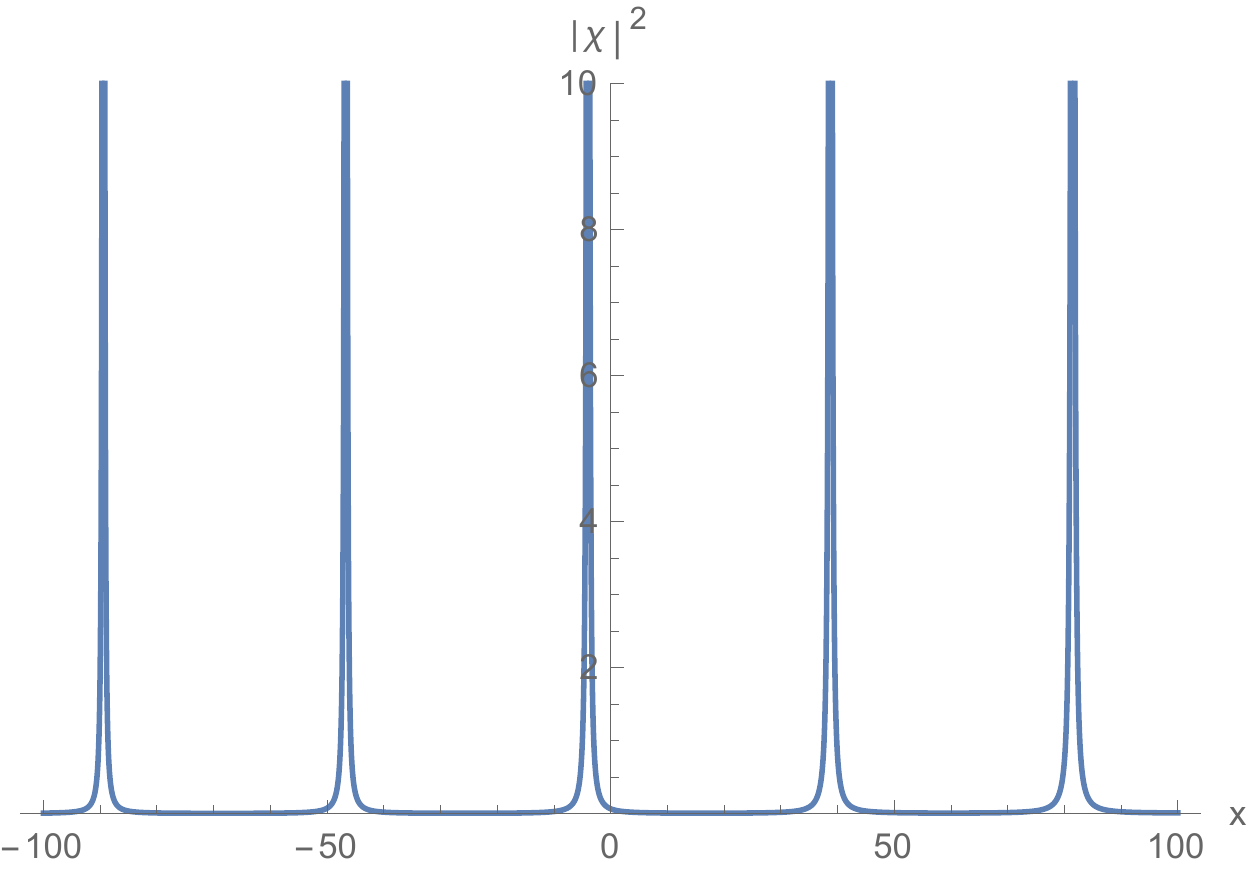}}}
\end{picture}
\end{center}
\vskip 20pt
\begin{center}
\begin{minipage}{15cm}{\footnotesize
\qquad\qquad\qquad\quad(a)\quad\qquad\qquad\qquad\qquad\qquad\qquad\qquad (b) \qquad\qquad\qquad\qquad\qquad\qquad\qquad\quad (c)\\
{\bf Fig. 12}. One-soliton solution $|\chi|^2$ given by \eqref{sdc-x-1ss-mo} with $\delta=1$ and $k_1=0.005+8i$:
(a) shape and movement with $c_1=0.5+0.4i$ and $d_1=0.3+0.5i$;
(b) 2D-plot of (a) at $m=1$;
(c) 2D-plot with $c_1=d_1=1+i$ at $m=1$.}
\end{minipage}
\end{center}

\vspace{.2cm}
\noindent{\bf Continuous limits:} Performing the continuous limits \eqref{rcl-nt} and \eqref{rcl-nx}, respectively,
to the cnsdsG-$x$ equation \eqref{cnsdsG-x} and the cnsdsG-$t$ equation \eqref{cnsdsG-t}, we get the
cncsG equation
\begin{align}	
\label{cncsG}
\cdd{\al}'+2\al \be'=\al, \quad \cdd{\be}=\eta \al\al^*_{\sigma},
\end{align}	
where $\be=\be^*_{\sigma}$. This equation
is preserved under transformations $\al\rightarrow -\al$ and $\al\rightarrow \pm i\al$.

\begin{Thm}
\label{Thm-cncsG-solu}
The functions $\al=g/f$ and $\be=\cdd{f}/f$ with
\begin{align}
\label{cncsG-solu}
f=|\wh{\phi^{(N)}};T\wh{\phi_{\sigma}^{*(N)}}|,\quad
g=2|\wh{\phi^{(N+1)}};T\wh{\phi_{\sigma}^{*(N-1)}}|,
\end{align}
solve the cncsG equation \eqref{cncsG}, where $\phi=e^{B x+(4B)^{-1}t}C^{+}$ and $T$ is a constant matrix of order $2(N+1)$, satisfying
\begin{align}
\label{cncsG-B-T}
B T+\sigma TB^*=\bm 0,\quad TT^*=-\sigma\eta I.
\end{align}
\end{Thm}

\section{Conclusions}

In this paper, local and nonlocal reductions of the dnAKNS equation \eqref{dnAKNS} are
investigated. As a result, real/complex local and nonlocal discrete sG equations are derived,
where exact solutions to the nonlocal discrete sG equation are determined by applying the so-called bilinearization reduction method.
In the real case, by solving the matrix equations \eqref{rndsG-AT}, we obtain one-, two-soliton solutions and the
simplest Jordan-block solution for the rndsG equation \eqref{rndsG}.
Dynamics of variable $u$ are given by asymptotic analysis as a demonstration.
Through the semi-continuous limit in $m$-direction, respectively, $n$-direction, two real nonlocal
semi-discrete sG equations (equations \eqref{rnsdsG-t} and \eqref{rnsdsG-x} with $\sigma=-1$)
are derived. For these two equations, we present their one-, two-soliton solutions and the simplest Jordan-block solution,
as well as the dynamical properties. Furthermore, by the continuous limits, we retrieve the real nonlocal sG equation \eqref{rncsG}.
In the complex case, we focus on the matrix equations \eqref{cndsG-AT}.
In the assumption of $L$ being a diagonal matrix with different complex nonzero discrete spectral parameters $\{k_j\}$,
the one-soliton solution and its dynamics behavior are discussed.
By using the same semi-continuous limits and continuous limits as the real case, the construction of the cnsdsG-$t$, cnsdsG-$x$ and
cncsG equations, i.e., equations \eqref{cnsdsG-t}, \eqref{cnsdsG-x} and \eqref{cncsG}, are presented.
One-soliton solution and corresponding dynamics to the first two equations are studied emphatically.
The one-soliton solutions of the rndsG/rnsdsG-$t$/rnsdsG-$x$ equations exhibit the usual bell-type structure, while of
the cndsG/cnsdsG-$t$/cnsdsG-$x$ equations behave quasi-periodically.

We end the paper with the following remarks.

First of all, for solutions to the matrix equations \eqref{rndsG-AT} with $\sigma=-1$, we have other choices, one of which reads (cf. \cite{CDLZ,LWZ-SAPM})
\begin{align}
\label{Ga-T-ex-oth}
\Ga=\left(
\begin{array}{cc}
\Lambda_1 & \bm 0  \\
\bm 0 & \Lambda_2
\end{array}\right),\quad
T=|e^{\Gamma}|\left(
\begin{array}{cc}
I & \bm 0  \\
\bm 0 & -I
\end{array}\right).
\end{align}	
In the simplest case, we take $\Lambda_1=k_1$ and $\Lambda_2=\bar{k}_1$ and have the one-soliton solution
\begin{subequations}
\begin{align}
\label{rndsG-1SS}
u=& (e^{\xi_1+\bar{\xi}_1}\sinh (k_{1}-\bar{k}_1)\sech (\xi_1-\bar{\xi}_1))/\epsilon, \\
w=& \cosh(k_{1}-\bar{k}_{1}+\xi_{1}-\bar{\xi}_{1})\cosh(\tau_{1}-\bar{\tau}_{1}+\xi_{1}-\bar{\xi}_{1})\sech(\xi_{1}-\bar{\xi}_{1}) \nn \\
& \sech(k_{1}-\bar{k}_{1}+\tau_{1}-\bar{\tau}_{1}+\xi_{1}-\bar{\xi}_{1}),
\end{align}
\end{subequations}
where $\bar{\xi}_{1}=\xi_{1}|_{k_1\rightarrow \bar{k}_{1}}$ and $\bar{\tau}_{1}=\tau_{1}|_{k_1\rightarrow \bar{k}_{1}}$.
Under the assumption $k_1\neq -\bar{k}_1$, solution $u$ in \eqref{rndsG-1SS} behaves totally different from the one \eqref{dr-1ss-u}.
In fact, for a given $m$, $u\rightarrow \pm(\mp) \infty$ under the limit $n\rightarrow \pm\infty$ with assumption $\epsilon(k_1-\bar{k}_1)>(<)0$.

What's more, we have shown that the bilinearization reduction method is valid in the study of nonlocal discrete integrable systems.
Compared with the Cauchy matrix reduction approach \cite{XFZ-TMPH}, there is a great advantage of the bilinearization reduction
method. The latter method allows one to construct exact solutions of the real reduced equations. While in the Cauchy matrix reduction
scheme, one can not achieve that.
This is because, in the Cauchy matrix reduction approach, solutions of the original before-reduction
AKNS system should satisfy two Sylvester equations \cite{XFZ-TMPH}.
In the real reduction case, the solvability of these two Sylvester equations usually conflicts with the
solvability of the Sylvester equation in the matrix equation set (For more detailed explanations,
one can refer to the conclusions in \cite{XZ-TMPH}).

In addition, let's go back to the sG equation.
Through suitable reciprocal transformation \cite{HT-JPSJ}, the sG equation can be transformed into the short
pulse \cite{USP}, which describes the propagation of short optical pulses in nonlinear media.
In \cite{CLZ-SP}, the hodograph transformations between nonlocal short pulse models and nonlocal sG system are revealed.
It is shown that the independent variables of the short pulse models and sG equation that are connected via
hodograph transformation are covariant in nonlocal reductions.
This gives us the motivation to consider the hodograph transformations between the discrete short pulse
models \cite{DSP} and the discrete sG equations \cite{Yu-SAPM,XFZ-TMPH}, as well as their nonlocal cases.

To summarize, from the dnAKNS equation \eqref{dnAKNS}, we consider its local and nonlocal reductions by utilizing the bilinearization reduction method.
This approach can also be used to discuss nonlocal reductions of the positive order discrete AKNS system, which admits double Casoratian solutions.
Very recently, we employ this method to study the nonlocal reductions of a discrete positive order AKNS equation \cite{Hirota-2000}.
Consequently, the real/complex local and nonlocal discrete mKdV equations are obtained.
We also investigated solutions and continuum limits of the resulting nonlocal discrete mKdV equations (cf. \cite{ZXS}).

\section*{Acknowledgments}

We are very grateful to the reviewers for their invaluable and expert comments.
This project is supported by the National Natural Science Foundation of
China (Nos. 12071432, 11301483) and the Natural Science Foundation of Zhejiang Province (No. LY17A010024).


{\small
\begin{thebibliography}{99}
\bibitem{AbMu-2013} Ablowitz MJ, Musslimani ZH.
        Integrable nonlocal nonlinear Schr\"{o}dinger equation.
        {\it Phys Rev Lett.} 2013;110:064105.	
\bibitem{Fokas} Fokas AS.
        Integrable multidimensional versions of the nonlocal nonlinear Schr\"{o}dinger equation.
        {\it Nonlinearity.} 2016;29:319--324.
\bibitem{Lou-JMP} Lou SY.
        Alice-Bob systems, $\hat{P}-\hat{T}-\hat{C}$ symmetry invariant and symmetry breaking soliton solutions.
        {\it J Math Phys.} 2018;59:083507.
\bibitem{Lou-CTP} Lou SY.
        Multi-place physics and multi-place nonlocal systems.
        {\it Commun Theor Phys.} 2020;72:057001.
\bibitem{MPW} Markum H, Pullirsch R, Wettig T.
        Non-hermitian random matrix theory and lattice QCD with chemical potential.
        {\it Phys Rev Lett.} 1999;83:484--487.
\bibitem{LSEK} Lin Z, Schindler J, Ellis FM, Kottos T.
        Experimental observation of the dual behavior of PT-symmetric scattering.
        {\it Phys Rev A.} 2012;85:050101.
\bibitem{RMGCSK} Ruter CE, Makris KG, EI-Ganainy R, Christodoulides DN, Segev M, Kip D.
        Observation of parity-time symmetry in optics.
        {\it Nat Phys.} 2010;6:192--195.
\bibitem{MMGC} Musslimani ZH, Makris KG, El-Ganainy R, Christodoulides DN.
        Optical Solitons in PT Periodic Potentials.
        {\it Phys Rev Lett.} 2008;100:030402.
\bibitem{DGPS} Dalfovo F, Giorgini S, Pitaevskii LP, Stringari S.
        Theory of Bose-Einstein condensation in trapped gases.
        {\it Rev Mod Phys.} 1999;71:463--512.
\bibitem{AbMu-2016} Ablowitz MJ, Musslimani ZH.
        Integrable nonlocal nonlinear equations.
        {\it Stud Appl Math.} 2016;139:7--59.
\bibitem{Lou-SR} Lou SY, Huang F.
        Alice-Bob physics: coherent solutions of nonlocal KdV systems.
        {\it Sci Rep}. 2017;7:869.
\bibitem{JZ-2017} Ji JL, Zhu ZN.
        On a nonlocal modified Korteweg-de Vries equation: Integrability, Darboux transformation and soliton solutions.
        {\it Commun Nonl Sci Numer Simulat.} 2017;42:699--708.
\bibitem{AFLM-SAPM} Ablowitz MJ, Feng BF, Luo XD, Musslimani ZH.
        Reverse space-time nonlocal sine-Gordon/sinh-Gordon equations with nonzero boundary conditions.
        {\it Stud Appl Math.} 2018;141:267--307.
\bibitem{AFLM-2018} Ablowitz MJ, Feng BF, Luo XD, Musslimani ZH.
        Inverse scattering transform for the nonlocal reverse space-time nonlinear Schr\"{o}dinger equation.
        {\it Theor Math Phys.} 2018;196:1241--1267.
\bibitem{SSZ-2019} Shi Y, Shen SF, Zhao SL.
        Solutions and connections of nonlocal derivative nonlinear Schr\"{o}dinger equations.
        {\it Nonlinear Dyn.} 2019;95:1257--1267.
\bibitem{Zhou-SAPM} Zhou ZX.
        Darboux transformations and global explicit solutions for nonlocal Davey-Stewartson I equation.
        {\it Stud Appl Math.} 2018;141:186--204.
\bibitem{RZFH} Rao JG, Zhang YS, Fokas AS, He JS.
        Rogue waves of the nonlocal Davey-Stewartson I equation.
        {\it Nonlinearity.} 2018;31:4090--4107.
\bibitem{AM-Nonl-2016} Ablowitz MJ, Musslimani ZH.
        Inverse scattering transform for the integrable nonlocal nonlinear Schr\"odinger equation.
        {\it Nonlinearity.} 2016;29:915--946.
\bibitem{LiX-PRE-2015} Li M, Xu T.
        Dark and antidark soliton interactions in the nonlocal nonlinear Schr\"odinger equation with the self-induced parity-time-symmetric potential.
        {\it Phys Rev E.} 2015;91:033202.
\bibitem{SXZ} Song CQ, Xiao DM, Zhu ZN.
        Reverse space-time nonlocal Sasa-Satsuma equation and its solutions.
        {\it J Phys Soc Japan.} 2017;86:054001.
\bibitem{XU-AML-2016} Xu ZX, Chow KW.
        Breathers and rogue waves for a third order nonlocal partial differential equation by a bilinear transformation.
        {\it Appl Math Lett.} 2016;56:72--77.
\bibitem{CDLZ} Chen K, Deng X, Lou SY, Zhang DJ.
        Solutions of local and nonlocal equations reduced from the AKNS hierarchy.
        {\it Stud Appl Math.} 2018;141:113--141.
\bibitem{GP-CNSNS} G\"{u}rses M, Pekcan A.
        Nonlocal modified KdV equations and their soliton solutions by Hirota method.
        {\it Commun Nonlinear Sci Numer Simulat.} 2019;67:427--448.
\bibitem{FZ-ROMP} Feng W, Zhao SL.
        Cauchy matrix type solutions for the nonlocal nonlinear Schr\"odinger equation.
        {\it Rep Math Phys.} 2019;84:75--83.
\bibitem{Yan} Yan Z.
        Integrable PT-symmetric local and nonlocal vector nonlinear Schr\"{o}dinger equations: A unified two-parameter model.
        {\it Appl Math Lett.} 2015;47:61--68.
\bibitem{YY-SAPM} Yang B, Yang JK.
        Transformations between nonlocal and local integrable equations.
        {\it Stud Appl Math.} 2017;40:178--201.
\bibitem{XCLM} Xu T, Chen Y, Li M, Meng DX.
        General stationary solutions of the nonlocal nonlinear Schr\"{o}dinger equation and their relevance to the PT-symmetric system.
        {\it Chaos.} 2019;29:123124.
\bibitem{SMMC} Sarma AK, Miri MA, Musslimani ZH, Christodoulides DN.
        Continuous and discrete Schr\"{o}dinger systems with parity-time-symmetric nonlinearities.
        {\it Phys Rev E.} 2014;89:052918.
\bibitem{AM-2014} Ablowitz MJ, Musslimani ZH.
        Integrable discrete PT symmetric model.
        {\it Phys Rev E.} 2014;90:1--5.
\bibitem{Ger} Gerdjikov VS.
        On the Integrability of Ablowitz-Ladik models with local and nonlocal reductions.
        {\it J Phys Conf Ser.} 2019;1205:012015.
\bibitem{DLZ-AMC} Deng X, Lou SY, Zhang DJ.
        Bilinearisation-reduction approach to the nonlocal discrete nonlinear Schr\"{o}dinger equations.
        {\it Appl Math Comput.} 2018;332:477--483.
\bibitem{MZ-JMP} Ma LY, Zhu ZN.
        Nonlocal nonlinear Schr\"{o}dinger equation and its discrete version: Soliton solutions and gauge equivalence.
        {\it J Math Phys.} 2016;57:083507.
\bibitem{MSZ-AML} Ma LY, Shen SF, Zhu ZN.
        From discrete nonlocal nonlinear Schr\"{o}dinger equation to coupled discrete Heisenberg ferromagnet equation.
        {\it Appl Math Lett.} 2022;130:108002.
\bibitem{N-2004-math} Hietarinta J, Joshi N, Nijhoff FW.
        {\it Discrete Systems and Integrability.}
        Cambridge: Cambridge University Press; 2016.
\bibitem{ZKZ} Zhang DD, Van Der Kamp PH, Zhang DJ.
        Multi-component extension of CAC systems.
        {\it SIGMA} 2020;16:060.
\bibitem{Bri} Bridgman T, Hereman W, Quispel GRW, van der Kamp PH.
        Symbolic computation of Lax pairs of partial difference equations using consistency around the cube.
        {\it Found Comput Math.} 2013;13:517--544.
\bibitem{XFZ-TMPH} Xiang XB, Feng W, Zhao SL.
        Local and nonlocal complex discrete sine-Gordon equation. Solutions and continuum limits.
        {\it Theor Math Phys.} 2022;211:758--774.
\bibitem{XZ-TMPH} Xu HJ, Zhao SL.
        Cauchy matrix solutions to some local and nonlocal complex equations.
        {\it Theor Math Phys.} 2022;213:1513--1542.
\bibitem{Yu-SAPM} Yu GF.
		Discrete analogues of a negative order AKNS equation.
		{\it Stud Appl Math.} 2015;135:117--138.
\bibitem{ZJZ-PD} Zhang DJ, Ji J, Zhao SL.
        Soliton scattering with amplitude changes of a negative order AKNS equation.
        {\it Physica D.} 2009;238:2361--2367.
\bibitem{KK} Kakuhata H, Konno K.
        A generalization of coupled integrable, dispersionless system.
        {\it J Phys Soc Japan.} 1996;65:340--341.
\bibitem{AKNS-1973} Ablowitz MJ, Kaup DJ, Newell AC, Segur H.
        Nonlinear-evolution equations of physical significance.
        {\it Phys Rev Lett.} 1973;31:125--127.
\bibitem{AKNS-1974} Ablowitz MJ, Kaup DJ, Newell AC, Segur H.
        The inverse scattering transform-Fourier analysis for nonlinear problems.
        {\it Stud Appl Math.} 1974;54:249--315.
\bibitem{FN-1983} Freeman NC, Nimmo JJC.
        Soliton solutions of the KdV and KP equations: The Wronskian technique.
        {\it Phys Lett A.} 1983;95:1--3.
\bibitem{Hall} Hall BC.
        {\it Lie groups, Lie algebras, and representations: An elementary introduction, Graduate Texts in Mathematics.}
        Berlin: Springer-Verlag; vol. 222 (2nd ed.); 2015.
\bibitem{SRH} Sahadevan R, Rasin OG, Hydon PE.
        Integrability conditions for nonautonomous quad-graph equations.
        {\it J Math Anal Appl.} 2007;331:712--726.
\bibitem{GR} Grammaticos B, Ramani A.
        Singularity confinement property for the (non-autonomous) Adler-Bobenko-Suris integrable lattice equations.
        {\it Lett Math Phys.} 2010;92:33--45.
\bibitem{Syl} Sylvester J.
        Sur l'equation en matrices $px=xq$.
        {\it C. R. Acad. Sci. Paris.} 1884;99:67--71, 115--116.
\bibitem{BR} Bhatia R, Rosenthal P.
        How and why to solve the operator equation $AX-XB=Y$.
        {\it Bull London Math Soc.} 1997;29:1--21.
\bibitem{ZZSZ} Zhang DJ, Zhao SL, Sun YY, Zhou J.
        Solutions to the modified Korteweg-de Vries equation (review).
        {\it Rev Math Phys.} 2014;26:14300064.
\bibitem{LWZ-SAPM} Liu SZ, Wang J, Zhang DJ.
        The Fokas-Lenells equations: Bilinear approach.
        {\it Stud Appl Math.} 2022;148:651--688.
\bibitem{HT-JPSJ} Sakovicha A, Sakovich S.
        The short pulse equation is integrable.
        {\it J Phys Soc Japan.} 2005;74:239--241.
\bibitem{USP} Sch\"{a}fer T, Wayne CE.
        Propagation of ultra-short optical pulses in nonlinear media.
        {\it Physica D.} 2004;196:90--105.
\bibitem{CLZ-SP} Chen K, Liu SM, Zhang DJ.
        Covariant hodograph transformations between nonlocal short pulse models and the AKNS(-1) system.
        {\it Appl Math Lett.} 2019;88:230--236.
\bibitem{DSP} Feng BF, Maruno K, Ohta Y.
        Integrable discretizations of the short pulse equation.
        {\it J Phys A: Math Theor.} 2010;43:085203.
\bibitem{Hirota-2000} Hirota R.
        Discretization of coupled modified KdV equations.
        {\it Chaos Solitons \& Fractals.} 2000;11:77--84.
\bibitem{ZXS} Zhao SL, Xiang XB, Shen SF.
        Solutions and continuum limits to nonlocal discrete modified Korteweg de-Vries equations.
        {\it preprint} 2022.	

\end{thebibliography}}
\end{document}